# Grounding FO and FO(ID) with Bounds

**Johan Wittocx** JOHAN.WITTOCX@CS.KULEUVEN.BE
**Maarten Mariën** MAARTEN.MARIEN@CS.KULEUVEN.BE
**Marc Denecker** MARC.DENECKER@CS.KULEUVEN.BE
*Katholieke Universiteit Leuven*
*Department of Computer Science*
*Celestijnenlaan 200A, 3001 Heverlee, Belgium*

## Abstract

Grounding is the task of reducing a first-order theory and finite domain to an equivalent propositional theory. It is used as preprocessing phase in many logic-based reasoning systems. Such systems provide a rich first-order input language to a user and can rely on efficient propositional solvers to perform the actual reasoning.

Besides a first-order theory and finite domain, the input for grounders contains in many applications also additional data. By exploiting this data, the size of the grounder's output can often be reduced significantly. A common practice to improve the efficiency of a grounder in this context is by manually adding semantically redundant information to the input theory, indicating where and when the grounder should exploit the data. In this paper we present a method to compute and add such redundant information automatically. Our method therefore simplifies the task of writing input theories that can be grounded efficiently by current systems.

We first present our method for classical first-order logic (FO) theories. Then we extend it to FO(ID), the extension of FO with inductive definitions, which allows for more concise and comprehensive input theories. We discuss implementation issues and experimentally validate the practical applicability of our method.

## 1. Introduction

Grounding, or propositionalization, is the task of reducing a first-order theory and finite domain to an equivalent propositional theory, called *a grounding*. Grounding is used as a preprocessing phase in many logic-based reasoning systems. It serves to provide the user with a rich input language, while enabling the system to rely on efficient propositional solvers to perform the actual reasoning.

Examples of systems that rely on grounding can be found in the area of finite first-order model generation (Claessen & Sörensson, 2003; McCune, 2003; East, Iakhiaev, Mikitiuk, & Truszczyński, 2006; Mitchell, Ternovska, Hach, & Mohebali, 2006; Torlak & Jackson, 2007; Wittocx, Mariën, & Denecker, 2008d). Such systems are in turn used as part of theorem provers (Claessen & Sörensson, 2003) and for lightweight software verification (Jackson, 2006). Currently, almost all Answer Set Programming (ASP) systems rely on grounding as a preprocessing phase (Gebser, Schaub, & Thiele, 2007; Perri, Scarcello, Catalano, & Leone, 2007; Syrjänen, 2000; Syrjänen, 2009). Also in planning systems (Kautz & Selman, 1996) and relational data mining (Krogel, Rawles, Zelezný, Flach, Lavrac, & Wrobel, 2003) grounding is frequently used. This large number of applications indicates the importance of grounding in logic-based reasoning systems and the need to develop efficient grounders.

A basic (naive) grounding method is by instantiating the variables in the input theory by all possible combinations of domain elements. Grounding in this way is polynomial in the size of the domain but exponential in the maximum width of a formula in the input theory, and may easily produce groundings of unwieldy size. Several techniques have been developed to efficiently produce smaller groundings. There are two main categories of such techniques. In the first, the input theory is rewritten such that the maximum width of the formulas decreases. Methods like *clause splitting* (Schulz, 2002) and *partitioning* (Ramachandran & Amir, 2005) belong to this category.

The second type of techniques is applicable when besides the finite domain, additional data is available. This is often the case in practical model generation problems, such as the ones that are typical in ASP. In a graph problem the data could be an encoding of the input graph; in the context of planning, it could be a description of the initial and goal state, etc. Sometimes the data is explicitly available, e.g., in the form of a database, sometimes it is implicit, e.g., as a set of ground facts in the input theory. The second type of techniques aims at efficiently computing small groundings by taking the data into account.

Observe that both types of techniques can be combined in a grounder. In this paper we mainly focus on a technique of the second category. To explain the intuition underlying our method, consider the following model generation problem.

**Example 1.** Let $T_1$ the first-order logic theory over the vocabulary $\{Edge, Sub\}$, consisting of the two sentences

$$\forall u \forall v \; (Sub(u,v) \supset Edge(u,v)) \tag{1}$$

$$\forall x \forall y \forall z \; (Sub(x,y) \land Sub(x,z) \supset y = z), \tag{2}$$

$T_1$ expresses that $Sub$ is a subgraph of $Edge$ with at most one outgoing edge in each vertex. Computing such a subgraph of a given graph $G = \langle V, E \rangle$ can be cast as a model generation problem with input theory $T_1$ and data $G$. The data can be represented as a structure $I_\sigma$ for the subvocabulary $\sigma_1 = \{Edge\}$ with domain $V$ and $Edge^{I_\sigma} = E$. A solution can be obtained by generating a model of $T_1$ that expands $I_\sigma$ with an interpretation of $Sub$.

Applying the naive grounding algorithm produces $|V|^2$ instantiations of (1) and $|V|^3$ instantiations of (2). By taking the data into account, atoms over '$Edge$' and '=' can be substituted by their truth value in $I_\sigma$. Simplifying the resulting grounding then eliminates $|E|$ instantiations of (1) and $|V|$ instantiations of (2). Smart grounding algorithms interleave this substitution and simplification with the grounding process in order to avoid creating unnecessary parts of the grounding.

Observe that substituting atoms over $\sigma_1$ and then simplifying still produces a grounding of size $\mathcal{O}(|V|^3)$. Indeed, the simplified grounding of (2) is the set of binary clauses $\neg Sub(i,j) \lor \neg Sub(i,k)$ such that $i,j,k \in V$ and $i \neq j$. This set has size $|V|^3 - |V|$.

Some grounders apply reasoning on the ground theory to reduce it even further. In the example, the simplified grounding of (1) consists of the clauses $\neg Sub(i,j)$ such that $(i,j) \notin E$. Since these are unit clauses, each of them is certainly true in every model of the ground theory. It follows that each binary clauses $\neg Sub(i,j) \lor \neg Sub(i,k)$ such that either $\neg Sub(i,j)$ or $\neg Sub(i,k)$ belongs to the simplified grounding of (1) is certainly true in every model of the ground theory and thus can be omitted from the simplified grounding of (2). The result is a grounding of size $|E \bowtie_{1=1} E|$, where $\bowtie_{1=1}$ denotes the natural join matching the first columns. For a sparse graph, $|E \bowtie_{1=1} E|$ is much smaller than $|V|^3$. However, since reasoning on the ground theory does not avoid creating all instantiations of a formula, it does not significantly speed up the grounding process.

One way to avoid a large grounding without relying on reasoning on the ground theory is by adding redundant information to formulas. This method is frequently used in ASP. For example,

$$\forall x \forall y \forall z (Edge(x,y) \land Sub(x,y) \land Edge(x,z) \land Sub(x,z) \supset y = z) \tag{3}$$

is equivalent to (2) given (1), but its grounding (without reasoning on the ground theory) is equal to the one obtained by the kind of reasoning on the ground theory illustrated above. This illustrates how adding redundant information may sometimes dramatically reduce the size of the grounding. Since current grounders are optimized to ground formulas like (3) without trying all instances, grounding may also speed up a lot.

However, manually adding redundancy to formulas has its disadvantages: it leads to more complex and hence, less readable theories. Worse, it might introduce errors. It requires a good understanding of the used grounder, since it depends on the grounder what information is beneficial to add and where. Also, a human developer could easily miss useful information.

The above motivates a study of automated methods for deriving such redundant information and of principled ways of adding it to formulas. We develop an algorithm that, given a model generation problem with input theory $T$ and input data $I_\sigma$, derives such redundant information, in the form of a pair of a *symbolic upper and lower bound* for each subformula of $T$. Each of these bounds is a formula over the vocabulary of $I_\sigma$. For instance, for Example 1, our algorithm will compute $Edge(x, y)$ as upper bound for $Sub(x, y)$, meaning that if $Edge(x, y)$ is not true, then $Sub(x, y)$ is not true either. We also show how to insert these bounds in the formulas of $T$. For example, inserting the upperbound $Edge(x, y)$ for $Sub(x, y)$ and the upperbound $Edge(x, z)$ for $Sub(x, z)$ transforms (2) into (3).

The rest of this paper is organized as follows. In the next section we recall some notions from first-order logic (FO) and we introduce the notations used throughout the paper. In Section 3 we formally define grounding and model generation with additional data. In Section 4 we introduce upper- and lowerbounds for formulas. We present an any-time algorithm to compute them in the context of FO input theories. We show how the bounds can be used to rewrite the input theory to an equivalent theory that has a smaller grounding.

Although many search problems can be cast concisely and naturally as FO model generation problems, some problems require richer logics than FO. One such logic is FO(ID), an extension of FO with inductive definitions. Such definitions can be used to represent, e.g., the concept of reachability in a graph. In Section 5 we extend our rewriting method to FO(ID).

In Section 6 we discuss how to implement our algorithm to compute bounds. As a case study, we show for one particular grounding algorithm how it can be adapted to exploit bounds directly. We also present experimental results that indicate the impact of our method on grounding size and time. We end with related work and conclusions.

The current paper extends our previous work (Wittocx, Mariën, & Denecker, 2008c). Besides proofs for all main propositions and a more thorough experimental validation, also the following parts were added:

- The theoretical result stating that our rewriting method certainly yields smaller groundings (Proposition 23);
- The extension of the rewriting method to FO(ID) (Section 5);
- The section about implementation issues (Section 6).

## 2. Preliminaries

In this section, we introduce the conventions and notations used in this paper. We assume the reader is familiar with FO.

### 2.1 First-Order Logic

A *vocabulary* $\Sigma$ is a tuple $\langle \Sigma_P, \Sigma_F, \Sigma_V \rangle$ where $\Sigma_P$, $\Sigma_F$ and $\Sigma_V$ are respectively sets of predicate symbols, function symbols and variables. We identify *constants* with zero-arity function symbols. Abusing notation, we will often leave out $\Sigma_V$ and simply write $\langle \Sigma_P, \Sigma_F \rangle$ to represent $\Sigma$. A vocabulary $\sigma$ is a subvocabulary of $\Sigma$, denoted $\sigma \subseteq \Sigma$, if $\sigma_P \subseteq \Sigma_P$, $\sigma_F \subseteq \Sigma_F$ and $\sigma_V \subseteq \Sigma_V$.

Throughout this paper variables are denoted by lowercase letters, predicate and function symbols by uppercase letters. Each predicate and function symbol has an associated arity $n \in \mathbb{N}$. We often denote a predicate symbol $P$ by $P/n$ and a function symbol $F$ by $F/n$ to indicate their arities.

Tuples and sets of variables are denoted by $\overline{x}$, $\overline{y}$, $\overline{z}$. A *term* over $\Sigma$ is inductively defined by

- A variable $x \in \Sigma$ is a term;
- If $F/n$ is a function symbol of $\Sigma$ and $t_1, \ldots, t_n$ are terms over $\Sigma$, then $F(t_1, \ldots, t_n)$ is a term.

Tuples of terms are denoted by $\bar{t}$, $\bar{t}_1$, $\bar{t}_2$, .... A first-order logic formula over $\Sigma$ is inductively defined by

- If $P/n$ is a predicate symbol and $t_1, \ldots, t_n$ are terms, then $P(t_1, \ldots, t_n)$ is a formula.
- If $t_1$ and $t_2$ are two terms, then $t_1 = t_2$ is a formula.
- If $\varphi$ and $\psi$ are formulas and $x$ is a variable, then $\neg\varphi$, $\varphi \wedge \psi$, $\varphi \vee \psi$, $\exists x\ \varphi$ and $\forall x\ \varphi$ are formulas.

We use $\varphi \supset \psi$, $\varphi \equiv \psi$ and $t_1 \neq t_2$ as a shorthands for respectively $\neg\varphi \vee \psi$, $(\varphi \supset \psi) \wedge (\psi \supset \varphi)$ and $\neg(t_1 = t_2)$. An *atom* is a formula of the form $P(\bar{t})$ or $t_1 = t_2$. A *literal* is an atom or the negation of an atom.

An occurrence of a formula $\varphi$ as subformula in a formula $\psi$ is *positive*, respectively *negative*, if it occurs in the scope of an even, respectively odd, number of negations.

For a formula $\varphi$, we often write $\varphi[\overline{x}]$ to indicate that $\overline{x}$ are its free variables. That is, if $y \in \overline{x}$, then $y$ occurs in $\varphi$, but not in the scope of a quantifier $\forall y$ or $\exists y$ in $\varphi$. For a variable $x$ and a term $t$, the formula $\varphi[x/t]$ denotes the result of replacing all free occurrences of $x$ in $\varphi$ by $t$. This notation is extended to tuples of variables and terms of the same length. A *sentence* is a formula without free variables. A *theory* is a finite set of sentences.

A $\Sigma$-*interpretation* $I$ consists of a domain $D$ and

- a domain element $x^I \in D$ for each variable $x \in \Sigma_V$;
- a relation $P^I \subseteq D^n$ for each predicate symbol $P/n \in \Sigma_P$;
- a function $F^I : D^n \to D$ for each function symbol $F/n \in \Sigma_F$.

A $\Sigma$-*structure* is an interpretation of only the relation and function symbols of $\Sigma$. The *restriction* of a $\Sigma$-interpretation $I$ to a vocabulary $\sigma \subseteq \Sigma$ is denoted by $I|_\sigma$. Vice versa, $I$ is called an *expansion of* $I|_\sigma$ *to* $\Sigma$. For a variable $x$ and domain element $d$, $I[x/d]$ is the interpretation that assigns $d$ to $x$ and corresponds to $I$ on all other symbols. This notation is extended to tuples of variables and domain elements of the same length. An interpretation $I$ is called *finite* if its domain is finite.

The value $t^I$ of a term $t$ in an interpretation $I$, and the satisfaction relation $\models$ are defined as usual (e.g., Enderton, 2001). $I$ is called a *model* of a formula $\varphi$ if $I \models \varphi$. We denote by $T_1 \models T_2$ that every model of theory $T_1$ is also a model of theory $T_2$.

A *query* is an expression of the form $\{\overline{x} \mid \varphi\}$, where the free variables of $\varphi$ are among $\overline{x}$. A tuple $\overline{d}$ of domain elements is an *answer to* $\{\overline{x} \mid \varphi\}$ *in a structure* $I$ if $I[\overline{x}/\overline{d}] \models \varphi$. The set of all answers to $\{\overline{x} \mid \varphi\}$ in $I$ is denoted by $\{\overline{x} \mid \varphi\}^I$.

### 2.2 Rewriting and Term Normal Form

In this paper we will use the following well-known equivalences to rewrite formulas to logically equivalent formulas.

1. Moving quantifiers

$$\forall x \forall y\ \varphi \quad\equiv\quad \forall y \forall x\ \varphi \tag{4}$$
$$\exists x \exists y\ \varphi \quad\equiv\quad \exists y \exists x\ \varphi \tag{5}$$
$$\forall x\ (\varphi \wedge \psi) \quad\equiv\quad (\forall x\ \varphi) \wedge (\forall x\ \psi) \tag{6}$$
$$\exists x\ (\varphi \vee \psi) \quad\equiv\quad (\exists x\ \varphi) \vee (\exists x\ \psi) \tag{7}$$
$$\forall x\ (\varphi \vee \psi) \quad\equiv\quad \varphi \vee (\forall x\ \psi) \qquad \text{if } x \text{ does not occur free in } \varphi \tag{8}$$
$$\exists x\ (\varphi \wedge \psi) \quad\equiv\quad \varphi \wedge (\exists x\ \psi) \qquad \text{if } x \text{ does not occur free in } \varphi \tag{9}$$

2. Moving negations

$$\neg(\varphi \wedge \psi) \equiv (\neg\varphi) \vee (\neg\psi) \tag{10}$$
$$\neg(\varphi \vee \psi) \equiv (\neg\varphi) \wedge (\neg\psi) \tag{11}$$
$$\neg(\forall x\ \varphi) \equiv \exists x\ (\neg\varphi) \tag{12}$$
$$\neg(\exists x\ \varphi) \equiv \forall x\ (\neg\varphi) \tag{13}$$

3. Flattening terms

$$P(t_1, \ldots, t_i, \ldots, t_n) \equiv \exists x\ (x = t_i \wedge P(t_1, \ldots, t_{i-1}, x, t_{i+1}, \ldots, t_n)) \tag{14}$$

where $x$ does not occur in $P(t_1, \ldots, t_n)$.

To facilitate the presentation, we will sometimes require that formulas are in *term normal form* (TNF). We say that a formula $\varphi$ is in TNF, if every atomic subformula of $\varphi$ is of the form $P(\overline{x})$, $F(\overline{x}) = y$ or $x = y$, and all negations occur directly in front of atoms. Using (10)–(14), every formula can be transformed in an equivalent formula in TNF. We say that a theory is in TNF if all its sentences are.

### 2.3 SAT

A vocabulary $\Sigma$ is *propositional* if $\Sigma_F = \emptyset$ and every predicate symbol in $\Sigma_P$ has arity zero. A *propositional theory* (PC theory) is a theory over a propositional vocabulary. A propositional *clause* is a disjunction of propositional literals. A PC theory is in *conjunctive normal form* (CNF) if all its sentences are clauses. The *Boolean satisfiability problem* (SAT) is the **NP**-complete problem of deciding for a PC theory whether it is satisfiable. The **NP** search problem corresponding to a SAT problem is the problem of computing a witness of the decision problem in the form of a model of the theory. SAT solvers typically operate by constructing such a model.

Contemporary SAT solvers exhibit impressive performance. As such, many **NP** problems can be solved efficiently by translating them to SAT. For instance, this is done in the areas of model generation (Claessen & Sörensson, 2003; McCune, 2003), planning (Kautz & Selman, 1996) and relational data mining (Krogel et al., 2003). Most modern SAT solvers expect a CNF theory as input, instead of a general PC theory. When the input is a satisfiable theory, they return a model as a witness to their answer.

## 3. Model Generation and Grounding

Model generation is the problem of computing a model of a logic theory $T$, usually in the context of a given finite domain, typically the Herbrand Universe. A model generator allows to decide the satisfiability of the theory in the context of this fixed domain. This is useful, e.g., in the context of lightweight verification (Jackson, 2006). Beyond determining satisfiability, there is a broad class of problems of which the answers are naturally given by the models of a declarative domain theory. For example, the model of a theory specifying a scheduling domain typically contains a (correct) schedule. Thus, a model generator applied to this theory will solve the scheduling problem for this domain.[1] This idea of model generation as a declarative problem solving paradigm has been pioneered in the area of ASP (Marek & Truszczyński, 1999; Niemelä, 1999). In this area, answers to a problem are given by the models of an ASP theory.

As mentioned in the introduction, many practical model generation problems contain additional data besides the input theory and finite domain. This data can be implicit in the input theory. For

1. For a set of problems of this kind, see, e.g., the benchmarks of the ASP-competition (`http://dtai.cs.kuleuven.be/events/ASP-competition`).

example, ASP problems can be split into two parts: a non-ground theory and a list of ground facts. The latter part essentially represents given data. In other contexts (Mitchell & Ternovska, 2005; Torlak & Jackson, 2007; Wittocx et al., 2008d), the data is given as a (partial) structure interpreting part of the vocabulary of the input theory. In this paper we assume without loss of generality that the data is represented by a structure. In practice, it is often the case that some preprocessing, e.g., materializing a view on a database, needs to be done before the data is in this format (see also Section 5.3.2).

### 3.1 The Model Expansion Search Problem

Model generation with an input theory and input structure is called *model expansion*. Model expansion for a logic $\mathcal{L}$, denoted MX($\mathcal{L}$), is defined as follows.

**Definition 1.** Let $T$ be an $\mathcal{L}$-theory over a vocabulary $\Sigma$, $\sigma$ a subvocabulary of $\Sigma$ and $I_\sigma$ a finite $\sigma$-structure. The *model expansion search problem with input* $\langle T, I_\sigma \rangle$ is the problem of computing a $\Sigma$-structure $M$ such that $M \models T$ and $M|_\sigma = I_\sigma$.

The vocabulary $\sigma$ is called the *input vocabulary* of the problem, the vocabulary $\Sigma \setminus \sigma$ the *expansion vocabulary*. $I_\sigma$ is called the *input structure*. We denote by $M \models_{I_\sigma} T$ that $M$ is a solution to the model expansion search problem with input $\langle T, I_\sigma \rangle$. Similarly, for a formula $\varphi$ over $\Sigma$ we denote by $M \models_{I_\sigma} \varphi$ that $M$ expands $I_\sigma$ to $\Sigma$ and satisfies $\varphi$.

Observe that if $\sigma = \Sigma$, model expansion reduces to model checking, while if $\sigma = \langle \emptyset, \emptyset \rangle$, it reduces to model generation for $T$ with a given finite size. Also, if $T$ is a theory over a vocabulary $\Sigma$ containing no function symbols of arity greater than zero, Herbrand model generation for $T$ can be simulated by model expansion. Indeed, let $\sigma = \langle \emptyset, \Sigma_F \rangle$, and $I_\sigma$ the structure with the Herbrand universe of $T$ such that $C^{I_\sigma} = C$ for every constant $C \in \Sigma_F$.

We illustrate model expansion by two examples. In the examples in this paper, we often use *many-sorted* FO, since this leads to more concise and readable sentences. In many-sorted FO, the domain of an interpretation is partitioned in sorts (or types), each variable has an associated sort, each $n$-ary predicate symbol has an $n$-tuple of associated sorts and each $n$-ary function symbol an associated $(n+1)$-tuple of sorts. If $I$ is an interpretation and variable $x$ has associated sort $s$, then $x^I \in s^I$, where $s^I$ denotes the set of domain elements of sort $s$. Similarly, if $P/n$ has associated sorts $(s_1, \dots, s_n)$, then $P^I \subseteq s_1^I \times \dots \times s_n^I$, if $F/n$ has associated sorts $(s_1, \dots, s_{n+1})$, then $F^I : s_1^I \times \dots \times s_n^I \to s_{n+1}^I$. We often denote $P$ by $P(s_1, \dots, s_n)$ and $F$ by $F(s_1, \dots, s_n) : s_{n+1}$ to indicate their associated sorts.

**Example 2** (Graph Colouring)**.** The graph colouring problem is the problem of colouring a given graph with a given set of colours such that adjacent vertices have different colours. To express this problem in MX(FO), let $Vtx$ and $Col$ be sorts and let $\sigma = \langle \{Edge(Vtx, Vtx)\}, \emptyset \rangle$. The sort $Col$ denotes the given set of colours, the given graph is represented by $Vtx$ and $Edge$. Let $\Sigma$ be the vocabulary $\langle \sigma_P, \{Colour(Vtx) : Col\} \rangle$ and $T$ the theory that consists of the sentence

$$\forall v_1 \forall v_2\ (Edge(v_1, v_2) \supset Colour(v_1) \neq Colour(v_2)).$$

Then model expansion with input theory $T$ and input vocabulary $\sigma$ expresses the graph colouring problem. Indeed, for any $M \models_{I_\sigma} T$, $Colour^M$ is a proper colouring of the graph represented by $I_\sigma$.

**Example 3** (SAT)**.** To represent the SAT problem in MX(FO), let $\sigma$ be a vocabulary containing the two sorts $Atom$ and $Clause$, representing the atoms and the clause of the input CNF theory, and the two predicates $PosIn(Atom, Clause)$ and $NegIn(Atom, Clause)$, to represent the positive, respectively negative, occurrences of atoms in clauses. The theory given by

$$\forall c\ \exists a\ ((PosIn(a, c) \land True(a)) \lor (NegIn(a, c) \land \neg True(a)))$$

over $\Sigma = \langle \sigma_P \cup \{True(Atom)\}, \emptyset \rangle$ expresses the SAT problem: for any $M \models_{I_\sigma} T$, the propositional structure represented by $True^M$ is a model of the CNF theory represented by $I_\sigma$. Indeed, the theory forces that every clause contains at least one true literal.

As shown by Mitchell and Ternovska (2005), it follows from Fagin's (1974) theorem that model expansion for FO *captures* **NP**, in the following sense:

- For any fixed $T$ and $\sigma$ the problem of deciding whether there exists a model of $T$ expanding an input structure $I_\sigma$ is in **NP**.
- Vice versa, for any **NP** decision problem $X$ on the class of finite $\sigma$-structures there is a vocabulary $\Sigma \supseteq \sigma$ and a first-order $\Sigma$-theory $T$ such that model expansion with input theory $T$ *expresses* $X$, i.e., $I_\sigma$ belongs to $X$ iff there exists a $\Sigma$-structure $M$ such that $M \models_{I_\sigma} T$.

This result proves that any **NP** problem $X$ can be expressed by an MX(FO) problem, and hence shows the broad applicability of MX(FO) solvers to solve **NP** problems.

As illustrated by the examples above, it is the intention that the theory $T$ is an intuitive representation of a problem $X$. Not all **NP** problems can be represented in a natural manner in MX(FO). For instance, the problem of deciding whether a graph is connected can be expressed in MX(FO), but this requires a non-trivial encoding of a fixpoint operator in FO. Model expansion for richer logics than FO is better suited for such problems. In Section 5 we consider MX for FO(ID), an extension of FO with inductive definitions.

### 3.2 Reducing MX(FO) to SAT

For the rest of this paper, let $T$ be a theory over a vocabulary $\Sigma$, $\sigma$ a subvocabulary of $\Sigma$ and $I_\sigma$ a finite $\sigma$-structure with domain $D$.

Since for every FO theory $T$, deciding whether $T$ has a model expanding $I_\sigma$ is in **NP**, this problem can be reduced to a SAT problem $T_{\text{prop}}$ in polynomial time. However, if we want to find models of $T$ expanding $I_\sigma$ by using a SAT solver, we need a method to translate models of $T_{\text{prop}}$ into models of $T$. Moreover, if we are interested in finding *all* models of $T$ expanding $I_\sigma$, a one-to-one correspondence between these models and the models of $T_{\text{prop}}$ is needed. In this paper we focus on reductions that preserve all models, which is the setting in the ASP paradigm (Marek & Truszczyński, 1999; Niemelä, 1999).

Let $\tau$ be the vocabulary of $T_{\text{prop}}$. To have a one-to-one correspondence between the models of $T$ expanding $I_\sigma$ and the models of $T_{\text{prop}}$, it should be possible to represent $\Sigma$-structures expanding $I_\sigma$ by $\tau$-structures. The most natural way to accomplish this is by choosing $\tau$ such that it contains a symbol $P_{\overline{d}}$ for every $P/n \in \Sigma_P$ and $\overline{d} \in D^n$, and a symbol $F_{\overline{d},d'}$ for every $F/n \in \Sigma_F$ and $(\overline{d}, d') \in D^{n+1}$. A $\tau$-structure making $P_{\overline{d}}$, respectively $F_{\overline{d},d'}$ true then corresponds to a $\Sigma$ structure $M$ such that $\overline{d} \in P^M$, respectively $F^M(\overline{d}) = d'$. In this manner, every $\Sigma$-structure expanding $I_\sigma$ has a corresponding $\tau$-structure. Vice versa, every $\tau$-structure $A$ satisfying the requirement that for every function symbol $F/n$ and $\overline{d} \in D^n$, there is exactly one $d' \in D$ such that $F_{\overline{d},d'}$ is true in $A$, corresponds to a $\Sigma$-structure with the same domains as $I_\sigma$. That is, there is a one-to-one correspondence between the $\tau$-structures satisfying for every function symbol $F/n$ and $\overline{d} \in D^n$ the formula

$$\left( \bigvee_{d' \in D} F_{\overline{d},d'} \right) \wedge \left( \bigwedge_{d'_1 \in D} \left( \bigwedge_{d'_2 \in D \setminus d'_1} \neg F_{\overline{d},d'_1} \vee \neg F_{\overline{d},d'_2} \right) \right) \tag{15}$$

and the $\Sigma$-structures with domain $D$.

Denote by $\Sigma^{\text{dom}(I_\sigma)}$ the vocabulary $\Sigma$ extended with a new constant symbol $\mathbf{d}$ for every $d \in D$. We call these new constants *domain constants.* Abusing notation, we will denote both domain elements and their corresponding domain constants by $d$. For a formula $\varphi[\overline{x}]$ and a tuple $\overline{d}$ of domain constants, we call $\varphi[\overline{x}/\overline{d}]$ an *instance of* $\varphi$. For a $\Sigma$-interpretation $M$ expanding $I_\sigma$ and a formula $\varphi$ containing domain constants, we denote by $M \models \varphi$ that the expansion of $M$ to $\Sigma^{\mathrm{dom}(I_\sigma)}$ defined by interpreting every domain constant by its corresponding domain element, satisfies $\varphi$.

**Definition 2.** Two formulas $\varphi_1$ and $\varphi_2$ over $\Sigma^{\mathrm{dom}(I_\sigma)}$ are *$I_\sigma$-equivalent* if $M \models_{I_\sigma} \varphi_1$ iff $M \models_{I_\sigma} \varphi_2$, for every $\Sigma$-interpretation $M$.

The following are some straightforward results about $I_\sigma$-equivalence.

**Lemma 3.** *1. Two logically equivalent formulas are $I_\sigma$-equivalent.*

*2. $\bigwedge_{d \in D} \varphi[x/d]$ is $I_\sigma$-equivalent to $\forall x\ \varphi[x]$.*

*3. $\bigvee_{d \in D} \varphi[x/d]$ is $I_\sigma$-equivalent to $\exists x\ \varphi[x]$.*

*4. If $\varphi'$ and $\psi'$ are $I_\sigma$-equivalent to respectively $\varphi$ and $\psi$, then $\neg\varphi'$, $\varphi' \land \psi'$, $\varphi' \lor \psi'$, $\exists x\ \varphi'$ and $\forall x\ \varphi'$ are $I_\sigma$-equivalent to respectively $\neg\varphi$, $\varphi \land \psi$, $\varphi \lor \psi$, $\exists x\ \varphi$ and $\forall x\ \varphi$.*

*5. If $\psi$ is a subformula of $\varphi$ and is $I_\sigma$-equivalent to $\psi'$, then the result of replacing $\psi$ by $\psi'$ in $\varphi$ is $I_\sigma$-equivalent to $\varphi$.*

A formula is in *ground normal form* (GNF) if it contains no quantifiers and all its atomic subformulas are of the form $P(d_1, \dots, d_n)$, $F(d_1, \dots, d_n) = d$ or $d_1 = d_2$, where $d_1, \dots, d_n, d$ are domain constants. A theory is in GNF if all its sentences are in GNF. A GNF theory is essentially propositional: by replacing in a GNF theory $T$ every atom $P(\overline{d})$ by $P_{\overline{d}}$, $F(\overline{d}) = d'$ by $F_{\overline{d},d'}$, $d_i = d_j$ by $\top$ or $\bot$ if, respectively, $i = j$ or $i \neq j$, and adding the formula (15) for every function symbol $F/n$ and $\overline{d} \in D^n$, we obtain a propositional theory $T_{\mathrm{prop}}$ such that the models of $T$ and $T_{\mathrm{prop}}$ correspond. Also note the similarity between GNF and TNF theories.

**Definition 4.** A *grounding for $T$ with respect to $I_\sigma$* is a GNF theory $T_{\mathrm{g}}$ over $\Sigma^{\mathrm{dom}(I_\sigma)}$ such that $T$ and $T_{\mathrm{g}}$ are $I_\sigma$-equivalent. $T_{\mathrm{g}}$ is called *reduced* if it does not contain symbols of $\sigma$.

#### 3.2.1 Grounding Algorithms

For the rest of this section, we assume that $T$ is a theory in TNF. As explained in Section 2.2, we can make this assumption without loss of generality. Below we introduce, as a reference, the grounding for $T$ with respect to $I_\sigma$ obtained by the naive grounding algorithm mentioned in the introduction. We call this grounding the *full grounding* and define it formally by induction.

**Definition 5.** The *full grounding* $\mathrm{Gr}_{\mathrm{full}}(\varphi, I_\sigma)$ of a TNF sentence $\varphi$ with respect to $I_\sigma$ is defined by

$$\mathrm{Gr}_{\mathrm{full}}(\varphi) = \begin{cases} \varphi & \text{if } \varphi \text{ is a literal} \\ \mathrm{Gr}_{\mathrm{full}}(\psi_1) \land \mathrm{Gr}_{\mathrm{full}}(\psi_2) & \text{if } \varphi \text{ is equal to } \psi_1 \land \psi_2 \\ \mathrm{Gr}_{\mathrm{full}}(\psi_1) \lor \mathrm{Gr}_{\mathrm{full}}(\psi_2) & \text{if } \varphi \text{ is equal to } \psi_1 \lor \psi_2 \\ \bigwedge_{d \in D} \mathrm{Gr}_{\mathrm{full}}(\psi[x/d]) & \text{if } \varphi \text{ is equal to } \forall x\ \psi[x] \\ \bigvee_{d \in D} \mathrm{Gr}_{\mathrm{full}}(\psi[x/d]) & \text{if } \varphi \text{ is equal to } \exists x\ \psi[x] \end{cases} \tag{16}$$

The *full grounding for $T$ with respect to $I_\sigma$* is the theory consisting of the full groundings of all sentences in $T$ with respect to $I_\sigma$.

We denote the full grounding by $\mathrm{Gr}_{\mathrm{full}}(T, I_\sigma)$, or by $\mathrm{Gr}_{\mathrm{full}}(T)$ if $I_\sigma$ is clear from the context. It follows directly from Lemma 3 that $\mathrm{Gr}_{\mathrm{full}}(T, I_\sigma)$ is indeed a grounding for $T$ with respect to $I_\sigma$. The size of the full grounding is exponential in the maximal nesting depth of quantifiers in sentences of $T$, and polynomial in the domain size of $I_\sigma$.

An inductive definition like (16) can be evaluated in a top-down or bottom-up way. Both approaches are applied in current grounders. On the one hand, there are grounders that go top-down through the syntax trees of the sentences in $T$. When a subformula $\varphi$ of the form $\forall x\ \psi[x]$, respectively $\exists x\ \psi[x]$ is reached, the grounding of $\psi[x/d]$ is constructed for every domain constant $d$, and then $\varphi$ is replaced by the conjunction, respectively disjunction, of all these groundings. The grounder of the DLV system (Perri et al., 2007) and the grounders GRINGO (Gebser et al., 2007) and GidL (Wittocx, Mariën, & Denecker, 2008b) take this approach.

Other grounders go bottom-up through the syntax trees. For each subformula $\varphi[\overline{x}]$ a table is computed consisting of tuples $\overline{d}$ and corresponding groundings of $\varphi[\overline{x}/\overline{d}]$. These tables are computed first for atomic formulas and subsequently for compound formulas. For example, let $\varphi[x, y, z]$ be the formula $\psi[x, y] \land \chi[y, z]$ and assume the tables for $\psi$ and $\chi$ have been computed. Then the table for $\varphi$ is computed by taking the natural join of the tables for $\psi$ and $\chi$ on the value for $y$, and constructing the grounding for $\varphi[x/d_x, y/d_y, z/d_z]$ as the (possibly simplified) conjunction of the groundings for $\psi[x/d_x, y/d_y]$ and $\chi[y/d_y, z/d_z]$. Examples of grounders with a bottom-up approach are LPARSE (Syrjänen, 2000; Syrjänen, 2009), KODKOD (Torlak & Jackson, 2007) and MXG (Mitchell et al., 2006).

To obtain a reduced grounding for $T$ with respect to $I_\sigma$ one could first construct the full grounding and then replace every subformula $\varphi$ over $\sigma^{\mathrm{dom}(I_\sigma)}$ in it by $\top$ if $I_\sigma \models \varphi$ and by $\bot$ otherwise. The result can further be simplified by recursively replacing $\bot \land \psi$ by $\bot$, $\top \land \psi$ by $\psi$, etc. The resulting grounding is the one computed by most current grounding algorithms and is often a lot smaller than the full grounding. We denote it by $\mathrm{Gr}_{\mathrm{red}}(T, I_\sigma)$, or by $\mathrm{Gr}_{\mathrm{red}}(T)$ if $I_\sigma$ is clear from the context.

Smart grounding algorithms do not use the approach outlined above, but try to avoid creating the full grounding by substituting ground formulas over the input vocabulary $\sigma$ as soon as possible. For example, a grounder with a top-down approach constructs the grounding of $\forall x\ \psi[x]$, by grounding all instances $\psi[x/d]$ one by one and then making the conjunction. During this process, all instances $\psi[x/d]$ that are detected to be certainly true are omitted. As soon as an instance $\psi[x/d]$ is detected to be certainly false, $\bot$ is returned as grounding for $\forall x\ \psi[x]$.

A grounder using the bottom-up approach can reduce the size of the tables it computes by not storing tuples that have some default value, e.g., $\top$, as corresponding grounding. In particular, if $\varphi[\overline{x}]$ is a formula over $\sigma$, it only stores the tuples $\overline{d}$ such that $I_\sigma \not\models \varphi[\overline{x}/\overline{d}]$. By reducing the size of the tables in this way, the reduced grounding can be obtained much more efficiently.

## 4. Grounding with Bounds

In this section we present our method for reducing grounding size. As mentioned in the introduction, it is based on computing bounds for subformulas of the input theory $T$. Each bound for a subformula $\varphi[\overline{x}]$ is a formula over the input vocabulary $\sigma$. It describes a set of tuples $\overline{d}$ for which $\varphi[\overline{x}/\overline{d}]$ is certainly true (false) in every model of $T$ expanding any $I_\sigma$. The larger the set described by a bound, the more precise the bound is. Observe that the fact that bounds are formulas over $\sigma$ means that they can be evaluated using the given structure $I_\sigma$.

In Section 4.1, we formally define bounds. Then we indicate how bounds can be inserted in $T$ to obtain a new theory $T'$. The reduced grounding of $T'$ is often a lot smaller than the reduced grounding of $T$. The more precise the inserted bounds are, the smaller the grounding of $T'$ becomes. However, we will see that $T'$ is in general weaker than $T$ and that additional axioms have to be added to $T'$ to obtain equivalence with $T$. These additional axioms need to be grounded as well so that, if we are not careful, the total size of the grounded theory does not decrease at all. In Section 4.3, we search for sufficient conditions on the bounds to guarantee a smaller grounding.

In Section 4.4, we show how to derive bounds. Our method works in two stages. First, bounds for all subformulas of $T$ are computed using an any-time algorithm. The longer the algorithm runs, the more precise bounds are derived. Often, the bounds derived at this stage do not lead to smaller groundings, for the reason explained in the previous paragraph. In the second stage, bounds that satisfy the conditions to guarantee smaller groundings are derived from the ones computed in the first stage.

### 4.1 Bounds

We distinguish between two kinds of bounds.

**Definition 6.** A *certainly true bound (ct-bound) over $\sigma$ with respect to $T$ for a formula $\varphi[\overline{x}]$* is a formula $\varphi_{\mathrm{ct}}[\overline{y}]$ over $\sigma$ such that $\overline{y} \subseteq \overline{x}$ and $T \models \forall \overline{x}\ (\varphi_{\mathrm{ct}}[\overline{y}] \supset \varphi[\overline{x}])$. Vice versa, a *certainly false bound (cf-bound) over $\sigma$ with respect to $T$ for* $\varphi[\overline{x}]$ is a formula $\varphi_{\mathrm{cf}}[\overline{z}]$ over $\sigma$ such that $\overline{z} \subseteq \overline{x}$ and $T \models \forall \overline{x}\ (\varphi_{\mathrm{cf}}[\overline{z}] \supset \neg\varphi[\overline{x}])$.

We do not mention $\sigma$ and $T$ if they are clear from the context.

Intuitively, a ct-bound $\varphi_{\mathrm{ct}}$ for $\varphi[\overline{x}]$ provides for every structure $I_\sigma$ a lower bound for the set of tuples for which $\varphi$ is true in every model of $T$ expanding $I_\sigma$. Indeed, for every $M \models_{I_\sigma} T$ we have that $\{\overline{x} \mid \varphi_{\mathrm{ct}}\}^{I_\sigma} \subseteq \{\overline{x} \mid \varphi\}^M$. Vice versa, a cf-bound $\varphi_{\mathrm{cf}}$ provides a lower bound on the set of tuples for which $\varphi$ is false: $\{\overline{x} \mid \varphi_{\mathrm{cf}}\}^{I_\sigma} \subseteq \{\overline{x} \mid \neg\varphi\}^M$ for every $M \models_{I_\sigma} T$. Observe that the negation of a ct-bound, respectively cf-bound, gives an upper bound on the set of tuples for which $\varphi$ is false, respectively true, in at least one model of $T$ expanding $I_\sigma$.

**Example 4** (Example 1 ctd.)**.** Let $\varphi_1$ be the subformula $\mathit{Sub}(x,y) \wedge \mathit{Sub}(x,z)$ of $T_1$. Then $\neg \mathit{Edge}(x,y) \vee \neg \mathit{Edge}(x,z)$ is a cf-bound over $\sigma_1$ with respect to $T_1$ for $\varphi_1$. Indeed, one can derive from (1) that $T_1$ entails

$$\forall x \forall y \forall z\ \left((\neg \mathit{Edge}(x,y) \vee \neg \mathit{Edge}(x,z)) \supset \neg\varphi_1\right).$$

Observe that $\top$ is a ct-bound for every *sentence* of $T$. Indeed, for every sentence $\varphi$ of $T$, $T \models \varphi$ and therefore $T \models \top \supset \varphi$. Also, $\bot$ is a ct-bound as well as a cf-bound for every *formula*. We call $\bot$ the *trivial bound*. Intuitively, the trivial bound contains no information at all: $\{\overline{x} \mid \bot\}^{I_\sigma} = \emptyset$ for every $I_\sigma$ and $\overline{x}$. According to the following definition, it is the least precise bound.

**Definition 7.** Let $\psi[\overline{y}]$ and $\chi[\overline{z}]$ be two (ct- or cf-) bounds for $\varphi[\overline{x}]$. We say that $\psi[\overline{y}]$ *is more precise than* $\chi[\overline{z}]$ if $\forall \overline{x}\ (\chi[\overline{z}] \supset \psi[\overline{y}])$ is valid.

If $\psi$ is a more precise bound for $\varphi[\overline{x}]$ than $\chi$, $\psi$ provides a larger lower bound because $\{\overline{x} \mid \chi\}^{I_\sigma} \subseteq \{\overline{x} \mid \psi\}^{I_\sigma}$ for every $I_\sigma$.

**Definition 8.** A *c-map* $\mathcal{C}$ for $T$ over $\sigma$ is a mapping from all subformulas $\varphi$ of $T$ to tuples $(\mathcal{C}^{\mathrm{ct}}(\varphi), \mathcal{C}^{\mathrm{cf}}(\varphi))$, where $\mathcal{C}^{\mathrm{ct}}(\varphi)$ and $\mathcal{C}^{\mathrm{cf}}(\varphi)$ are respectively a ct- and cf-bound for $\varphi$ over $\sigma$ with respect to $T$.

The notion of precision pointwise extends to c-maps. That is, if $\mathcal{C}_1$ and $\mathcal{C}_2$ are two c-maps for $T$, then $\mathcal{C}_1$ is more precise than $\mathcal{C}_2$ iff for every subformula $\varphi$ of $T$, $\mathcal{C}_1^{\mathrm{ct}}(\varphi)$ is more precise than $\mathcal{C}_2^{\mathrm{ct}}(\varphi)$ and $\mathcal{C}_1^{\mathrm{cf}}(\varphi)$ is more precise than $\mathcal{C}_2^{\mathrm{cf}}(\varphi)$.

Let $M$ be a model of $T$ and $\mathcal{C}$ a c-map for $T$ over $\sigma$. From the definition of ct- and cf-bounds it follows immediately that for every subformula $\varphi[\overline{x}]$ of $T$, both $M \models \forall \overline{x}\ (\mathcal{C}^{\mathrm{ct}}(\varphi) \supset \varphi)$ and $M \models \forall \overline{x}\ (\mathcal{C}^{\mathrm{cf}}(\varphi) \supset \neg\varphi)$ hold. We say that a structure *satisfies* $\mathcal{C}$ if it has precisely this property.

**Definition 9.** Let $\mathcal{C}$ be a c-map for $T$ over $\sigma$. Then the theory $\overline{\mathcal{C}}$ is defined by

$$\begin{aligned}\overline{\mathcal{C}} = &\{\forall \overline{x}\ (\mathcal{C}^{\mathrm{ct}}(\varphi) \supset \varphi) \mid \varphi[\overline{x}] \text{ is a subformula of } T\}\\ &\cup \{\forall \overline{x}\ (\mathcal{C}^{\mathrm{cf}}(\varphi) \supset \neg\varphi) \mid \varphi[\overline{x}] \text{ is a subformula of } T\}.\end{aligned}$$

A structure $I$ *satisfies* $\mathcal{C}$ if $I \models \overline{\mathcal{C}}$.

Clearly, if $\mathcal{C}$ is a c-map for $T$ over $\sigma$ and $M \models_{I_\sigma} T$, then $M \models \overline{\mathcal{C}}$. We call two formulas $\varphi[\overline{x}]$ and $\psi[\overline{x}]$ *$\mathcal{C}$-equivalent* if $\{\overline{x} \mid \varphi\}^I = \{\overline{x} \mid \psi\}^I$ for each structure $I$ that satisfies $\mathcal{C}$. Equivalently, $\varphi$ and $\psi$ are $\mathcal{C}$-equivalent if $\overline{\mathcal{C}} \models \forall \overline{x}\ (\varphi \equiv \psi)$.

A c-map is *inconsistent* if some formula $\varphi$ is both certainly true and false for some tuple, according to that c-map:

**Definition 10.** A c-map $\mathcal{C}$ for $T$ over $\sigma$ is *inconsistent* if $\exists \overline{x}\ (\mathcal{C}^{\mathrm{ct}}(\varphi) \wedge \mathcal{C}^{\mathrm{cf}}(\varphi))$ is valid for some subformula $\varphi[\overline{x}]$ of $T$. A c-map $\mathcal{C}$ is *$I_\sigma$-inconsistent* if $I_\sigma \models \exists \overline{x}\ (\mathcal{C}^{\mathrm{ct}}(\varphi) \wedge \mathcal{C}^{\mathrm{cf}}(\varphi))$ for some subformula of $T$.

**Proposition 11.** *If there exists an $I_\sigma$-inconsistent c-map for $T$ over $\sigma$, then $M \not\models_{I_\sigma} T$ for every $M$. If there exists an inconsistent c-map for $T$ over $\sigma$, then $M \not\models_{I_\sigma} T$ for every $M$ and $I_\sigma$.*

*Proof.* Let $\mathcal{C}$ be an $I_\sigma$-inconsistent c-map for $T$ over $\sigma$ and $\varphi[\overline{x}]$ a subformula of $T$ such that $I_\sigma \models \exists \overline{x}\ (\mathcal{C}^{\mathrm{ct}}(\varphi) \wedge \mathcal{C}^{\mathrm{cf}}(\varphi))$. Then there exists a tuple of domain elements $\overline{d}$ such that $I_\sigma[\overline{x}/\overline{d}] \models \mathcal{C}^{\mathrm{ct}}(\varphi)$ and $I_\sigma[\overline{x}/\overline{d}] \models \mathcal{C}^{\mathrm{cf}}(\varphi)$. Assume towards a contradiction that $M \models_{I_\sigma} T$. Then $M \models \overline{\mathcal{C}}$, and hence $M[\overline{x}/\overline{d}] \models \mathcal{C}^{\mathrm{ct}}(\varphi) \supset \varphi$ and $M[\overline{x}/\overline{d}] \models \mathcal{C}^{\mathrm{cf}}(\varphi) \supset \neg\varphi$. Since $M|_\sigma = I_\sigma$, it follows that $M[\overline{x}/\overline{d}] \models \varphi$ and $M[\overline{x}/\overline{d}] \models \neg\varphi$. This is a contradiction.

To prove the second statement, let $\mathcal{C}$ be an inconsistent c-map for $T$ over $\sigma$. Then $\mathcal{C}$ is a also an $I_\sigma$-inconsistent c-map for every $\sigma$-structure $I_\sigma$. As such, for any $I_\sigma$ there is no model of $T$ expanding $I_\sigma$. □

### 4.2 C-Transformation

For the rest of this section, fix a c-map $\mathcal{C}$ for $T$ over $\sigma$. We now show how to insert the bounds of $\mathcal{C}$ into the sentences of $T$. This insertion is based on the following lemma.

**Lemma 12.** *Let $\varphi[\overline{x}]$ be a subformula of $T$. Then $\varphi$ is $\mathcal{C}$-equivalent to $\varphi \vee \mathcal{C}^{\mathrm{ct}}(\varphi)$ and to $\varphi \wedge \neg\mathcal{C}^{\mathrm{cf}}(\varphi)$.*

*Proof.* We have to prove that $\overline{\mathcal{C}} \models \forall \overline{x}\ (\varphi \equiv (\varphi \vee \mathcal{C}^{\mathrm{ct}}(\varphi)))$ and $\overline{\mathcal{C}} \models \forall \overline{x}\ (\varphi \equiv (\varphi \wedge \neg\mathcal{C}^{\mathrm{cf}}(\varphi)))$. The former immediately follows from the fact that $\overline{\mathcal{C}} \models \forall \overline{x}\ (\mathcal{C}^{\mathrm{ct}}(\varphi) \supset \varphi)$, the latter from the fact that $\overline{\mathcal{C}} \models \forall \overline{x}\ (\mathcal{C}^{\mathrm{cf}}(\varphi) \supset \neg\varphi)$ . □

As a corollary of lemma 12 we have the following lemma.

**Lemma 13.** *Let $\psi$ be a sentence of $T$ and $\varphi$ a subformula of $\psi$. If $\psi'$ is the result of replacing the subformula $\varphi$ in $\psi$ by $\varphi \vee \mathcal{C}^{\mathrm{ct}}(\varphi)$, by $\varphi \wedge \neg\mathcal{C}^{\mathrm{cf}}(\varphi)$ or by $(\varphi \wedge \neg\mathcal{C}^{\mathrm{cf}}(\varphi)) \vee \mathcal{C}^{\mathrm{ct}}(\varphi)$, then $M \models \psi$ iff $M \models \psi'$ for every $M$ that satisfies $\mathcal{C}$.*

Observe that if $\mathcal{C}^{\mathrm{ct}}(\varphi) = \mathcal{C}^{\mathrm{cf}}(\varphi) = \bot$, then both $\varphi \vee \mathcal{C}^{\mathrm{ct}}(\varphi)$ and $\varphi \wedge \neg\mathcal{C}^{\mathrm{cf}}(\varphi)$ are logically equivalent to $\varphi$. Hence, in this case the sentence $\psi'$ in Lemma 13 is essentially the sentence $\psi$. Intuitively, adding trivial bounds to a sentence $\psi$ does not change the sentence at all.

The bounds assigned by $\mathcal{C}$ can be "inserted" in $T$ by applying the transformation of Lemma 13 to all subformulas of $T$. The result is called a c-transformation of $T$, and is formally defined as follows.

**Definition 14** (c-transformation)**.** A *c-transformation* of a subformula $\varphi$ of $T$ with respect to $\mathcal{C}$, denoted $\mathcal{C}\langle\varphi\rangle$, is the formula $(\varphi' \wedge \neg\mathcal{C}^{\mathrm{cf}}(\varphi)) \vee \mathcal{C}^{\mathrm{ct}}(\varphi)$ where $\varphi'$ is defined by

$$\varphi' := \begin{cases} \varphi & \text{if } \varphi \text{ is an atom} \\ \neg\mathcal{C}\langle\psi\rangle & \text{if } \varphi \text{ is equal to } \neg\psi \\ \mathcal{C}\langle\psi\rangle \wedge \mathcal{C}\langle\chi\rangle & \text{if } \varphi \text{ is equal to } \psi \wedge \chi \\ \mathcal{C}\langle\psi\rangle \vee \mathcal{C}\langle\chi\rangle & \text{if } \varphi \text{ is equal to } \psi \vee \chi \\ \exists x\ \mathcal{C}\langle\psi\rangle & \text{if } \varphi \text{ is equal to } \exists x\ \psi \\ \forall x\ \mathcal{C}\langle\psi\rangle & \text{if } \varphi \text{ is equal to } \forall x\ \psi \end{cases}$$

A *c-transformation* $\mathcal{C}\langle T\rangle$ *of* $T$ *with respect to* $\mathcal{C}$ consists of a c-transformation with respect to $\mathcal{C}$ of every sentence of $T$.

From Lemma 13, we derive the following.

**Lemma 15.** *$T$ and $\mathcal{C}\langle T\rangle$ are $\mathcal{C}$-equivalent.*

In general $T$ and $\mathcal{C}\langle T\rangle$ are not logically equivalent. $\mathcal{C}\langle T\rangle$ may have models that do not satisfy $\mathcal{C}$, and therefore cannot be models of $T$. For example, let $\mathcal{C}$ be the c-map that assigns $(\top, \bot)$ to every sentence and $(\bot, \bot)$ to every other subformula of $T$. Then all sentences in $\mathcal{C}\langle T\rangle$ are of the form $\varphi \vee \top$ and hence $\mathcal{C}\langle T\rangle$ simplifies to $\top$, which is in general not equivalent to $T$. To obtain from $\mathcal{C}\langle T\rangle$ a theory that is equivalent to $T$, we must add $\overline{\mathcal{C}}$.

**Theorem 16.** *If $\mathcal{C}$ is a c-map for $T$ over $\sigma$ and $\overline{\mathcal{C}}$ the theory defined in Definition 9, then $\mathcal{C}\langle T\rangle \cup \overline{\mathcal{C}}$ is equivalent to $T$.*

*Proof.* Let $M$ be a model of $T$. Then $M \models \overline{\mathcal{C}}$, and because of Lemma 13, $M \models \mathcal{C}\langle T\rangle \cup \overline{\mathcal{C}}$. On the other hand, if $M \models \mathcal{C}\langle T\rangle \cup \overline{\mathcal{C}}$, then by Lemma 13, $M \models T$. $\square$

**Corollary 17.** *If $\mathcal{C}$ is a c-map for $T$ over $\sigma$, then $T$ and $\mathcal{C}\langle T\rangle \cup \overline{\mathcal{C}}$ are $I_\sigma$-equivalent for any $\sigma$-structure $I_\sigma$.*

### 4.3 Atom-Based and Atom-Equal C-Maps

Corollary 17 implies that we can compute a grounding for $T$ with respect to $I_\sigma$ by first computing a c-map $\mathcal{C}$ for $T$ over $\sigma$ and then grounding $\mathcal{C}\langle T\rangle \cup \overline{\mathcal{C}}$. This approach is beneficial if the reduced grounding of $\mathcal{C}\langle T\rangle \cup \overline{\mathcal{C}}$ is smaller than the reduced grounding of $T$, and can be constructed at least as fast. In general these conditions are not satisfied. The more precise c-map $\mathcal{C}$ is, the smaller the reduced grounding of $\mathcal{C}\langle T\rangle$ becomes, but the larger the reduced grounding of $\overline{\mathcal{C}}$ is:

**Proposition 18.** *If $\mathcal{C}_1$ is more precise than $\mathcal{C}_2$, then $Gr_{red}(\mathcal{C}_1\langle T\rangle)$ is smaller than $Gr_{red}(\mathcal{C}_2\langle T\rangle)$. Moreover, every subformula that occurs in $Gr_{red}(\mathcal{C}_1\langle T\rangle)$ also occurs in $Gr_{red}(\mathcal{C}_2\langle T\rangle)$.*

*Proof.* (Sketch) Let $\varphi[\overline{x}]$ be a subformula of $T$ and $\overline{d}$ a tuple of domain elements. It suffices to show that if $\mathcal{C}_2\langle\varphi\rangle[\overline{x}/\overline{d}]$ is replaced by $\top$, respectively $\bot$, when grounding, then this is also the case for $\mathcal{C}_1\langle\varphi\rangle[\overline{x}/\overline{d}]$. This can be proven by induction. For the base case, assume $\varphi$ is an atom. Then $\mathcal{C}_2\langle\varphi\rangle[\overline{x}/\overline{d}]$ is the formula $((\varphi \wedge \neg\mathcal{C}_2^{\mathrm{cf}}(\varphi)) \vee \mathcal{C}_2^{\mathrm{ct}}(\varphi))[\overline{x}/\overline{d}]$. If this formula is replaced by $\top$ or $\bot$ when grounding, there are three possibilities: $\varphi$ is a formula over $\sigma$, $I_\sigma[\overline{x}/\overline{d}] \models \mathcal{C}_2^{\mathrm{ct}}(\varphi)$ or $I_\sigma[\overline{x}/\overline{d}] \models \mathcal{C}_2^{\mathrm{cf}}(\varphi)$. Since $\mathcal{C}_1$ is more precise than $\mathcal{C}_2$, $I_\sigma[\overline{x}/\overline{d}] \models \mathcal{C}_2^{\mathrm{ct}}(\varphi)$ implies $I_\sigma[\overline{x}/\overline{d}] \models \mathcal{C}_1^{\mathrm{ct}}(\varphi)$ and $I_\sigma[\overline{x}/\overline{d}] \models \mathcal{C}_2^{\mathrm{cf}}(\varphi)$ implies $I_\sigma[\overline{x}/\overline{d}] \models \mathcal{C}_1^{\mathrm{cf}}(\varphi)$. We conclude that if $\mathcal{C}_2\langle\varphi\rangle[\overline{x}/\overline{d}]$ is replaced by $\top$ or $\bot$ when grounding, then this is also the case for $\mathcal{C}_1\langle\varphi\rangle[\overline{x}/\overline{d}]$. The inductive case is similar. $\square$

**Proposition 19.** *If $\mathcal{C}_1$ is more precise than $\mathcal{C}_2$, then $Gr_{red}(\overline{\mathcal{C}_1})$ is larger than $Gr_{red}(\overline{\mathcal{C}_2})$.*

*Proof.* (Sketch) Every sentence in $\overline{\mathcal{C}_1}$ is of the form $\forall\overline{x}\ (\mathcal{C}_1^{\mathrm{ct}}(\varphi) \supset \varphi)$ or $\forall\overline{x}\ (\mathcal{C}_1^{\mathrm{cf}}(\varphi) \supset \neg\varphi)$. The number of instances of $\mathcal{C}_1^{\mathrm{ct}}(\varphi) \supset \varphi$ in the reduced grounding of $\overline{\mathcal{C}_1}$ is equal to the number of $\overline{d}$ such that $I_\sigma[\overline{x}/\overline{d}] \models \mathcal{C}_1^{\mathrm{ct}}(\varphi)$. Similarly for $\mathcal{C}_1^{\mathrm{cf}}(\varphi) \supset \neg\varphi$. Since $\mathcal{C}_2$ is less precise than $\mathcal{C}_1$, the number of instances in $\mathrm{Gr}_{\mathrm{red}}(\overline{\mathcal{C}_2})$ of the corresponding sentences $\forall\overline{x}\ (\mathcal{C}_2^{\mathrm{ct}}(\varphi) \supset \varphi)$ and $\forall\overline{x}\ (\mathcal{C}_2^{\mathrm{cf}}(\varphi) \supset \neg\varphi)$ is smaller. $\square$

A c-map that is useful to reduce grounding size should therefore not be too precise, in order to avoid a large theory $\mathrm{Gr}_{\mathrm{red}}(\overline{\mathcal{C}})$, but still be precise enough to decrease the size of $\mathrm{Gr}_{\mathrm{red}}(\mathcal{C}\langle T\rangle)$. In this section, we present sufficient conditions to ensure these properties. We first define a class of c-maps that "avoid" a blow-up of $\mathrm{Gr}_{\mathrm{red}}(\overline{\mathcal{C}})$ by ensuring $\overline{\mathcal{C}}$ can be replaced by an equivalent, smaller and easy-to-find theory $\overline{\mathcal{C}}_A$. As such, $\mathrm{Gr}_{\mathrm{red}}(\overline{\mathcal{C}})$ can be replaced by the smaller theory $\mathrm{Gr}_{\mathrm{red}}(\overline{\mathcal{C}}_A)$. In the class we present, $\overline{\mathcal{C}}_A$ is a subset of $\overline{\mathcal{C}}$, namely the set of sentences in $\overline{\mathcal{C}}$ that stem from the atomic subformulas of $T$:

**Definition 20.** Define the theory $\overline{\mathcal{C}}_A$ by

$$\begin{aligned}\overline{\mathcal{C}}_A = &\{\forall \overline{x}\ (\mathcal{C}^{\mathrm{ct}}(\varphi) \supset \varphi)\ \mid\ \varphi[\overline{x}] \text{ is an atomic subformula of } T\}\\ &\cup \{\forall \overline{x}\ (\mathcal{C}^{\mathrm{cf}}(\varphi) \supset \neg\varphi)\ \mid\ \varphi[\overline{x}] \text{ is an atomic subformula of } T\}.\end{aligned}$$

We call $\mathcal{C}$ *atom-based* if $\overline{\mathcal{C}}_A \models \overline{\mathcal{C}}$.

**Example 5** (Example 1 ctd.)**.** Let $\mathcal{C}_2$ be the c-map that assigns $(\bot, \neg(Edge(x,y) \wedge Edge(x,z)))$ to $Sub(x,y) \wedge Sub(x,z)$ and $(\bot,\bot)$ to every other subformula. $\mathcal{C}_2$ is not atom-based, since $(\overline{\mathcal{C}}_2)_A$ is equivalent to $\top$, while $\overline{\mathcal{C}}_2$ contains the sentence

$$\forall x \forall y \forall z\ (\neg(Edge(x,y) \wedge Edge(x,z)) \supset \neg(Sub(x,y) \wedge Sub(x,z))). \tag{17}$$

Let $\mathcal{C}_3$ be the c-map that assigns $(\bot, \neg Edge(x,y))$ to $Sub(x,y)$, $(\bot, \neg Edge(x,z))$ to $Sub(x,z)$ and corresponds to $\mathcal{C}_2$ on all other subformulas of $T_1$. $\mathcal{C}_3$ is atom-based. Indeed, $(\overline{\mathcal{C}}_3)_A$ consists of the (equivalent) sentences

$$\forall x \forall y\ (\neg Edge(x,y) \supset \neg Sub(x,y)) \tag{18}$$

$$\forall x \forall z\ (\neg Edge(x,z) \supset \neg Sub(x,z)) \tag{19}$$

and $\overline{\mathcal{C}}_3$ consists of the sentences (17), (18) and (19). Both (18) and (19) imply (17), and therefore, $(\overline{\mathcal{C}}_3)_A \models \overline{\mathcal{C}}_3$.

Clearly, a c-map assigning $(\bot,\bot)$ to every non-atomic subformula of $T$ is an example of an atom-based c-map. As such, any c-map can be transformed into an atom-based one by replacing every bound assigned to a non-atomic subformula by $\bot$. In the next section, we show how to compute more interesting atom-based c-maps.

Observe that $\mathrm{Gr}_{\mathrm{red}}(\overline{\mathcal{C}}_A)$ contains only unit clauses. Combining the definition of atom-based c-map and Theorem 16 immediately gives the following result.

**Proposition 21.** *Let $\mathcal{C}$ be an atom-based c-map for $T$ over $\sigma$. Then $T$ and $\mathcal{C}\langle T\rangle \cup \overline{\mathcal{C}}_A$ are equivalent, and hence $I_\sigma$-equivalent for every $\sigma$-structure $I_\sigma$.*

To obtain small groundings using bounds, it is important that the information in the bounds is exploited wherever possible. In particular, if a ct- or cf-bound $\psi$ is assigned to an atom $P(\overline{x})$, then a similar bound should be assigned to every other atom of the form $P(\overline{y})$. We call a c-map *atom-equal* if it has exactly this property for all atomic subformulas of $T$. That is, $\mathcal{C}$ is atom-equal if it assigns essentially the same bounds to atomic subformulas over the same predicate or function symbol:

**Definition 22.** A c-map $\mathcal{C}$ for a TNF theory $T$ over $\sigma$ is *atom-equal* if for every predicate symbol $P/n$ there exist formulas $\varphi_P^{\mathrm{ct}}[x_1,\dots,x_n]$ and $\varphi_P^{\mathrm{cf}}[x_1,\dots,x_n]$ such that for every atom $P(y_1,\dots,y_n)$ that occurs in $T$, $\mathcal{C}^{\mathrm{ct}}(P(y_1,\dots,y_n))$ is equal to $\varphi_P^{\mathrm{ct}}[x_1/y_1,\dots,x_n/y_n]$ and $\mathcal{C}^{\mathrm{cf}}(P(y_1,\dots,y_n))$ is equal to $\varphi_P^{\mathrm{cf}}[x_1/y_1,\dots,x_n/y_n]$, and similarly for function symbols.

Note that if no predicate or function symbol occurs more than once in a theory $T$, then every c-map for $T$ is atom-equal.

**Example 6** (Example 1 ctd.)**.** Let $T_2$ be the theory obtained by adding the sentence $\exists w\ Sub(w,w)$ to $T_1$. The only predicate that occurs more than once in $T_2$ is the predicate $Sub$. Let $\mathcal{C}_4$ be a c-map for $T_2$ that assigns the following bounds to the atomic subformulas of $T_2$ over $Sub$: $(\bot, \neg Edge(u,v))$ to $Sub(u,v)$, $(\bot, \neg Edge(x,y))$ to $Sub(x,y)$, $(\bot, \neg Edge(x,z))$ to $Sub(x,z)$ and $(\bot, \neg Edge(w,w))$ to $Sub(w,w)$. Then $\mathcal{C}_4$ is atom-equal. Indeed, if we take $\varphi_{Sub}^{\mathrm{ct}} = \bot$ and $\varphi_{Sub}^{\mathrm{cf}} = \neg Edge(x_1,x_2)$, then the conditions of Definition 22 are satisfied for predicate $Sub$.

For an atom-equal c-map $\mathcal{C}$, $\overline{\mathcal{C}}_A$ in general contains many equivalent sentences. For example, for the c-map $\mathcal{C}_4$ as in Example 6, $(\overline{\mathcal{C}}_4)_A$ contains amongst others, the equivalent sentences (18) and (19). It also contains $\forall w\ \neg Edge(w,w) \supset \neg Sub(w,w)$, which is implied by (18). As a result, if $\mathcal{C}$ is an atom-equal c-map, grounding $\overline{\mathcal{C}}_A$ in a naive way yields a grounding that contains several formulas more than once. In the following proposition, we assume this redundancy is removed. In other words, we assume a grounding algorithm for $\overline{\mathcal{C}}_A$ that never adds the same GNF formula more than once to the grounding. This can be accomplished by grounding instead of $\overline{\mathcal{C}}_A$ the sentences $\forall \overline{x}\ (\varphi_P^{\mathrm{ctb}} \supset P(\overline{x}))$ and $\forall \overline{x}\ (\varphi_P^{\mathrm{cfb}} \supset \neg P(\overline{x}))$ for every predicate symbol $P$, where $\varphi_P^{\mathrm{ctb}}$ and $\varphi_P^{\mathrm{cfb}}$ are as in Definition 22, and similarly for function symbols.

**Proposition 23.** *Let $\mathcal{C}$ be an atom-based, atom-equal c-map for a TNF theory $T$. If $T$ has a model expanding $I_\sigma$, then $Gr_{red}(\mathcal{C}\langle T\rangle \cup \overline{\mathcal{C}}_A)$ is at most as large as $Gr_{red}(T)$.*

In the proof, we denote the size of a theory $T_{\mathrm{g}}$ by $|T_{\mathrm{g}}|$.

*Proof.* The outline of this proof is as follows. First, we show that every subformula that occurs in $\mathrm{Gr}_{\mathrm{red}}(\mathcal{C}\langle T\rangle)$, occurs in $\mathrm{Gr}_{\mathrm{red}}(T)$. Then, we prove that no atom occurring in $\mathrm{Gr}_{\mathrm{red}}(\overline{\mathcal{C}}_A)$ occurs in $\mathrm{Gr}_{\mathrm{red}}(\mathcal{C}\langle T\rangle)$. Next, we show that every atom occurring in $\mathrm{Gr}_{\mathrm{red}}(\overline{\mathcal{C}}_A)$ occurs at least once in $\mathrm{Gr}_{\mathrm{red}}(T)$. Since we assumed $\mathrm{Gr}_{\mathrm{red}}(\overline{\mathcal{C}}_A)$ does not contain any formula more than once, it follows that $|\mathrm{Gr}_{\mathrm{red}}(\mathcal{C}\langle T\rangle)| \leq |\mathrm{Gr}_{\mathrm{red}}(T)| - |\mathrm{Gr}_{\mathrm{red}}(\overline{\mathcal{C}}_A)|$, which concludes the proof.

We can directly apply Proposition 18 to show that every subformula of $\mathrm{Gr}_{\mathrm{red}}(\mathcal{C}\langle T\rangle)$ occurs in $\mathrm{Gr}_{\mathrm{red}}(T)$: if $\mathcal{C}'$ is the trivial c-map, then $\mathrm{Gr}_{\mathrm{red}}(T)$ is equal to $\mathrm{Gr}_{\mathrm{red}}(\mathcal{C}'\langle T\rangle)$, and clearly $\mathcal{C}$ is more precise than $\mathcal{C}'$.

We now show that none of the atoms occurring in $\mathrm{Gr}_{\mathrm{red}}(\overline{\mathcal{C}}_A)$ occur in $\mathrm{Gr}_{\mathrm{red}}(\mathcal{C}\langle T\rangle)$. Let $P(\overline{d})$ be an atom occurring in $\mathrm{Gr}_{\mathrm{red}}(\mathcal{C}\langle T\rangle)$. Then there is an atomic subformula $P(\overline{x})$ of $T$ such that $\overline{d} \notin \{\overline{x} \mid \mathcal{C}^{\mathrm{ct}}(P(\overline{x}))\}^{I_\sigma}$ and $\overline{d} \notin \{\overline{x} \mid \mathcal{C}^{\mathrm{cf}}(P(\overline{x}))\}^{I_\sigma}$. Because $\mathcal{C}$ is atom-equal, it follows that for any subformula $P(\overline{y})$ occurring in $T$, neither $\overline{d} \in \{\overline{y} \mid \mathcal{C}^{\mathrm{ct}}(P(\overline{y}))\}^{I_\sigma}$ nor $\overline{d} \in \{\overline{y} \mid \mathcal{C}^{\mathrm{cf}}(P(\overline{y}))\}^{I_\sigma}$. Therefore $P(\overline{d})$ does not occur in $\mathrm{Gr}_{\mathrm{red}}(\overline{\mathcal{C}}_A)$.

It remains to show that every atom that occurs in $\mathrm{Gr}_{\mathrm{red}}(\overline{\mathcal{C}}_A)$ also occurs in $\mathrm{Gr}_{\mathrm{red}}(T)$. Let $M$ be a model of $\mathrm{Gr}_{\mathrm{red}}(T)$. Such a model exists because we assumed that $T$ has a model expanding $I_\sigma$. Let $P(\overline{d})$ be an atom that does not occur in $\mathrm{Gr}_{\mathrm{red}}(T)$. If $P$ is a predicate of the input vocabulary, then $P(\overline{d})$ does not occur in $\mathrm{Gr}_{\mathrm{red}}(\overline{\mathcal{C}}_A)$ either. If on the other hand, $P$ is in the expansion vocabulary, then the structure $M'$ obtained from $M$ by swapping the truth value of $P(\overline{d})$ is also a model of $\mathrm{Gr}_{\mathrm{red}}(T)$. Since $\mathrm{Gr}_{\mathrm{red}}(\mathcal{C}\langle T\rangle \cup \overline{\mathcal{C}}_A)$ is $I_\sigma$-equivalent to $\mathrm{Gr}_{\mathrm{red}}(T)$ and $P \notin \sigma$, it follows that $M \models \mathrm{Gr}_{\mathrm{red}}(\overline{\mathcal{C}}_A)$ and $M' \models \mathrm{Gr}_{\mathrm{red}}(\overline{\mathcal{C}}_A)$. Because $\mathrm{Gr}_{\mathrm{red}}(\overline{\mathcal{C}}_A)$ only contains unit clauses, we conclude that $P(\overline{d})$ does not occur in $\mathrm{Gr}_{\mathrm{red}}(\overline{\mathcal{C}}_A)$. □

We now have the following algorithm to create a small grounding for $T$ with respect to $I_\sigma$: first compute an atom-based, atom-equal c-map $\mathcal{C}$ for $T$ over $\sigma$ (We will present an algorithm for this in Section 4.4). If $\mathcal{C}$ is $I_\sigma$-inconsistent, output $\bot$ and stop. Else, output $\mathrm{Gr}_{\mathrm{red}}(\mathcal{C}\langle T\rangle \cup \overline{\mathcal{C}}_A)$.

It follows from Propositions 11 and 21 that the result of this algorithm is indeed a grounding for $T$ with respect to $I_\sigma$. Observe that the first step of this algorithm is independent of $I_\sigma$. If one has to solve several model expansion problems with a fixed input theory $T$ and input vocabulary $\sigma$, but varying $I_\sigma$, it suffices to compute $\mathcal{C}$ only once.

To perform the last step of the algorithm, one could apply any off-the-shelf grounder on input $\mathcal{C}\langle T\rangle \cup \overline{\mathcal{C}}_A$.

### 4.4 Computing Bounds

We now present an algorithm to compute a (non-trivial) c-map $\mathcal{C}$. It is based on our work on approximate reasoning for FO (Wittocx, Mariën, & Denecker, 2008a). In general the resulting c-map is neither atom-based nor atom-equal, but an atom-based, atom-equal c-map can be derived from it.

### 4.4.1 Refining C-Maps

Constructing a non-trivial c-map can be done by starting from the least precise c-map, i.e., the one that assigns $(\bot, \bot)$ to every subformula of $T$, and then gradually refining it. Each refinement step consists of three operations:

1. Choose a subformula $\varphi$ of $T$.
2. Compute from the current c-map $\mathcal{C}$ a new ct-bound $\varphi^r_{\mathrm{ct}}$ or cf-bound $\varphi^r_{\mathrm{cf}}$ for $\varphi$. Below, we elaborate on this step: we present six different ways to obtain new ct- or cf-bounds, called *refinement bounds*, from $T$ and $\mathcal{C}$. If the sentences of $T$ are represented by their "syntax trees", each node corresponds to a subformula of $T$. *Bottom-up refinement bounds* are bounds for a node computed by considering the bounds assigned by $\mathcal{C}$ to its children. Vice versa, *top-down refinement bounds* are computed by looking at the parents and siblings of a node. *Axiom refinement bounds* are bounds for the roots, i.e., for the sentences of $T$, while *input*, *copy* and *functional refinement bounds* are in practice mainly bounds for atomic subformulas of $T$.
3. Substitute $\mathcal{C}^{\mathrm{ct}}(\varphi)$ by $\mathcal{C}^{\mathrm{ct}}(\varphi) \vee \varphi^r_{\mathrm{ct}}$, respectively $\mathcal{C}^{\mathrm{cf}}(\varphi)$ by $\mathcal{C}^{\mathrm{cf}}(\varphi) \vee \varphi^r_{\mathrm{cf}}$.

According to the following lemma, a refinement step yields a new bound for $\varphi$ that is more precise than the one assigned by $\mathcal{C}$.

**Lemma 24.** *If $\psi$ and $\chi$ are two ct-bounds for $\varphi$ with respect to $T$, then $\psi \vee \chi$ is also a ct-bound for $\varphi$. Moreover, $\psi \vee \chi$ is more precise than $\psi$ and more precise than $\chi$. The same holds for cf-bounds.*

*Proof.* Let $\psi$ and $\chi$ be two ct-bounds for $\varphi[\overline{x}]$. By definition, $T \models \forall \overline{x}\ (\psi \supset \varphi)$ and $T \models \forall \overline{x}\ (\chi \supset \varphi)$. Therefore $T \models \forall \overline{x}\ ((\psi \vee \chi) \supset \varphi)$, which proves that $\psi \vee \chi$ is a ct-bound for $\varphi$. Since $\models \psi \supset (\psi \vee \chi)$ and $\models \chi \supset (\psi \vee \chi)$, $\psi \vee \chi$ is a more precise bound than $\psi$ and $\chi$. The proof for cf-bounds is similar. □

We conclude that repeatedly applying refinement steps leads to a more and more precise c-map. The resulting algorithm is an any-time algorithm. In Section 6 we will discuss a stop criterion for the algorithm. We will also give examples where it can reach a fixpoint, and examples where it cannot.

We now present the different ways to obtain refinement bounds.

**Input Refinement** Let $\varphi[\overline{x}]$ be a formula over the input vocabulary $\sigma$. Since $T \models \forall \overline{x}\ (\varphi[\overline{x}] \supset \varphi[\overline{x}])$ and $T \models \forall \overline{x}\ (\neg\varphi[\overline{x}] \supset \neg\varphi[\overline{x}])$, it is clear that $\varphi[\overline{x}]$ is a ct-bound and $\neg\varphi[\overline{x}]$ a cf-bound for $\varphi[\overline{x}]$. We call these *input* refinement ct- and cf-bounds.

**Axiom Refinement** If $\varphi$ is a sentence of $T$, then $\top$ is an *axiom* refinement ct-bound for $\varphi$. This refinement bound states that a sentence of $T$ is true in every model of $T$.

**Bottom-Up Refinement** For a compound subformula $\varphi$, depending on its structure, Table 1 gives the *bottom-up* refinement ct-bound $\varphi^r_{\mathrm{ct}}$ and cf-bound $\varphi^r_{\mathrm{cf}}$ for $\varphi$ with respect to $\mathcal{C}$. It is rather straightforward to obtain these formulas. For instance, the formula in the bottom-right of the table indicates that if $\varphi$ is the formula $\psi \vee \chi$, then $\varphi$ is certainly false for those tuples for which both $\psi$ and $\chi$ are certainly false. Or, more formally, if both $T \models \mathcal{C}^{\mathrm{cf}}(\psi) \supset \neg\psi$ and $T \models \mathcal{C}^{\mathrm{cf}}(\chi) \supset \neg\chi$, then $T \models \mathcal{C}^{\mathrm{cf}}(\psi) \wedge \mathcal{C}^{\mathrm{cf}}(\chi) \supset \neg(\psi \vee \chi)$.

**Top-Down Refinement** In the case of *top-down* refinements, the bounds of a formula $\psi$ are used to construct refinement bounds for one of its direct subformulas $\varphi$ (i.e., $\varphi$ is one of $\psi$'s children in the syntax tree). The top-down refinement ct-bounds $\varphi^r_{\mathrm{ct}}$ and cf-bounds $\varphi^r_{\mathrm{cf}}$ for $\varphi$ are given in Table 2. In this table, the tuple $\overline{y}$ denotes the free variables of $\psi$ that do not occur in $\varphi$ and $x'$ denotes a new variable. We illustrate some of these refinement bounds. For further explanation why these bounds are in a certain sense the most precise ones that can be obtained, we refer to our work on approximate reasoning (Wittocx et al., 2008a).

| $\varphi$ | $\varphi^r_{\mathrm{ct}}$ | $\varphi^r_{\mathrm{cf}}$ |
|---|---|---|
| $\neg\psi$ | $\mathcal{C}^{\mathrm{cf}}(\psi)$ | $\mathcal{C}^{\mathrm{ct}}(\psi)$ |
| $\forall x\ \psi$ | $\forall x\ \mathcal{C}^{\mathrm{ct}}(\psi)$ | $\exists x\ \mathcal{C}^{\mathrm{cf}}(\psi)$ |
| $\exists x\ \psi$ | $\exists x\ \mathcal{C}^{\mathrm{ct}}(\psi)$ | $\forall x\ \mathcal{C}^{\mathrm{cf}}(\psi)$ |
| $\psi \land \chi$ | $\mathcal{C}^{\mathrm{ct}}(\psi) \land \mathcal{C}^{\mathrm{ct}}(\chi)$ | $\mathcal{C}^{\mathrm{cf}}(\psi) \lor \mathcal{C}^{\mathrm{cf}}(\chi)$ |
| $\psi \lor \chi$ | $\mathcal{C}^{\mathrm{ct}}(\psi) \lor \mathcal{C}^{\mathrm{ct}}(\chi)$ | $\mathcal{C}^{\mathrm{cf}}(\psi) \land \mathcal{C}^{\mathrm{cf}}(\chi)$ |

Table 1: Bottom-up refinement bounds

| $\psi$ | $\varphi^r_{\mathrm{ct}}$ |
|---|---|
| $\neg\varphi$ | $\mathcal{C}^{\mathrm{cf}}(\psi)$ |
| $\forall x\ \varphi$ | $\mathcal{C}^{\mathrm{ct}}(\psi)$ |
| $\exists x\ \varphi$ | $\mathcal{C}^{\mathrm{ct}}(\psi) \land \forall x'\ (x \neq x' \supset \mathcal{C}^{\mathrm{cf}}(\varphi)[x/x'])$ |
| $\varphi \land \chi$ or $\chi \land \varphi$ | $\exists \overline{y}\ \mathcal{C}^{\mathrm{ct}}(\psi)$ |
| $\varphi \lor \chi$ or $\chi \lor \varphi$ | $\exists \overline{y}\ (\mathcal{C}^{\mathrm{ct}}(\psi) \land \mathcal{C}^{\mathrm{cf}}(\chi))$ |

| $\psi$ | $\varphi^r_{\mathrm{cf}}$ |
|---|---|
| $\neg\varphi$ | $\mathcal{C}^{\mathrm{ct}}(\psi)$ |
| $\forall x\ \varphi$ | $\mathcal{C}^{\mathrm{cf}}(\psi) \land \forall x'\ (x \neq x' \supset \mathcal{C}^{\mathrm{ct}}(\varphi)[x/x'])$ |
| $\exists x\ \varphi$ | $\mathcal{C}^{\mathrm{cf}}(\psi)$ |
| $\varphi \land \chi$ or $\chi \land \varphi$ | $\exists \overline{y}\ (\mathcal{C}^{\mathrm{cf}}(\psi) \land \mathcal{C}^{\mathrm{ct}}(\chi))$ |
| $\varphi \lor \chi$ or $\chi \lor \varphi$ | $\exists \overline{y}\ \mathcal{C}^{\mathrm{cf}}(\psi)$ |

Table 2: Top-down refinement bounds

Let $\psi$ be the formula $\forall x\ P(x,y)$. Recall that intuitively, the ct-bound $\mathcal{C}^{\text{ct}}(\psi)$ indicates for which domain elements $d$, $\forall x\ P(x,d)$ is certainly true. For such a $d$ and an arbitrary $d' \in D$, $P(d',d)$ must be true. Hence, $\mathcal{C}^{\text{ct}}(\psi)$ is a ct-bound for $\varphi$. Indeed, since $x$ does not occur free in $\mathcal{C}^{\text{ct}}(\psi)$, $T \models \forall x \forall y\ (\mathcal{C}^{\text{ct}}(\psi) \supset P(x,y))$ follows from $T \models \forall y\ (\mathcal{C}^{\text{ct}}(\psi) \supset \forall x\ P(x,y))$.

Now let $\psi$ be the formula $P(x) \wedge Q(x,y)$. If we know that $P(d_1) \wedge Q(d_1,d_2)$ is certainly false, but $Q(d_1,d_2)$ is certainly true, then $P(d_1)$ must be certainly false. Hence, $\exists y\ \mathcal{C}^{\text{cf}}(\psi) \wedge \mathcal{C}^{\text{ct}}(\chi)$ is a cf-bound for $P(x)$.

Let $\psi$ be the formula $\exists x\ P(x,y)$ and assume that $\exists x\ P(x,d_y)$ is certainly true, but for all $d'_x$, except $d_x$, $P(d'_x,d_y)$ is certainly false. Then we can conclude that $P(d_x,d_y)$ must be true. This is precisely what is expressed by the formula $\mathcal{C}^{\text{ct}}(\psi) \wedge \forall x'\ (x \neq x' \supset \mathcal{C}^{\text{cf}}(\varphi)[x/x'])$.

**Functional Refinement** If $\varphi[\overline{x},y]$ is the formula $F(\overline{x}) = y$, functional refinement bounds for $\varphi$ take into account that $F$ is a function. The functional refinement ct-bound $\varphi^r_{\text{ct}}$ and cf-bound $\varphi^r_{\text{cf}}$ are given by:

$$\varphi^r_{\text{ct}} := \forall y'\ (y' \neq y \supset \mathcal{C}^{\text{cf}}(\varphi)[y/y'])$$

$$\varphi^r_{\text{cf}} := \exists y'\ (\mathcal{C}^{\text{ct}}(\varphi)[y/y'] \wedge y \neq y')$$

where $y'$ is a new variable. Informally, the first of these formulas indicates that $F(\overline{x})$ is certainly equal to $y$ if for every $y' \neq y$, $F(\overline{x})$ is certainly not equal to $y'$. The second one says that $F(\overline{x})$ is certainly not equal to $y$ if $F(\overline{x})$ is certainly equal to $y'$ for some $y' \neq y$.

**Copy Refinement** Let $\varphi[x_1,\dots,x_n]$ and $\psi[y_1,\dots,y_m]$ be two formulas such that $\varphi[x_1/z,\dots,x_n/z]$ and $\psi[y_1/z,\dots,y_m/z]$ are the same, modulo a renaming of their non-free variables. That is, $\varphi$ and $\psi$ have exactly the same syntax tree, but their variables may differ. Denote by $E(\varphi,\psi)$ the set of all equalities $x_i = y_j$ such that for some occurrence of $x_i$ in $\varphi$, $y_j$ occurs in the corresponding position in $\psi$. Then the formula $\exists y_1 \dots \exists y_m\ (\mathcal{C}^{\text{ct}}(\psi) \wedge \bigwedge E(\varphi,\psi))$ is a copy refinement ct-bound for $\varphi$ and the formula $\exists y_1 \dots \exists y_m\ (\mathcal{C}^{\text{cf}}(\psi) \wedge \bigwedge E(\varphi,\psi))$ is a copy refinement cf-bound for $\varphi$. We also say that these are the copy-refinement bounds *from $\psi$ to $\varphi$*.

**Example 7.** Let $\varphi$ be the formula $P(x_1,x_1) \wedge \forall s\ Q(x_2,s)$ and $\psi$ the formula $P(y_1,y_2) \wedge \forall t\ Q(y_2,t)$. Because $\varphi[x_1/z,x_2/z]$ is equal to $\psi[y_1/z,y_2/z]$ modulo the renaming of $s$ by $t$, these formulas satisfy the requirement for copy refinement. The set $E(\varphi,\psi)$ is given by $\{x_1 = y_1, x_1 = y_2, x_2 = y_2\}$ and hence,

$$\exists y_1 \exists y_2\ (\mathcal{C}^{\text{ct}}(\psi) \wedge x_1 = y_1 \wedge x_1 = y_2 \wedge x_2 = y_2)$$

is a copy refinement ct-bound for $\varphi$. Observe that if $\mathcal{C}^{\text{ct}}(\psi)$ does not contain bounded occurrences of $x_1$ or $x_2$, this formula is equivalent to the simpler formula $\mathcal{C}^{\text{ct}}(\psi)[y_1/x_1,y_2/x_1] \wedge x_1 = x_2$.

**One-Step Refinements** We call $\varphi^r_{\text{ct}}$ ($\varphi^r_{\text{cf}}$) a *refinement ct-bound (cf-bound) for $\varphi$ with respect to $\mathcal{C}$* if it is an input, axiom, bottom-up, top-down, functional or copy refinement ct-bound (cf-bound) for $\varphi$ with respect to $\mathcal{C}$. Lemma 25 states that a refinement ct-bound (cf-bound) is indeed a ct-bound (cf-bound).

**Lemma 25.** *If $\varphi^r_{\text{ct}}$ is a refinement ct-bound for $\varphi$ with respect to $\mathcal{C}$, then it is a ct-bound for $\varphi$. Similarly for cf-bounds.*

*Proof.* The proof consists of a simple analysis of all cases. We proved some of the cases when we introduced input, bottom-up and top-down refinement. The proof of the other cases is similar. □

**Definition 26.** Let $\mathcal{C}$ be a c-map for $T$ over $\sigma$, $\varphi$ a subformula of $T$, $\varphi^r_{\text{ct}}$ a refinement ct-bound and $\varphi^r_{\text{cf}}$ a refinement cf-bound for $\varphi$ with respect to $\mathcal{C}$. An assignment $\mathcal{C}^r$ that corresponds to $\mathcal{C}$, except that it assigns $\mathcal{C}^r(\varphi) = (\mathcal{C}^{\text{ct}}(\varphi) \vee \varphi^r_{\text{ct}}, \mathcal{C}^{\text{cf}}(\varphi))$ or $\mathcal{C}^r(\varphi) = (\mathcal{C}^{\text{ct}}(\varphi), \mathcal{C}^{\text{cf}}(\varphi) \vee \varphi^r_{\text{cf}})$ is called a *one-step refinement of $\mathcal{C}$*.

From Lemma 24 and 25 we obtain the following result.

**Proposition 27.** *Every one-step refinement of a c-map for $T$ over $\sigma$ is a c-map for $T$ over $\sigma$.*

As already mentioned at the beginning of this section, one can compute a c-map for $T$ over $\sigma$ by first assigning $(\bot, \bot)$ to every subformula of $T$ and then repeatedly applying one-step refinements. We call this nondeterministic any-time algorithm the *refinement algorithm*.

**Example 8** (Example 1 ctd.)**.** Figure 1 shows a possible run of the refinement algorithm for input $T$ and $\sigma$. Here, the sentences of $T_1$ are represented by their syntax trees. The numbers indicate at which step the bounds are refined. The trivial bounds are not shown.

In step (1), ct-bound $\bot$ for the first sentence is replaced by $\bot \lor \top$ using axiom refinement. Of course, this new bound can be simplified to $\top$. For all following steps, the figure shows simplified bounds. In step (2) and (3) the bounds of subformula $Edge(u, v)$ are refined by input refinement. Then, top-down refinement is used to set the ct-bound of $\neg Sub(u, v) \lor Edge(u, v)$ to $\top$. Next, by top-down refinement, $\neg Edge(u, v)$ becomes the ct-bound for $\neg Sub(u, v)$ and then the cf-bound for $Sub(u, v)$.

In a similar way, the cf-bound $y \neq z$ is derived for subformula $Sub(x, y) \land Sub(x, z)$ (step (7) – (12)). Then, by copy refinement, the cf-bounds for $Sub(x, y)$ becomes $\exists u \exists v\ (\neg Edge(u, v) \land u = x \land v = y)$, wich simplifies to $\neg Edge(x, y)$. Likewise, after simplification, $\neg Edge(x, z)$ is the copy refinement cf-bound for $Sub(x, z)$. Finally, two steps of bottom-up refinement are used to set the ct-bound of $\neg(Sub(x, y) \land Sub(x, z))$ to $y \neq z \lor \neg Edge(x, y) \lor \neg Edge(x, z)$.

At this step, a fixpoint is reached: every one-step refinement that can be performed yields a bound that is logically equivalent to the one it tries to refine.

**Example 9.** Consider a simplified planning problem, where actions should be scheduled such that if an action $a_p$ is a precondition of an action $a_0$, then $a_p$ is performed at an earlier time point than $a_0$. This problem is described by the theory $T_3$, consisting of the sentence

$$\forall a_0 \forall a_p \forall t_0\ Prec(a_p, a_0) \land Do(a_0, t_0) \supset (\exists t_p\ t_p < t_0 \land Do(a_p, t_p)).$$

From this sentence, it follows that if a chain of $i$ actions must be executed before $a_0$ can be executed, then $a_0$ cannot be executed before the $i$th timepoint. Therefore, for any $i > 0$, the following formula is a cf-bound for $Do(a_0, t_0)$ over $\sigma_2 = \{Prec, <\}$:

$$\exists a_1 \cdots \exists a_i\ (Prec(a_1, a_0) \land \ldots \land Prec(a_i, a_{i-1})) \land \neg \exists t_1 \cdots \exists t_i\ (t_1 < t_0 \land \ldots \land t_i < t_{i-1}).$$

Denote this formula by $\chi_i$. For any $n > 0$ and a sufficient number of steps, the refinement algorithm can derive that $\psi_n := \chi_1 \lor \ldots \lor \chi_n$ is a cf-bound for $Do(a_0, t_0)$. Clearly, for $n_1 \neq n_2$, $\psi_{n_1}$ is not logically equivalent to $\psi_{n_2}$. This indicates that the refinement algorithm will not reach a fixpoint for input $T_3$ and $\sigma_2$.

As shown by the examples, there are several issues concerning the practical implementation of the refinement algorithm.

1. Due to the non-deterministic nature of the algorithm, a heuristic is needed to choose which bounds to refine and which kind of refinement to apply. A reasonable choice is to first apply all possible axiom and input refinements. Then, top-down refinement for formula $\varphi$ is applied only if a bound for its parent or one of its siblings in the syntax tree has recently been refined. Similarly, bottom-up refinement is applied if a bound for one of $\varphi$'s children has been refined. Such a strategy was used in Example 1.

2. The bounds should be simplified at regular time points, i.e., they should be replaced by equivalent but smaller formulas. If bounds are not simplified, they can only grow in size, rapidly leading to formulas of unwieldy size. A simplification algorithm is discussed in Section 6.

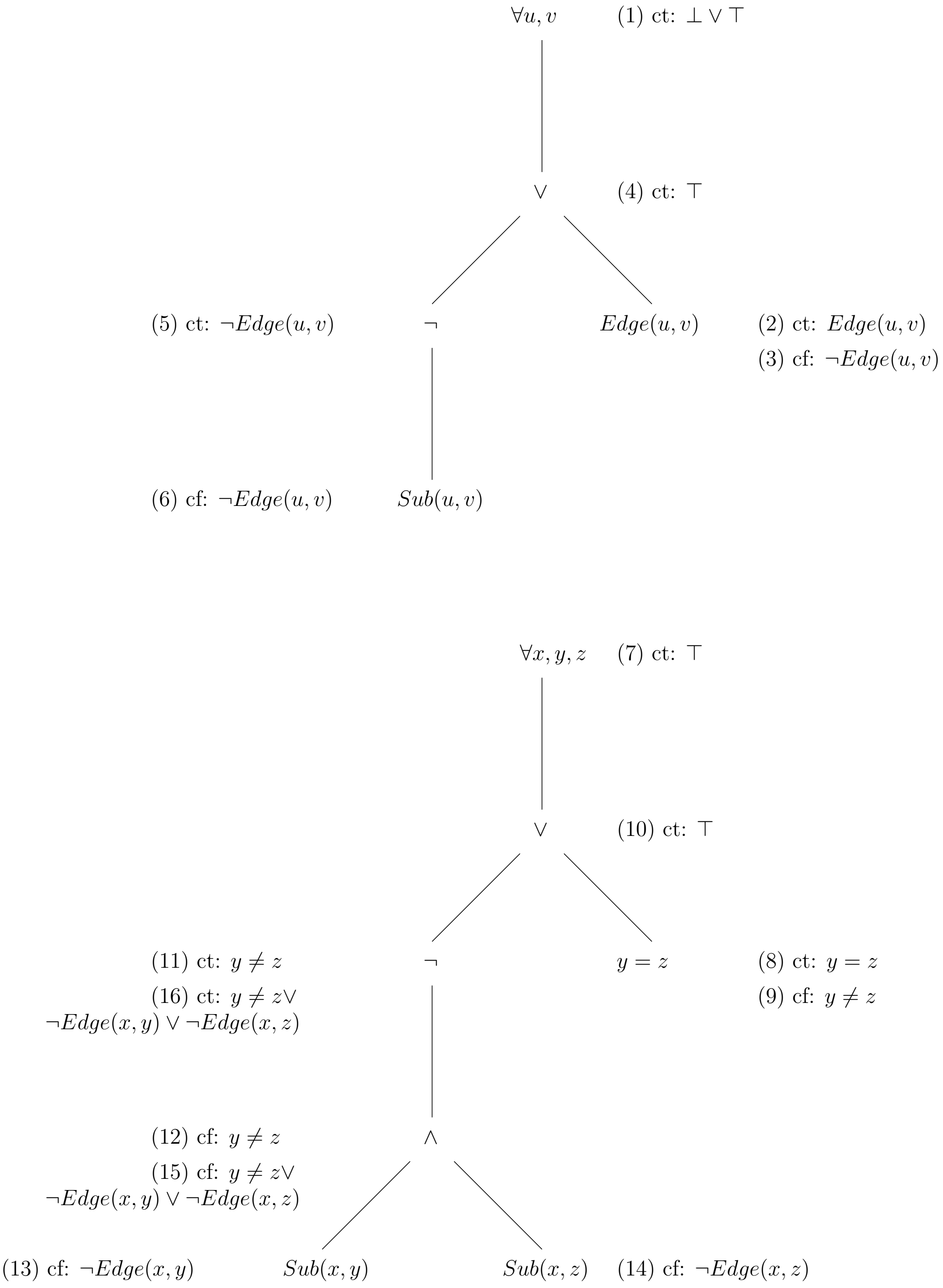

Figure 1: Refining a c-map

3. To be able to detect that a fixpoint has been reached, one needs to find out that two bounds are equivalent. In general this is undecidable. To detect a fixpoint in at least some cases, one could use an FO theorem prover (and restrict its running time).

   In case a fixpoint cannot be reached or detected, another stop criterion is needed. For example, one could restrict the number of one-step refinements, or the total time the refinement algorithm can use. Another stop criterion, and a simple fixpoint check are discussed in Section 6.

#### 4.4.2 Extracting an Atom-Based and Atom-Equal C-Map

The c-maps obtained by the refinement algorithm are in general neither atom-based nor atom-equal. To derive from an arbitrary c-map $\mathcal{C}$ an atom-equal c-map that is at least as precise as $\mathcal{C}$, we first collect for each predicate $P$ all bounds that are assigned to occurrences of $P$ in the theory. Then the disjunction of these bounds is assigned as new bound to each occurrence of $P$. Because all bounds assigned to atoms over $P$ are then essentially the same, we have an atom-equal c-map. We now present this method more formally:

**Definition 28.** Let $\mathcal{C}$ be a c-map for a TNF theory $T$ and $P/n$ a predicate. Let $P(x_{11},\dots,x_{1n})$, $\dots$, $P(x_{m1},\dots,x_{mn})$ be all occurrences of $P$ in $T$ and let $y_1,\dots,y_n$ be $n$ new variables. Denote by $\varphi^i_{\mathrm{ct}}$, respectively $\varphi^i_{\mathrm{cf}}$, the formulas

$$\exists x'_{i1}\cdots\exists x'_{in}\ (\mathcal{C}^{\mathrm{ct}}(P(x_{i1},\dots,x_{in}))[x_{i1}/x'_{i1},\dots,x_{in}/x'_{in}] \land y_1 = x'_{i1} \land \ldots \land y_n = x'_{in})$$

and

$$\exists x'_{i1}\cdots\exists x'_{in}\ (\mathcal{C}^{\mathrm{cf}}(P(x_{i1},\dots,x_{in}))[x_{i1}/x'_{i1},\dots,x_{in}/x'_{in}] \land y_1 = x'_{i1} \land \ldots \land y_n = x'_{in}),$$

where the variables $x'_{ij}$ are new variables. The *ct-copy closure of* $P(x_{k1},\dots,x_{kn})$ *with respect to* $\mathcal{C}$ is the disjunction $\bigvee_{1\le i\le m}\varphi^i_{\mathrm{ct}}[y_1/x_{k1},\dots,y_n/x_{kn}]$. The *cf-copy closure of* $P(x_{k1},\dots,x_{kn})$ is the formula $\bigvee_{1\le i\le n}\varphi^i_{\mathrm{cf}}[y_1/x_{k1},\dots,y_n/x_{kn}]$. The copy-closure for atoms of the form $F(\overline{x}) = y$ is defined similarly.

We denote the ct-copy closure of an atom $\varphi$ by $\mathrm{copy}^{\mathcal{C}}_{\mathrm{ct}}(\varphi)$, and its cf-copy closure by $\mathrm{copy}^{\mathcal{C}}_{\mathrm{cf}}(\varphi)$.

**Definition 29.** The *copy-closure of* $\mathcal{C}$ is the c-map that assigns $(\mathrm{copy}^{\mathcal{C}}_{\mathrm{ct}}(\varphi), \mathrm{copy}^{\mathcal{C}}_{\mathrm{cf}}(\varphi))$ to every atomic subformula $\varphi$ of $T$, and corresponds to $\mathcal{C}$ on all other subformulas.

**Example 10.** Let $T_4$ be the theory consisting of the sentences $\forall x\ (P(x) \supset R(x))$ and $\forall y\ (Q(y) \supset R(y))$ and let $\mathcal{C}_5$ be a c-map over $\sigma_3 = \{P, R\}$ that assigns $(P(x), \bot)$ to $R(x)$ and $(Q(y), \bot)$ to $R(y)$. The copy-closure of $\mathcal{C}_5$ assigns

$$((\exists x'\ (P(x') \land x' = x)) \lor (\exists x'\ (Q(x') \land x' = x)), (\exists x'\ (\bot \land x' = x)) \lor (\exists x'\ (\bot \land x' = x)))$$

to $R(x)$. These bounds simplify to $(P(x) \lor Q(x), \bot)$. Likewise, the copy-closure of $\mathcal{C}_5$ assigns to $R(y)$ bounds that simplify to $(P(y) \lor Q(y), \bot)$.

**Proposition 30.** *The copy-closure of a c-map is an atom-equal c-map.*

*Proof.* This follows immediately from the definition of atom-equal c-map since for every predicate symbol $P$ (or function symbol $F$), the same bounds, namely the formulas $\bigvee_{1\le i\le n}\varphi^i_{\mathrm{ct}}$ and $\bigvee_{1\le i\le n}\varphi^i_{\mathrm{cf}}$ mentioned in definition 28, are assigned to every atom over $P$ (respectively $F$). □

Recall that a c-map $\mathcal{C}$ is atom-based if $\overline{\mathcal{C}}$ is implied by $\overline{\mathcal{C}}_A$, i.e., by all sentences in $\overline{\mathcal{C}}$ that stem from bounds for atomic subformulas of $T$. A method to derive an atom-based c-map from an arbitrary c-map is based on the following observation. Let $\mathcal{C}$ be a c-map for $T$ over $\sigma$ and let $\varphi[\overline{x}]$ be the subformula $\chi \land \psi$ of $T$. If $\mathcal{C}^{\mathrm{ct}}(\varphi)$ is the formula $\mathcal{C}^{\mathrm{ct}}(\chi) \land \mathcal{C}^{\mathrm{ct}}(\psi)$, i.e., it is the bottom-up refinement ct-bound for $\varphi$ with respect to $\mathcal{C}$, then $T \models \forall\overline{x}\ (\mathcal{C}^{\mathrm{ct}}(\varphi) \supset \varphi)$ is implied by $T \models \forall\overline{x}\ (\mathcal{C}^{\mathrm{ct}}(\chi) \supset \chi)$ and $T \models \forall\overline{x}\ (\mathcal{C}^{\mathrm{ct}}(\psi) \supset \psi)$. It is easy to check that the same property holds for all other bottom-up refinement bounds:

**Lemma 31.** *Let $\mathcal{C}$ be a c-map for $T$ over $\sigma$ and $\varphi[\overline{x}]$ a subformula of $T$, and let $\varphi^r_{\mathrm{ct}}$ and $\varphi^r_{\mathrm{cf}}$ be the bottom-up refinement bounds for $\varphi$ with respect to $\mathcal{C}$. If $S$ is the set of direct subformulas of $\varphi$, i.e., its children in the syntax tree, and $T'$ is the theory given by*

$$T' := \{\forall \overline{y}\ \mathcal{C}^{\mathrm{ct}}(\psi) \supset \psi \mid \psi[\overline{y}] \in S\} \cup \{\forall \overline{y}\ \mathcal{C}^{\mathrm{cf}}(\psi) \supset \neg\psi \mid \psi[\overline{y}] \in S\},$$

*then $T' \models \forall \overline{x}\ \varphi^r_{\mathrm{ct}} \supset \varphi$ and $T' \models \forall \overline{x}\ \varphi^r_{\mathrm{cf}} \supset \neg\varphi$.*

**Definition 32.** A c-map $\mathcal{C}$ for $T$ is called a *bottom-up c-map* if for every non-atomic subformula $\varphi$ of $T$, $\mathcal{C}^{\mathrm{ct}}(\varphi)$ is the bottom-up ct-refinement bound for $\varphi$ with respect to $\mathcal{C}$, and $\mathcal{C}^{\mathrm{cf}}(\varphi)$ is the bottom-up cf-refinement bound for $\varphi$ with respect to $\mathcal{C}$.

The next proposition follows directly from Lemma 31.

**Proposition 33.** *A bottom-up c-map $\mathcal{C}$ is atom-based.*

Observe that a bottom-up c-map $\mathcal{C}$ for $T$ is completely determined by the bounds it assigns to the atomic subformulas of $T$. Hence, given a c-map, one can derive a bottom-up c-map from it by retaining the bounds for the atomic subformulas and then computing the corresponding bottom-up c-map. We conclude that we can derive an atom-based, atom-equal c-map from an arbitrary c-map by deriving an atom-based c-map from its copy-closure.

**Example 11** (Example 1 ctd.)**.** Let $\mathcal{C}_6$ be the fixpoint shown in Figure 1. This c-map is atom-equal (and equivalent to its copy-closure). The bottom-up c-map derived from $\mathcal{C}_6$ is shown in Figure 2. Observe that this c-map is less precise than $\mathcal{C}_6$. For instance, the cf-bound assigned by $\mathcal{C}_6$ to the conjunction $Sub(x,y) \wedge Sub(x,z)$ is a disjunction of two bounds, namely bound $y \neq z$, obtained by top-down refinement, and bound $\neg Edge(x,y) \vee \neg Edge(x,z)$, obtained by bottom-up refinement. In the c-map of Figure 2, only the latter bound is present.

For the c-map in Figure 2, the c-transformation of $Sub(x,y) \wedge Sub(x,z)$ is given by

$$((Sub(x,y) \wedge Edge(x,y)) \wedge (Sub(x,z) \wedge Edge(x,z))) \wedge (Edge(x,y) \wedge Edge(x,z)).$$

This formula contains repeated constraints $Edge(x,y)$ and $Edge(x,z)$ on the variables $x, y$ and $z$. In general bottom-up c-maps produce many such repetitions. These could easily be eliminated to speed up the grounding process, but it depends on the used grounding algorithm which ones are best deleted.

## 5. Inductive Definitions

Although all NP problems can be cast as MX(FO) problems, modelling such problems using pure FO can be extremely complex. In practice, modelling is often enhanced considerably by using extensions of FO with constructs such as inductive definitions, subsorts, aggregates, partial functions and arithmetic. For this enriched language we have implemented the model generator IDP (Wittocx et al., 2008b; Wittocx & Mariën, 2008).[2]

In this paper we focus on grounding of the extension of FO with *inductive definitions*. It is well-known that in arbitrary domains, inductively definable concepts such as "reachability" are not FO-expressible. In finite domains however, they can be encoded (e.g., by encoding the fixpoint construction), but the process is tedious and leads to large theories. In this section we will extend the refinement algorithm to FO(ID) (Denecker, 2000; Denecker & Ternovska, 2008). This language extends FO with a construct for representing some of the most common types of inductive definitions: monotone induction and non-monotone induction such as induction over a well-founded order and iterated inductive definitions. Such definitions have many applications in real-life computational problems, e.g., in planning problems or problems involving reachability or dynamic systems (Denecker & Ternovska, 2008, 2007). At the same time, FO(ID) is also an integration of FO and logic programming.

2. IDP can be downloaded from `http://dtai.cs.kuleuven.be/krr/software.html`

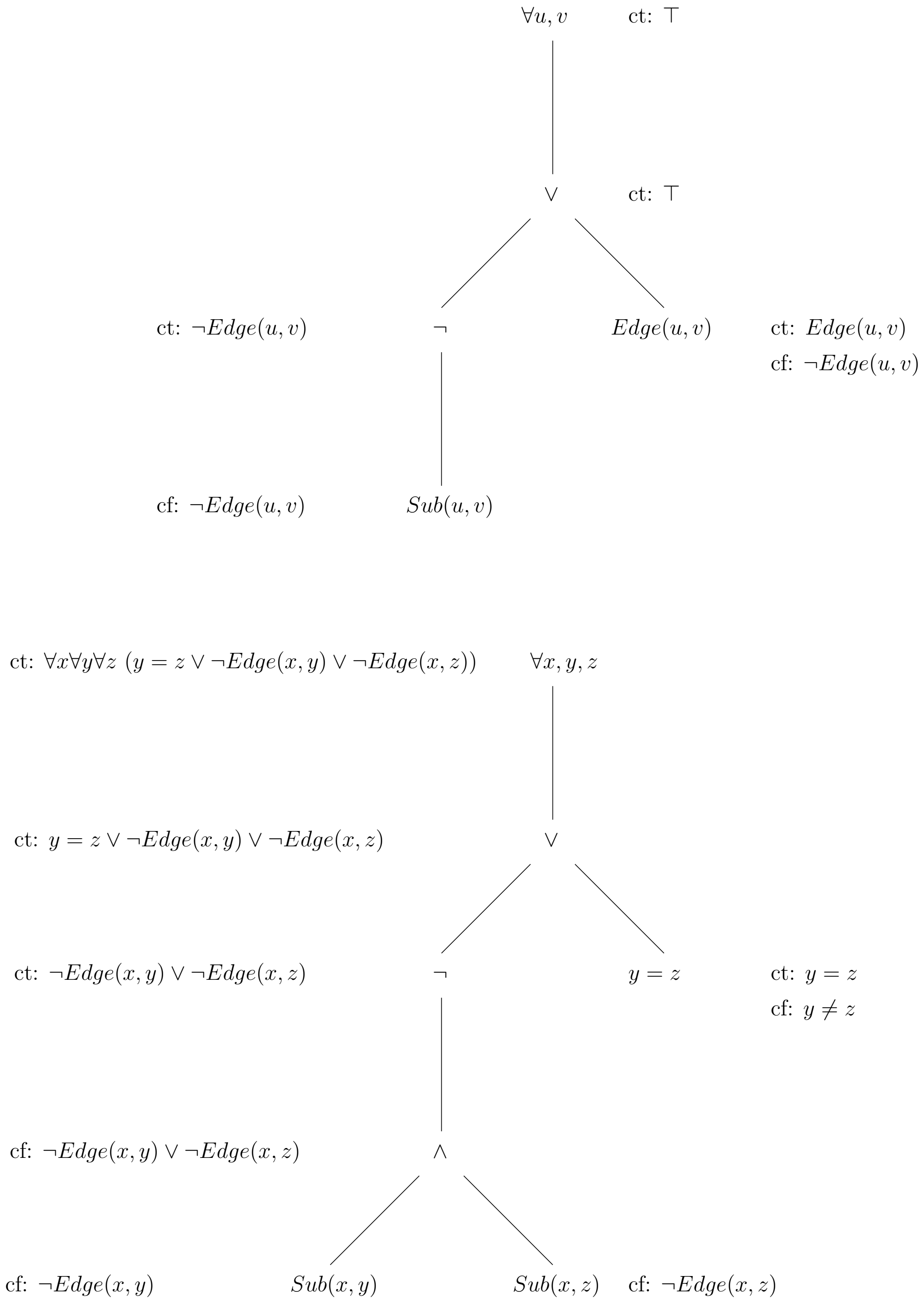

Figure 2: A bottom-up c-map

### 5.1 Three-Valued Structures

While FO(ID) has a standard two-valued semantics, three-valued structures are used in the formal semantics of definitions. Indeed, an inductive definition defines a set by describing how to *construct* it. In the semantics, the intermediate stages of the construction are recorded by three-valued sets, representing for any object whether it belongs to the set or not, or whether this has not yet been derived. We therefore recall the basic concepts of three-valued logic.

We denote the truth values *true, false* and *unknown* by respectively $\mathbf{t}$, $\mathbf{f}$ and $\mathbf{u}$. A three-valued $\Sigma$-interpretation $\tilde{I}$ consists of a domain $D$ and

- a domain element $x^{\tilde{I}} \in D$ for each variable $x$;
- a function $P^{\tilde{I}} : D^n \to \{\mathbf{t}, \mathbf{f}, \mathbf{u}\}$ for each predicate symbol $P/n$;
- a function $F^{\tilde{I}} : D^n \to D$ for each function symbol $F/n$.

If $P^{\tilde{I}}(\overline{d}) \neq \mathbf{u}$ for every tuple $\overline{d}$ of domain elements and predicate symbol $P$, then $\tilde{I}$ is two-valued: it corresponds to the interpretation $I$ that assigns $\overline{d} \in P^I$ iff $P^{\tilde{I}}(\overline{d}) = \mathbf{t}$ for every predicate $P$ and corresponds to $\tilde{I}$ on all other symbols.

The *truth order* $\leq$ on the set of truth values is induced by $\mathbf{f} < \mathbf{u} < \mathbf{t}$, the *precision order* $\leq_p$ is induced by $\mathbf{u} <_p \mathbf{f}$ and $\mathbf{u} <_p \mathbf{t}$. These orders are extended to three-valued $\Sigma$-structures: if $\tilde{I}$ and $\tilde{J}$ correspond on $\Sigma_F$, then we define

- $\tilde{I} \leq \tilde{J}$ iff $P^{\tilde{I}}(\overline{d}) \leq P^{\tilde{J}}(\overline{d})$ for every $\overline{d}$ and $P$;
- $\tilde{I} \leq_p \tilde{J}$ iff $P^{\tilde{I}}(\overline{d}) \leq_p P^{\tilde{J}}(\overline{d})$ for every $\overline{d}$, $P$.

Observe that two-valued structures are maximally precise three-valued structures. On the other hand, the least precise three-valued structure assigns $P^{\tilde{I}}(\overline{d}) = \mathbf{u}$ for every $\overline{d}$ and $P$.

We define the truth value $\tilde{I}(\varphi)$ of a formula $\varphi$ in a three-valued interpretation $\tilde{I}$ with domain $D$ by the standard Kleene semantics:

- $\tilde{I}(P(t_1, \ldots, t_n)) := P^{\tilde{I}}(t_1^{\tilde{I}}, \ldots, t_n^{\tilde{I}})$;
- $\tilde{I}(\varphi_1 \lor \varphi_2) := \text{lub}_{\leq}\{\tilde{I}(\varphi_1), \tilde{I}(\varphi_2)\}$;
- $\tilde{I}(\varphi_1 \land \varphi_2) := \text{glb}_{\leq}\{\tilde{I}\varphi_1, \tilde{I}(\varphi_2)\}$;
- $\tilde{I}(\exists x\ \varphi) := \text{lub}_{\leq}\{\tilde{I}[x/d](\varphi) \mid d \in D\}$;
- $\tilde{I}(\forall x\ \varphi) := \text{glb}_{\leq}\{\tilde{I}[x/d](\varphi) \mid d \in D\}$.

An atom of the form $P(\overline{d})$, where $\overline{d}$ is a tuple of domain constants, is called a *domain atom*. For a truth value $\mathbf{v}$ and a domain atom $P(\overline{d})$, we denote by $\tilde{I}[P(\overline{d})/\mathbf{v}]$ the interpretation that assigns $\mathbf{v}$ to $P(\overline{d})$ and corresponds to $\tilde{I}$ on all other symbols. This notation is extended to sets of domain atoms.

### 5.2 Inductive Definitions

An FO(ID) theory is a set of FO sentences and definitions. A *definition* $\Delta$ is a finite set of rules of the form[3]

$$\forall \overline{x}\ (P(\overline{x}) \leftarrow \varphi),$$

3. Usually, nested terms are allowed as arguments of $P$, but to facilitate the presentation, we only allow variables as arguments in this paper.

where $P$ is a predicate and $\varphi$ an FO formula. The free variables of $\varphi$ should be among $\overline{x}$. $P(\overline{x})$ is called the *head* of the rule, $\varphi$ the *body*. Predicates that occur in the head of a rule of $\Delta$ are called *defined* predicates of $\Delta$. The set of all defined predicates of $\Delta$ is denoted $\mathrm{Def}(\Delta)$. All other symbols are called *open* with respect to $\Delta$. The set of open symbols of $\Delta$ is denoted by $\mathrm{Open}(\Delta)$.

Observe that an FO(ID) theory has the appearance of an FO theory augmented with a collection of *logic programs*. As illustrated by Denecker and Ternovska (2008), this entails that FO(ID)'s definitions can not only be used to represent mathematical concepts, but also for the sort of common sense knowledge that is often represented by logic programs, such as (local forms of) the closed world assumption, inheritance, exceptions, defaults, causality, etc.

The semantics of definitions is given by their well-founded model (Van Gelder, Ross, & Schlipf, 1991). As argued by Denecker and Ternovska (2008), the well-founded semantics correctly formalizes the semantics of all of the above mentioned types of inductive definitions in mathematics. We borrow the presentation of this semantics from Denecker and Vennekens (2007).

**Definition 34.** Let $\Delta$ be a definition and $\tilde{I}$ a three-valued structure. A *well-founded induction for* $\Delta$ *above* $\tilde{I}$ is a sequence $\langle \tilde{J}_\xi \rangle_{0 \leq \xi \leq \alpha}$ of three-valued structures such that

1. $\tilde{J}_0$ assigns $P^{\tilde{J}_0}(\overline{d}) = \mathbf{u}$, if $P$ is a defined predicate and corresponds to $\tilde{I}$ on the open symbols;
2. For each limit ordinal $\lambda \leq \alpha$, $\tilde{J}_\lambda = \mathrm{lub}_{\leq_p}\{\tilde{J}_\xi \mid \xi < \lambda\}$;
3. For every ordinal $\xi$, $\tilde{J}_{\xi+1}$ relates to $\tilde{J}_\xi$ in one of the following ways:
   (a) $\tilde{J}_{\xi+1} = \tilde{J}_\xi[P(\overline{d})/\mathbf{t}]$ for some domain atom $P(\overline{d})$ such that $P^{\tilde{J}_\xi}(\overline{d}) = \mathbf{u}$ and for some rule $\forall \overline{x}\ (P(\overline{x}) \leftarrow \varphi)$ in $\Delta$, $\tilde{J}_\xi[\overline{x}/\overline{d}](\varphi) = \mathbf{t}$.
   (b) $\tilde{J}_{\xi+1} = \tilde{J}_\xi[U/\mathbf{f}]$, where $U$ is a set of domain atoms, such that for each $P(\overline{d}) \in U$, $P^{\tilde{J}_\xi}(\overline{d}) = \mathbf{u}$ and for all rules $\forall \overline{x}\ (P(\overline{x}) \leftarrow \varphi)$ in $\Delta$, $\tilde{J}_{\xi+1}[\overline{x}/\overline{d}](\varphi) = \mathbf{f}$.

Intuitively, (a) says that a domain atom $P(\overline{d})$ can be made true if there is a rule with $P(\overline{x})$ as head and body $\varphi$ such that $\varphi[\overline{x}/\overline{d}]$ is already true. On the other hand (b) explains that $P(\overline{d})$ can be made false if there is no possibility of making a corresponding body true, except by circular reasoning. The set $U$, commonly called an *unfounded set*, is a witness to this: making all atoms in $U$ false also makes all corresponding bodies false.

A well-founded induction is called *terminal* if it cannot be extended anymore. The limit of a terminal well-founded induction is its last element. Denecker and Vennekens (2007) show that each terminal well-founded induction for $\Delta$ above $\tilde{I}$ has the same limit, which corresponds to the well-founded model of $\Delta$ extending $\tilde{I}|_{\mathrm{Open}(\Delta)}$, and is denoted by $\mathrm{wfm}_\Delta(\tilde{I})$. The well-founded model is three-valued in general.

A two-valued structure $I$ satisfies a definition $\Delta$ if $I = \mathrm{wfm}_\Delta(I)$. An FO(ID) theory $T$ is a finite set of FO sentences and definitions. $I$ satisfies $T$ if it satisfies all definitions and sentences in $T$. If $\Delta$ is a definition over $\Sigma$ and $J$ a $\Sigma|_{\mathrm{Open}(\Delta)}$-structure, there exists at most one expansion $I$ of $J$ to $\Sigma$ such that $I \models \Delta$. A definition is called *total* if for any $\Sigma|_{\mathrm{Open}(\Delta)}$-structure $J$ there is precisely one expansion $I$ of $J$ to $\Sigma$ that satisfies $\Delta$. Intuitively, total definitions correspond to well-formed definitions: for every defined predicate $P$, they define for each tuple of domain elements whether $\overline{d}$ belongs to the relation denoted by $P$ or not. If a definition is not total, this typically indicates an error. Hence in practice, all definitions that occur in MX(FO(ID)) specifications are total. For example, this is the case for all MX(FO(ID)) specifications used in the second ASP-competition (Denecker, Vennekens, Bond, Gebser, & Truszczyński, 2009). In general, checking whether a definition is total is undecidable. However, there are several broad and easily recognizable classes of total definitions. For example, all monotone and stratified definitions are total.

We give some examples of definitions and MX(FO(ID)) problems.

**Example 12.** Definition $\Delta_1$ defines relation $TC$ to be the transitive closure of relation $R$.

$$\Delta_1 = \left\{ \begin{array}{l} \forall x \forall y \ (TC(x,y) \leftarrow R(x,y)). \\ \forall x \forall y \ (TC(x,y) \leftarrow \exists z \ (TC(x,z) \land TC(z,y))). \end{array} \right\}$$

**Example 13.** To cast the problem of finding a Hamiltonian path in a given graph as an MX(FO(ID)) problem, let

$$\sigma = \langle \{Edge/2\}, \emptyset \rangle$$
$$\Sigma = \{\sigma_P \cup \{Ham/2, Reached/1\}, \{Start/0\}\rangle.$$

Predicate $Ham$ represents the edges that form the path and $Reached$ the vertices that are in the path. The constant $Start$ represents the first vertex of the path. Let $T$ be the theory

$$\forall v_1 \forall v_2 \ (Ham(v_1,v_2) \supset Edge(v_1,v_2)).$$
$$\forall v_1 \forall v_2 \forall v_3 \ (Ham(v_1,v_2) \land Ham(v_1,v_3) \supset v_2 = v_3).$$
$$\forall v_1 \forall v_2 \forall v_3 \ (Ham(v_1,v_3) \land Ham(v_2,v_3) \supset v_1 = v_2).$$
$$\forall v \ \neg Ham(v, Start).$$
$$\forall v \ Reached(v).$$
$$\left\{ \begin{array}{l} \forall v \ (Reached(v) \leftarrow v = Start). \\ \forall v \ (Reached(v) \leftarrow \exists w \ (Reached(w) \land Ham(w,v))). \end{array} \right\}.$$

Then model expansion for input structure $T$ and input vocabulary $\sigma$ expresses the Hamiltonian path problem: in every model $M \models_{I_\sigma} T$, the collection of edges $(v_1, v_2) \in Ham^M$ forms a Hamiltonian path in the graph represented by $Edge^{I_\sigma}$.

A well-known concept that we will use later on in this section is the *completion* of a definition. The completion of a definition $\Delta$ is an FO theory that is weaker than $\Delta$, and is defined as follows.

**Definition 35.** The *completion* of a definition $\Delta$ is the FO theory that contains for every $P \in \mathrm{Def}(\Delta)$ the sentence

$$\forall \overline{x} \ (P(\overline{x}) \equiv ((\overline{x} = \overline{y}_1 \land \varphi_1) \lor \ldots \lor (\overline{x} = \overline{y}_n \land \varphi_n))),$$

where $\forall \overline{y}_1 \ (P(\overline{y}_1) \leftarrow \varphi_1), \ldots, \forall \overline{y}_n \ (P(\overline{y}_n) \leftarrow \varphi_n)$ are the rules in $\Delta$ with $P$ in the head.

We denote the completion of $\Delta$ by $\mathrm{Comp}(\Delta)$. Clearly, every body of a rule in $\Delta$ occurs in $\mathrm{Comp}(\Delta)$. If $T$ is a theory then we denote by $\mathrm{Comp}(T)$ the result of replacing in $T$ all definitions by their completion. The following result states that the completion of $T$ is weaker than $T$.

**Theorem 36** (Denecker & Ternovska, 2008)**.** $\Delta \models \mathrm{Comp}(\Delta)$ *and* $T \models \mathrm{Comp}(T)$ *for every definition* $\Delta$ *and FO(ID) theory* $T$*.*

The *SAT(ID) problem* is the problem of deciding whether a given propositional FO(ID) theory is satisfiable. Currently there exist three SAT(ID) solvers. IDsat (Pelov & Ternovska, 2005) works by translating a SAT(ID) problem into an equivalent SAT problem and then calls a SAT solver. MidL (Mariën, Wittocx, & Denecker, 2007) and MiniSAT(ID) (Mariën, Wittocx, Denecker, & Bruynooghe, 2008) take a native approach. Mariën (2009) provides details on the specific form of propositional FO(ID) theories accepted by these solvers, and a method to transform arbitrary propositional FO(ID) theories into this form.

### 5.3 Grounding Inductive Definitions

Like MX(FO) problems, MX(FO(ID)) problems can be reduced to SAT(ID) problems by grounding. In this section we extend grounding and the refinement algorithm of Section 4 to FO(ID). Without loss of generality (Mariën, Gilis, & Denecker, 2004), we assume that none of the predicates of the input vocabulary $\sigma$ is defined by a definition in $T$, and no predicate is defined by more than one definition. Moreover, we assume that every rule body is in TNF.

### 5.3.1 Full and Reduced Grounding

Let $T$ be an FO(ID) theory. As for FO, a grounding $T_{\mathrm{g}}$ for $T$ with respect to $I_\sigma$ is a propositional FO(ID) theory that is $I_\sigma$-equivalent to $T$. We extend the notion of full and reduced grounding to definitions.

**Definition 37.** The *full grounding of a rule* $\forall \overline{x}\ P(\overline{x}) \leftarrow \varphi$ with respect to $I_\sigma$ is the set $\{P(\overline{d}) \leftarrow \mathrm{Gr}_{\mathrm{full}}(\varphi[\overline{x}/\overline{d}]) \mid \overline{d} \in D^n\}$, where $n$ is the number of variables in $\overline{x}$. Similarly, the *reduced grounding of* $\forall \overline{x}\ (P(\overline{x}) \leftarrow \varphi)$ is the set $\{P(\overline{d}) \leftarrow \mathrm{Gr}_{\mathrm{red}}(\varphi[\overline{x}/\overline{d}]) \mid \overline{d} \in D^n\}$. The *full (reduced) grounding of a definition* $\Delta$ is the union of the full (reduced) groundings of all rules in $\Delta$.

The full (reduced) grounding of an FO(ID) theory $T$ is the set of the full (reduced) groundings of all sentences and definitions in $T$.

### 5.3.2 Definitions Depending Only on $\sigma$

We say that a definition $\Delta$ *depends on expansion symbols* if $\mathrm{Open}(\Delta) \not\subseteq \sigma$. If $\Delta$ does not depend on expansion symbols, then the interpretation of every predicate in $\mathrm{Def}(\Delta)$ is the same in every model $M$ of $T$ expanding $I_\sigma$. Indeed, for such a definition and any $M \models_{I_\sigma} T$, $M|_{\mathrm{Open}(\Delta)}$ is completely determined by $I_\sigma$. Therefore also $\mathrm{wfm}_\Delta(M)$ only depends on $I_\sigma$.

The deductive database literature describes several algorithms to compute $\mathrm{wfm}_\Delta(M)$ for a definition that does not depend on expansion symbols. Most of them are only defined for definitions where every rule body is a conjunction of atoms. But some of them, such as the Rete algorithm (Forgy, 1982) and the semi-naive evaluation technique (Ullman, 1988), can easily be adapted to handle full FO bodies.

Assume $\Delta$ is a definition that does not depend on expansion symbols. Let $\tau$ be the vocabulary $\langle \sigma_P \cup \mathrm{Def}(\Delta), \sigma_F\rangle$ and $I_\tau$ the $\tau$-structure such that $I_\tau|_\sigma = I_\sigma$ and $I_\tau \models \Delta$. Then clearly, $M \models_{I_\sigma} T$ iff $M \models_{I_\tau} T$ for any structure $M$. However, a grounding for $T \setminus \Delta$ with respect to $\tau$ can be obtained more efficiently, since $\mathrm{Gr}_{\mathrm{red}}(T \setminus \Delta, I_\tau)$ is necessarily smaller than $\mathrm{Gr}_{\mathrm{red}}(T, I_\sigma)$. Indeed, $T \setminus \Delta$ is a subtheory of $T$, and $\mathrm{Gr}_{\mathrm{red}}(T \setminus \Delta, I_\tau)$ does not contain symbols of $\mathrm{Def}(\Delta)$, while $\mathrm{Gr}_{\mathrm{red}}(T, I_\sigma)$ does.

Observe also that the set of c-maps for $T$ over $\tau$ is a superset of the set of c-maps for $T$ over $\sigma$, since the bounds assigned by the former c-maps are formulas over $\tau$, instead of only over $\sigma$. As such, c-maps computed by the refinement algorithm for $T$ over $\tau$ might yield more efficient grounding compared to c-maps computed for $T$ over $\sigma$.

### 5.3.3 Bounds for Definitions

We now extend the refinement algorithm to FO(ID).

**Definition 38.** A formula $\varphi$ is a *subformula* of an FO(ID) theory $T$ if it is a subformula of a sentence in $T$ or a subformula of a rule body in a definition of $T$. A c-map for $T$ over $\sigma$ is an assignment of a ct- and cf-bound over $\sigma$ to every subformula of $T$.

Note that a c-map does not assign bounds to heads of rules in a definition.

Our strategy to compute a c-map for an FO(ID) theory $T$ is simple: construct the completion of $T$ and apply the refinement algorithm on $\mathrm{Comp}(T)$ to obtain a c-map $\mathcal{C}$ for $\mathrm{Comp}(T)$. The restriction of $\mathcal{C}$ to the subformulas of $T$ is a c-map for $T$. Indeed, every subformula $\varphi$ of $T$ occurs in $\mathrm{Comp}(T)$ and since $T \models \mathrm{Comp}(T)$, $\mathrm{Comp}(T) \models \forall \overline{x}\ (\mathcal{C}^{\mathrm{ct}}(\varphi) \supset \varphi)$ and $\mathrm{Comp}(T) \models \forall \overline{x}\ (\mathcal{C}^{\mathrm{cf}}(\varphi) \supset \neg\varphi)$, also $T \models \forall \overline{x}\ (\mathcal{C}^{\mathrm{ct}}(\varphi) \supset \varphi)$ and $T \models \forall \overline{x}\ (\mathcal{C}^{\mathrm{cf}}(\varphi) \supset \neg\varphi)$.

In order to use a c-map for grounding, we lift the definition of c-transformation to FO(ID) theories.

**Definition 39.** Let $\mathcal{C}$ be a c-map for a theory $T$ and $\Delta$ a definition in $T$. The *c-transformation of a rule* $\forall \overline{x}\ (P(\overline{t}) \leftarrow \varphi)$ *of* $\Delta$ is given by $\forall \overline{x}\ (P(\overline{t}) \leftarrow \mathcal{C}\langle\varphi\rangle)$. The c-transformation $\mathcal{C}\langle\Delta\rangle$ of a definition $\Delta$ is the set of c-transformations of rules in $\Delta$. The c-transformation of $T$ is the set of the c-transformations of the formulas and definitions in $T$.

We also lift the notion of $\mathcal{C}$-equivalence to definitions.

**Definition 40.** Two definitions $\Delta_1$ and $\Delta_2$ are *$\mathcal{C}$-equivalent* if for every structure $I$ that satisfies $\overline{\mathcal{C}}$, $I \models \Delta_1$ iff $I \models \Delta_2$.

However, Lemma 15 does not hold for FO(ID) theories: for a definition $\Delta$, $\mathcal{C}\langle\Delta\rangle$ is not necessarily $\mathcal{C}$-equivalent to $\Delta$.

**Example 14.** Let $\sigma$ be the empty vocabulary and $T$ the theory

$$P$$
$$\{P \leftarrow P\}.$$

This theory is unsatisfiable because the definition $\{P \leftarrow P\}$ has only one model, in which $P$ is false. This contradicts the sentence in $T$. Clearly, $\top$ is a ct-bound for $P$. If $\mathcal{C}$ is a c-map for $T$ over $\sigma$ assigning $(\top, \bot)$ to $P$, then $\mathcal{C}\langle\{P \leftarrow P\}\rangle = \{P \leftarrow (P \wedge \neg\bot) \vee \top\}$, which is equivalent to $\{P \leftarrow \top\}$. This definition has only a model that assigns true to $P$. Since this model also satisfies $\mathcal{C}$, we conclude that $\{P \leftarrow P\}$ and $\mathcal{C}\langle\{P \leftarrow P\}\rangle$ are not $\mathcal{C}$-equivalent.

**Definition 41.** Let $\Delta$ a definition of $T$. We call c-map $\mathcal{C}$ for $T$ *$\Delta$-tolerant* if $\mathcal{C}\langle\Delta\rangle$ and $\Delta$ are $\mathcal{C}$-equivalent. We call $\mathcal{C}$ *$T$-tolerant* if it is $\Delta$-tolerant for every definition $\Delta$ of $T$.

In the following, we say that a formula occurs positively (negatively) in a definition $\Delta$ if it occurs positively (negatively) in a body of a rule in $\Delta$.

**Proposition 42.** *Let $\Delta$ be a definition of a theory $T$. Then a c-map $\mathcal{C}$ for $T$ over $\sigma$ is $\Delta$-tolerant if for every subformula $\varphi$ of $\Delta$ that contains a predicate $P \in \mathrm{Def}(\Delta)$, the following hold:*

1. *If $\Delta$ is not total, then $\mathcal{C}^{\mathrm{ct}}(\varphi) = \mathcal{C}^{\mathrm{cf}}(\varphi) = \bot$.*
2. *If $\varphi$ occurs positively in $\Delta$ and $P$ occurs positively in $\varphi$, then $\mathcal{C}^{\mathrm{ct}}(\varphi) = \bot$.*
3. *If $\varphi$ occurs negatively in $\Delta$ and $P$ occurs negatively in $\varphi$, then $\mathcal{C}^{\mathrm{cf}}(\varphi) = \bot$.*

Note that the c-map of Example 14 violates the second condition. We will prove Proposition 42 by inductively constructing for any structure $I$ that satisfies $\mathcal{C}$, a sequence of three-valued structures that is a well-founded induction above $I$ for both $\Delta$ and $\mathcal{C}\langle\Delta\rangle$. If $I \models \Delta$, we show that a terminal sequence with this property can be constructed, proving that $I$ also satisfies $\mathcal{C}\langle\Delta\rangle$. If $I \not\models \Delta$, a sequence with this property can be constructed such that its last element is not less precise than $I$. This shows that $I$ does not satisfy $\mathcal{C}\langle\Delta\rangle$ either. To construct a well-founded induction for both $\Delta$ and $\mathcal{C}\langle\Delta\rangle$, we prove that each step that extends a well-founded induction for $\Delta$ is also a valid step to extend it for $\mathcal{C}\langle\Delta\rangle$. Step (3a) in Definition 34 is covered by Lemma 43, step (3b) by Lemma 44.

**Lemma 43.** *Let $I$ be a structure that satisfies a c-map $\mathcal{C}$ for $T$ over $\sigma$ and let $\tilde{J} \leq_p I$ be a three-valued interpretation such that $\tilde{J}|_\sigma$ is two-valued. Then $\tilde{J}(\varphi) \leq_p \tilde{J}(\mathcal{C}\langle\varphi\rangle)$ for every subformula $\varphi$ of $T$.*

*Proof.* We prove this lemma by induction. First assume $\varphi[\overline{x}]$ is an atom. Then $\mathcal{C}\langle\varphi\rangle$ is the formula $(\varphi \wedge \neg\mathcal{C}^{\mathrm{cf}}(\varphi)) \vee \mathcal{C}^{\mathrm{ct}}(\varphi)$. If $\tilde{J}(\varphi) = \mathbf{u}$, then clearly $\tilde{J}(\mathcal{C}\langle\varphi\rangle) \geq_p \tilde{J}(\varphi)$. If $\tilde{J}(\varphi) = \mathbf{f}$, then $\tilde{J}(\mathcal{C}^{\mathrm{ct}}(\varphi))$ must be false, since $I \models \mathcal{C}$. Therefore $\tilde{J}(\mathcal{C}\langle\varphi\rangle) = \mathbf{f}$. If on the other hand, $\tilde{J}(\varphi) = \mathbf{t}$, then $\tilde{J}(\mathcal{C}^{\mathrm{cf}}(\varphi)) = \mathbf{f}$ and hence, $\tilde{J}(\mathcal{C}\langle\varphi\rangle) = \mathbf{t}$.

The inductive cases are all very similar to the base case. We prove one of them. Assume $\varphi$ is the formula $\psi \vee \chi$. Then $\mathcal{C}\langle\varphi\rangle$ is the formula $((\mathcal{C}\langle\psi\rangle \vee \mathcal{C}\langle\chi\rangle) \wedge \neg\mathcal{C}^{\mathrm{cf}}(\varphi)) \vee \mathcal{C}^{\mathrm{ct}}(\varphi)$. If $\tilde{J}(\varphi) = \mathbf{f}$, then $\tilde{J}(\psi) = \tilde{J}(\chi) = \mathbf{f}$, and by induction $\tilde{J}(\mathcal{C}\langle\psi\rangle) = \tilde{J}(\mathcal{C}\langle\chi\rangle) = \mathbf{f}$. Since also $\tilde{J}(\mathcal{C}^{\mathrm{ct}}(\varphi)) = \mathbf{f}$, we conclude that $\tilde{J}(\mathcal{C}\langle\varphi\rangle) = \mathbf{f}$. If on the other hand $\tilde{J}(\varphi) = \mathbf{t}$, then $\tilde{J}(\mathcal{C}^{\mathrm{cf}}(\varphi)) = \mathbf{f}$. Also $\tilde{J}(\psi) = \mathbf{t}$ or $\tilde{J}(\chi) = \mathbf{t}$, and therefore $\tilde{J}(\mathcal{C}\langle\psi\rangle) = \mathbf{t}$ or $\tilde{J}(\mathcal{C}\langle\chi\rangle) = \mathbf{t}$. Hence $\tilde{J}(\mathcal{C}\langle\varphi\rangle) = \mathbf{t}$. ☐

**Lemma 44.** *Let $\Delta$ be a definition of $T$ and $\mathcal{C}$ a c-map for $T$ over $\sigma$ that satisfies the three conditions of Proposition 42. Let $I$ be a structure that satisfies $\mathcal{C}$ and $\tilde{J} \leq_p I$ a three-valued interpretation such that $\tilde{J}|_\sigma$ is two-valued. If $U$ is a set of domain atoms defined in $\Delta$ and unknown in $\tilde{J}$, then for every subformula $\varphi$ of $\Delta$ such that $\tilde{J}[U/\mathbf{f}](\varphi) \neq \mathbf{u}$, the following hold:*

- $\tilde{J}[U/\mathbf{f}](\varphi) \leq \tilde{J}[U/\mathbf{f}](\mathcal{C}\langle\varphi\rangle)$ *if $\varphi$ occurs negatively in $\Delta$;*
- $\tilde{J}[U/\mathbf{f}](\varphi) \geq \tilde{J}[U/\mathbf{f}](\mathcal{C}\langle\varphi\rangle)$ *if $\varphi$ occurs positively in $\Delta$;*

*Proof.* Denote $\tilde{H} := \tilde{J}[U/\mathbf{f}]$. If $\tilde{J}(\varphi) \neq \mathbf{u}$, the result follows immediately from Lemma 43.

We prove the case where $\tilde{J}(\varphi) = \mathbf{u}$ by induction. Assume that $\varphi$ is an atom $P(\overline{x})$. Since $\tilde{J}(\varphi) = \mathbf{u}$ and $\tilde{H}(\varphi) \neq \mathbf{u}$, we know that $P(\overline{x}^{\tilde{J}}) \in U$ and $\tilde{H}(\varphi) = \mathbf{f}$. Therefore $\tilde{H}(\mathcal{C}\langle\varphi\rangle) = \tilde{H}((\varphi \land \neg\mathcal{C}^{\mathrm{cf}}(\varphi)) \lor \mathcal{C}^{\mathrm{ct}}(\varphi)) = \tilde{H}(\mathcal{C}^{\mathrm{ct}}(\varphi))$. If $\varphi$ occurs negatively in $\Delta$, then we have to prove that $\tilde{H}(\varphi) \leq \tilde{H}(\mathcal{C}\langle\varphi\rangle)$. Since $\tilde{H}(\varphi) = \mathbf{f}$, this inequality holds regardless the value of $\mathcal{C}^{\mathrm{ct}}(\varphi)$ and $\mathcal{C}^{\mathrm{cf}}(\varphi)$ in $\tilde{H}$. If on the other hand, $\varphi$ occurs positively, we have to prove that $\tilde{H}(\varphi) \geq \tilde{H}(\mathcal{C}\langle\varphi\rangle)$. Since $\tilde{H}(\varphi) = \mathbf{f}$ and $\tilde{H}(\mathcal{C}\langle\varphi\rangle) = \tilde{H}(\mathcal{C}^{\mathrm{ct}}(\varphi))$, this inequality can only hold if $\tilde{H}(\mathcal{C}^{\mathrm{ct}}(\varphi)) = \mathbf{f}$. Because the conditions on $\mathcal{C}$ ensure that $\mathcal{C}^{\mathrm{ct}}(\varphi) = \bot$, we can conclude that indeed $\tilde{H}(\mathcal{C}^{\mathrm{ct}}(\varphi)) = \mathbf{f}$.

We omit the inductive cases, since they are very similar to the base case. □

*Proof of Proposition 42.* Let $I$ be a structure that satisfies $\mathcal{C}$. We have to prove that $I \models \Delta$ iff $I \models \mathcal{C}\langle\Delta\rangle$. If $\Delta$ is not total, the proof is trivial, since then $\Delta$ and $\mathcal{C}\langle\Delta\rangle$ are equivalent.

Now assume that $\Delta$ is total and let $\langle\tilde{J}_\xi\rangle_{0\leq\xi\leq\alpha}$ be a well-founded induction for both $\Delta$ and $\mathcal{C}\langle\Delta\rangle$ above $I$. We will prove that if $\tilde{J}_\alpha$ is not two-valued, and $\tilde{J}_\alpha <_p I$, there exists a $\tilde{J}_{\alpha+1}$ such that $\langle\tilde{J}_\xi\rangle_{0\leq\xi\leq\alpha+1}$ is again a well-founded induction for $\Delta$ and $\mathcal{C}\langle\Delta\rangle$. Also observe that if $\lambda$ is a limit ordinal and $\langle\tilde{J}_\xi\rangle_{0\leq\xi<\lambda}$ is a well-founded induction for both $\Delta$ and $\mathcal{C}\langle\Delta\rangle$, then the same holds for $\langle\tilde{J}_\xi\rangle_{0\leq\xi\leq\lambda}$.

This is sufficient to conclude the proof. Indeed, if $I \models \Delta$, we can keep on extending the sequence until we end up in $I$, and derive that $I \models \mathcal{C}\langle\Delta\rangle$. If $I \not\models \Delta$, then we will eventually extend the well-founded induction with a structure $\tilde{J}_{\alpha+1} \not\leq_p I$. But then, the well-founded model of $\mathcal{C}\langle\Delta\rangle$ will also be more precise than $\tilde{J}_{\alpha+1}$, which shows that $I \not\models \mathcal{C}\langle\Delta\rangle$.

Assume that $\tilde{J}_\alpha$ is not two-valued and $\tilde{J}_\alpha <_p I$. Because $\Delta$ is total, there exists a $\tilde{J}_{\alpha+1}$ such that $\langle\tilde{J}_\xi\rangle_{0\leq\xi\leq\alpha+1}$ is a well-founded induction for $\Delta$. We have to prove that it is also a well-founded induction for $\mathcal{C}\langle\Delta\rangle$. There are two possibilities:

- $\tilde{J}_{\alpha+1} = \tilde{J}_\alpha[P(\overline{d})/\mathbf{t}]$ for some domain atom $P(\overline{d})$ and there is a rule $\forall\overline{x}\ (P(\overline{x}) \leftarrow \varphi)$ in $\Delta$ such that $\tilde{J}_\alpha[\overline{x}/\overline{d}](\varphi) = \mathbf{t}$. By Lemma 43, also $\tilde{J}_\alpha[\overline{x}/\overline{d}](\mathcal{C}\langle\varphi\rangle) = \mathbf{t}$. Hence, $\langle\tilde{J}_\xi\rangle_{0\leq\xi\leq\alpha+1}$ is a well-founded induction for $\mathcal{C}\langle\Delta\rangle$.
- $\tilde{J}_{\alpha+1} = \tilde{J}_\alpha[U/\mathbf{f}]$ and for every $P(\overline{d}) \in U$ and rule $\forall\overline{x}\ (P(\overline{x}) \leftarrow \varphi)$ in $\Delta$, $\tilde{J}_{\alpha+1}[\overline{x}/\overline{d}](\varphi) = \mathbf{f}$. By Lemma 44, we conclude that also $\tilde{J}_{\alpha+1}[\overline{x}/\overline{d}](\mathcal{C}\langle\varphi\rangle) = \mathbf{f}$. Therefore, $\langle\tilde{J}_\xi\rangle_{0\leq\xi\leq\alpha+1}$ is a well-founded induction for $\mathcal{C}\langle\Delta\rangle$.

□

From Proposition 42 we derive the following procedure to compute a $T$-tolerant c-map for a theory $T$. First compute a c-map $\mathcal{C}$ for $T$ that is not necessarily $T$-tolerant. Then, for every definition $\Delta$ of $T$ and every subformula $\varphi$ of $\Delta$, replace $\mathcal{C}^{\mathrm{ct}}(\varphi)$ and $\mathcal{C}^{\mathrm{cf}}(\varphi)$ by $\bot$, if this is required to satisfy the conditions of Proposition 42.

We conclude that the following algorithm produces a correct grounding for FO(ID) theory $T$:

1. Compute a c-map $\mathcal{C}$ for $T$ over $\sigma$.
2. If $\mathcal{C}$ is inconsistent with respect to $I_\sigma$, output $\bot$ and stop.
3. Else, derive an atom-based, T-tolerant c-map $\mathcal{C}'$ from $\mathcal{C}$.
4. Output $\mathrm{Gr}_{\mathrm{red}}(\mathcal{C}'\langle T\rangle \cup \overline{\mathcal{C}'}_A)$, using any off-the-shelf grounder for FO(ID).

## 6. Implementation and Experiments

So far we have focussed mostly on grounding size. Proposition 23 guaranteed that grounding with bounds produces smaller groundings. In this section we are concerned with the efficiency and practical implementation of grounding with bounds. A first issue was mentioned at the end of Section 4.4.2: an atom-based c-map $\mathcal{C}$ computed by the refinement algorithm contains many repeated constraints on variables. To ground $\mathcal{C}\langle T\rangle$ efficiently, such repetitions should be avoided as much as possible. Secondly, an efficient grounder consults bounds as soon as possible. In particular, it should use bounds to avoid unnecessary instantiations of variables, rather than to remove these instantiations afterwards. As a case study, we will show in detail how to adapt a basic "top-down style" grounding algorithm to efficiently exploit bounds. We sketch how the same principles can be applied for a "bottom-up style" grounder.

In the second part of this section we discuss some aspects of implementing the refinement algorithm. As we mentioned in Section 4.4.1, there are several issues concerning the practical implementation of this algorithm. In particular, a method to simplify bounds is needed, as well as a good stop criterion. We show how these issues can be addressed by representing bounds as *first-order binary decision diagrams.*

Finally, we report on our implementation, called GIDL, of the refinement and grounding algorithm. We present experimental results that show the impact of using bounds on grounding size and time.

### 6.1 Case Study: Top-Down Grounding with Bounds

For the rest of this section, assume $T$ is in TNF and fix an $I_\sigma$-consistent, atom-based c-map $\mathcal{C}$ for $T$ over $\sigma$. We call a formula of the form $\varphi \vee \psi$ or $\exists x\ \varphi$ a *disjunctive formula.* Vice versa, a *conjunctive formula* is a formula of the form $\varphi \wedge \psi$ or $\forall x\ \varphi$.

We now present a simple "top-down style" grounding algorithm that exploits bounds without constructing $\mathcal{C}\langle T\rangle \cup \overline{\mathcal{C}}_A$ explicitly. The algorithm is shown in Algorithm 1. Basically, it consults the bounds assigned by $\mathcal{C}$ whenever it substitutes the free variables of a formula $\varphi[\overline{x}]$ by domain constants $\overline{d}$. If according to the bounds, $\varphi[\overline{x}/\overline{d}]$ is certainly true, i.e., $I_\sigma[\overline{x}/\overline{d}] \models \mathcal{C}^{\text{ct}}(\varphi)$, then the grounding of $\varphi[\overline{x}/\overline{d}]$ is not computed. Instead, the algorithm then proceeds as if $\varphi[\overline{x}/\overline{d}]$ is equal to $\top$. Similarly if $\varphi[\overline{x}/\overline{d}]$ is certainly false. In this way, the algorithm avoids creating unnecessary instantiations. One can check that if $\mathcal{C}$ is the trivial c-map, Algorithm 1 reduces to a straightforward top-down style grounding algorithm that produces $\text{Gr}_{\text{full}}(T)$.

Line 1 of Algorithm 1 checks whether one of the sentences of $T$ is certainly false. If this is the case, then clearly $T$ is unsatisfiable (cf. Definition 10), and this can be reported immediately. Before a sentence is grounded, line 4 checks whether this sentence is certainly true according to $\mathcal{C}$. Only sentences that are not certainly true are grounded. Observe that both checks are simple syntactic checks and can be executed in constant time.

Function `groundConj` gets as input a formula $\varphi[\overline{x}]$ and returns a grounding for $\forall \overline{x}\ \varphi[\overline{x}]$. In particular, if $\varphi$ is a sentence, then the result of applying `groundConj` to $\varphi$ is a grounding for $\varphi$.

In `groundConj`, universal quantifiers are implicitly pushed inside conjunctions. That is, if $\varphi[\overline{x}]$ is a conjunction $\psi_1 \wedge \ldots \wedge \psi_n$, then for every $i \in [1, n]$, the grounding of $\forall \overline{x}\ \psi_i$ is computed by applying `groundConj` to $\psi_i$. The conjunction of these groundings is returned as grounding for $\forall \overline{x}\ \varphi$. According to equivalence (6) of Section 2.2, this transformation yields an equivalent formula.

Function `groundConj` only consults the c-map when variables are substituted by domain constants or when the input formula is an atom. As such, `groundConj` ignores ("eliminates") the bounds assigned to conjunctive formulas. As we mentioned at the end of Section 4.4.2, this is important to avoid repeated constraints on a variable.

In `groundConj`$(\varphi[\overline{x}])$, only those substitutions $\varphi[\overline{x}/\overline{d}]$ for which $I_\sigma[\overline{x}/\overline{d}] \not\models \mathcal{C}^{\text{ct}}(\varphi)$ are grounded (see, e.g., line 12). Indeed, the other substitutions yield a formula that is certainly true in all models of $T$ expanding $I_\sigma$, and can therefore be omitted from the ground conjunction $C$ that is computed.

**Algorithm 1**: Ground with Bounds

**Input**: $T$, $\sigma$, $I_\sigma$ and $\mathcal{C}$
**Output**: A grounding $T_{\mathrm{g}}$ for $T$ with respect to $I_\sigma$

**if** $\mathcal{C}^{\mathrm{cf}}(\varphi) = \top$ *for some sentence* $\varphi$ *of* $T$ **then return** $\bot$;
$T_{\mathrm{g}} := \emptyset$;
`// Ground all sentences of` $T$
**for** *every sentence* $\varphi$ *of* $T$ **do**
  **if** $\mathcal{C}^{\mathrm{ct}}(\varphi) \neq \top$ **then** Add `groundConj(`$\varphi$`)` to $T_{\mathrm{g}}$;
`// Ground all definitions of` $T$
**for** *every definition* $\Delta$ *of* $T$ **do**
  Add `groundDef(`$\Delta$`)` to $T_{\mathrm{g}}$;
`// Add the grounding of` $\overline{\mathcal{C}}_A$
**for** *every atomic subformula* $\varphi[\overline{x}]$ *of* $T$ **do**
  **for** *every* $\overline{d}$ *such that* $I_\sigma[\overline{x}/\overline{d}] \models \mathcal{C}^{\mathrm{ct}}(\varphi)$ **do**
    Add $\varphi[\overline{x}/\overline{d}]$ to $T_{\mathrm{g}}$;
  **for** *every* $\overline{d}$ *such that* $I_\sigma[\overline{x}/\overline{d}] \models \mathcal{C}^{\mathrm{cf}}(\varphi)$ **do**
    Add $\neg\varphi[\overline{x}/\overline{d}]$ to $T_{\mathrm{g}}$;
**return** $T_{\mathrm{g}}$;

**Function** `groundConj(`$\varphi[\overline{x}]$`)`

$C := \emptyset$;
**switch** $\varphi[\overline{x}]$ **do**
  **case** $\varphi$ *is a literal*
    **for** *all* $\overline{d}$ *such that* $I_\sigma \not\models \mathcal{C}^{\mathrm{ct}}(\varphi)[\overline{x}/\overline{d}]$ **do**
      **if** $I_\sigma \models \mathcal{C}^{\mathrm{cf}}(\varphi)[\overline{x}/\overline{d}]$ **then return** $\bot$;
      **else** Add $\varphi[\overline{x}/\overline{d}]$ to $C$;
  **case** $\varphi = \forall y\ \psi[\overline{x}, y]$
    **return** `groundConj(`$\psi[\overline{x}, y]$`)`;
  **case** $\varphi = \bigwedge_i \psi_i$
    $C := \bigcup_i$ `groundConj(`$\psi_i$`)`;
  **case** $\varphi$ *is a disjunctive formula*
    **for** *all* $\overline{d}$ *such that* $I_\sigma \not\models \mathcal{C}^{\mathrm{ct}}(\varphi)[\overline{x}/\overline{d}]$ **do**
      **if** $I_\sigma \models \mathcal{C}^{\mathrm{cf}}(\varphi)[\overline{x}/\overline{d}]$ **then return** $\bot$;
      **else** Add `groundDisj(`$\varphi[\overline{x}/\overline{d}]$`)` to $C$;
**return** $\bigwedge C$;

**Function** `groundDisj`($\varphi[\overline{x}]$)

$D := \emptyset$;<br>
**switch** $\varphi[\overline{x}]$ **do**<br>
    **case** $\varphi$ *is a literal*<br>
        **for** *all* $\overline{d}$ *such that* $I_\sigma \not\models \mathcal{C}^{\text{cf}}(\varphi)[\overline{x}/\overline{d}]$ **do**<br>
            **if** $I_\sigma \models \mathcal{C}^{\text{ct}}(\varphi)[\overline{x}/\overline{d}]$ **then** **return** $\top$;<br>
            **else** Add $\varphi[\overline{x}/\overline{d}]$ to $D$;<br>
    **case** $\varphi = \exists y\ \psi[\overline{x}, y]$<br>
        **return** `groundDisj`($\psi[\overline{x}, y]$);<br>
    **case** $\varphi = \bigvee_i \psi_i$<br>
        $D := \bigcup_i$ `groundDisj`($\psi_i$);<br>
    **case** $\varphi$ *is a conjunctive formula*<br>
        **for** *all* $\overline{d}$ *such that* $I_\sigma \not\models \mathcal{C}^{\text{cf}}(\varphi)[\overline{x}/\overline{d}]$ **do**<br>
            **if** $I_\sigma \models \mathcal{C}^{\text{ct}}(\varphi)[\overline{x}/\overline{d}]$ **then** **return** $\top$;<br>
            **else** Add `groundConj`($\varphi[\overline{x}/\overline{d}]$) to $D$;<br>
**return** $\bigvee D$;

**Function** `groundDef`($\Delta$)

$\Delta_g := \emptyset$;<br>
**for** *every rule* $\forall \overline{x}\ (P(\overline{x}) \leftarrow \varphi[\overline{y}])$ *in* $\Delta$ **do**<br>
    $\overline{z} := \overline{x} \setminus \overline{y}$;<br>
    **for** *every* $\overline{d}$ *such that* $I_\sigma \not\models \mathcal{C}^{\text{cf}}(\varphi[\overline{y}/\overline{d}])$ **do**<br>
        **if** $I_\sigma \models \mathcal{C}^{\text{ct}}(\varphi[\overline{y}/\overline{d}])$ **then** $\varphi_g := \top$;<br>
        **else** $\varphi_g :=$ `groundConj`($\varphi[\overline{y}/\overline{d}]$);<br>
        $n :=$ the number of variables in $\overline{z}$;<br>
        Add $P(\overline{x})[\overline{y}/\overline{d}, \overline{z}/\overline{d}'] \leftarrow \varphi_g$ to $\Delta_g$ for every $\overline{d}' \in D^n$;<br>
**return** $\Delta_g$;

Before $\varphi[\overline{x}/\overline{d}]$ is grounded, it is checked whether this substitution yields a formula that is certainly false (see, e.g., line 13). If this is the case, the whole conjunction $C$ will certainly be false, and therefore $\bot$ is returned immediately. Observe that *implicitly* the formula $\mathcal{C}^{\mathrm{ct}}(\varphi) \lor (\neg\mathcal{C}^{\mathrm{cf}}(\varphi) \land \varphi)$ is grounded. Hence the correctness of `groundConj` follows from Lemma 13.

Function `groundDisj` is dual to `groundConj`. On input $\varphi[\overline{x}]$, it returns a grounding for $\exists \overline{x}\ \varphi[\overline{x}]$. It implicitly pushes existential quantifiers through disjunctions and eliminates the bounds assigned to disjunctive formulas.

Function `groundDef` returns a grounding for its input definition $\Delta$. It grounds the rules of $\Delta$ one-by-one. For each rule $\forall \overline{x}\ (P(\overline{x}) \leftarrow \varphi[\overline{y}])$, only those substitutions $\varphi[\overline{y}/\overline{d}]$ that are possibly true are tried (line 4). If $\varphi[\overline{y}/\overline{d}]$ is certainly true, it is replaced by $\top$ (line 5).

In lines 7-11 of Algorithm 1, the theory $\overline{\mathcal{C}}_A$ is grounded. Recall that this is necessary to obtain a grounding that is $I_\sigma$-equivalent to $T$ (see Proposition 21). Observe that if $\mathcal{C}$ is the trivial c-map, no output is produced when lines 7-11 are executed.

The computationally expensive steps in Algorithm 1 are the steps where the truth values in $I_\sigma$ of (some of the) bounds assigned by $\mathcal{C}$ are computed. For large bounds, these steps can become infeasible. Indeed, the expression complexity of FO is PSPACE-complete (Stockmeyer, 1974). As such, grounding with too complex bounds may take more time and space than constructing the full grounding and simplifying it afterwards. The stop criterion of Section 6.2.3 for the refinement algorithm is designed to avoid too complex bounds. Our experiments in Section 6.3 show that carefully restricting the complexity of the bounds leads to faster grounding.

We stress that Algorithm 1 is just one example of a grounding algorithm that exploits bounds.[4] The principle of consulting bounds as soon as possible can be applied to adapt other grounding algorithms as well. For example, recall that a bottom-up style grounder starts by storing all instances of atomic subformulas of $T$ in a table. To exploit bounds efficiently, a bottom-up grounder should consult the bounds while constructing these tables and leave out, e.g., all instances that are certainly false. As such, it avoids unnecessary large tables, which in turn improves the speed of the subsequent grounding steps.

### 6.2 Implementing the Refinement Algorithm and Querying Bounds

In this section we discuss some aspects of implementing the refinement algorithm. As mentioned above, applying a simplification method for first-order formulas to simplify the bounds at regular time points is essential for a good implementation. One can use Goubault's (1995) method for this purpose. To this end, the bounds need to be represented by first-order binary decision diagrams. We show in this section that such a representation can be applied without too much overhead when applying one-step refinements. Moreover, using binary decision diagrams leads to extra benefits: we obtain a cheap equivalence check for bounds and an elegant algorithm to *query* bounds, which is needed to implement Algorithm 1. At the end of this section we discuss a stop criterion for the refinement algorithm and we discuss an implementation.

#### 6.2.1 First-Order Binary Decision Trees and Diagrams

We borrow the definition of first-order BDDs from Goubault (1995). Let $\varphi$, $\psi_1$ and $\psi_2$ be three formulas. The ternary *if-then-else operator* is denoted by "$\twoheadrightarrow$", and defined by $\varphi \twoheadrightarrow \psi_1; \psi_2 := (\varphi \land \psi_1) \lor (\neg\varphi \land \psi_2)$. The formula $\varphi \twoheadrightarrow \psi_1; \psi_2$ is also represented by the graph shown in Figure 3.

**Definition 45** (Goubault, 1995)**.** *FO binary decision trees* (BDTs) and *kernels* are defined by simultaneous induction:

- An atom is a kernel;

4. The question whether top-down grounders can be made more efficient than bottom-up grounders is outside the scope of this paper, and still undecided.

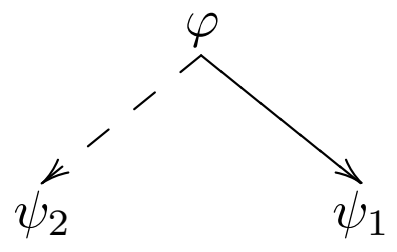

Figure 3: Graph representation of the formula $\varphi \rightharpoonup \psi_1; \psi_2$

- If $\varphi$ is a BDT and $x$ a variable, then $\exists x\ \varphi$ is a kernel;
- $\top$ and $\bot$ are BDTs;
- If $\varphi$ is a kernel and $\psi_1$ and $\psi_2$ are BDTs, then $\varphi \rightharpoonup \psi_1; \psi_2$ is a BDT.

Observe that the graph representation of a BDT is a tree whose nodes are atoms or existentially quantified BDTs.

Goubault (1995) showed that for every FO formula $\varphi$ there exists a BDT $\varphi'$ such that $\varphi$ and $\varphi'$ are equivalent. In an actual implementation, *sharing*, *reducing* and *ordering* are applied to obtain a simplified and compact representation of BDTs. Such representations are called *reduced ordered binary decision diagrams* (BDDs). Sharing means that isomorphic subtrees are stored at the same address in memory. Reducing involves exhaustively replacing subtrees of the form $\varphi \rightharpoonup \psi; \psi$ by $\psi$. A BDT $\varphi$ is ordered if the kernels appear in some fixed order on every path in the graph representation of $\varphi$.

As mentioned above, there are several important benefits of using BDDs to represent bounds for a formula:

- An implementation of the refinement algorithm using BDDs allows us to use the simplification algorithm for BDDs of Goubault (1995).
- As explained in Section 4.4, to detect that the refinement algorithm has reached a fixpoint, one needs to check the equivalence of bounds. Often, the BDDs representing two equivalent formulas will be equal.[5] Hence, a cheap (but necessarily incomplete) equivalence check for two bounds consists of checking the syntactic equality of the two BDDs representing them. Since equal BDDs are stored at the same address, this check is done in constant time.
- As we will show in Section 6.2.2, *querying* a bound $\varphi[\overline{x}]$, i.e., finding all tuples $\overline{d}$ such that $I_\sigma[\overline{x}/\overline{d}] \models \varphi$, can easily be implemented directly on a BDD representation of $\varphi$. Querying a bound is one of the main operations performed by a grounding algorithm that exploits bounds directly (such as Algorithm 1).

On the other hand, using BDDs does not result in too much overhead when computing a c-map. If $\varphi$, $\psi$ and $\chi[x, \overline{y}]$ are represented by BDDs, then a BDD representing $\neg\varphi$, $\exists x\ \varphi$, $\forall x\ \varphi$, $\varphi \land \psi$, $\varphi \lor \psi$ and $\chi[x/x', \overline{y}]$ can be computed efficiently (Bryant, 1986; Goubault, 1995). This implies that every one-step refinement on a c-map $\mathcal{C}$ can be implemented efficiently, even if the bounds assigned by $\mathcal{C}$ are BDDs.

### 6.2.2 Querying a Bound

In Algorithm 1, the main operation performed on a bound $\varphi[\overline{x}]$ is querying: finding tuples $\overline{d}$ of domain constants such that $I_\sigma \models \varphi[\overline{x}/\overline{d}]$. Finding a tuple $\overline{d}$ such that $I_\sigma \not\models \varphi[\overline{x}/\overline{d}]$ corresponds to querying $\neg\varphi$. We now show that querying a bound $\varphi[\overline{x}]$ can be done directly on the BDD representation by a simple backtracking algorithm.

5. For propositional BDDs, this is always the case.

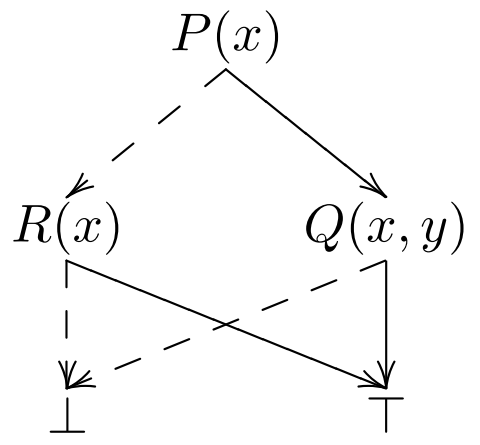

Figure 4: A BDD representing the formula $(P(x) \wedge Q(x, y)) \vee (\neg P(x) \wedge R(x))$

The idea is to traverse the BDD, starting from the root, and trying to end up in the leaf $\top$. At each inner node $\psi[\overline{y}] \twoheadrightarrow \psi_1; \psi_2$, the free variables in that node are replaced by domain constants $\overline{d}_y$. If $I_\sigma \models \psi[\overline{y}/\overline{d}_y]$, the algorithm continues via $\psi_1$, otherwise via $\psi_2$. If it ends up in $\bot$, it backtracks. If on the other hand, it ends up in $\top$, the performed substitutions constitute an answer for $\varphi$.

Function `query` implements the sketched query algorithm. It gets a bound $\varphi[\overline{x}]$ as input and returns a substitution $[\overline{x}/\overline{d}]$ such that $I_\sigma \models \varphi[\overline{x}/\overline{d}]$. If no such substitution exists, it returns FAIL. This algorithm can easily be adapted to return all answers to $\varphi[\overline{x}]$ instead of just one.

**Function** `query(`$\varphi[\overline{x}]$`)`

1 **if** $\varphi = \top$ **then return** *the empty substitution*;
2 **else if** $\varphi = \psi[\overline{y}] \twoheadrightarrow \psi_1; \bot$ **then**
3     **for** *every tuple* $\overline{d}$ *such that* $I_\sigma \models \psi[\overline{y}/\overline{d}]$ **do**
4         $\theta :=$ `query(`$\psi_1[\overline{y}/\overline{d}]$`)`;
5         **if** $\theta \neq$ *FAIL* **then return** $\theta \cup [\overline{y}/\overline{d}]$
6 **else if** $\varphi = \psi[\overline{y}] \twoheadrightarrow \bot; \psi_2$ **then**
7     **for** *every tuple* $\overline{d}$ *such that* $I_\sigma \not\models \psi[\overline{y}/\overline{d}]$ **do**
8         $\theta :=$ `query(`$\psi_2[\overline{y}/\overline{d}]$`)`;
9         **if** $\theta \neq$ *FAIL* **then return** $\theta \cup [\overline{y}/\overline{d}]$
10 **else if** $\varphi$ *is of the form* $\psi[\overline{y}] \twoheadrightarrow \psi_1; \psi_2$ **then**
11     **for** *every tuple* $\overline{d} \in D^{|\overline{y}|}$ **do**
12         **if** $I_\sigma \models \psi[\overline{y}/\overline{d}]$ **then** $\theta :=$ `query(`$\psi_1[\overline{y}/\overline{d}]$`)`;
13         **else** $\theta :=$ `query(`$\psi_2[\overline{y}/\overline{d}]$`)`;
14         **if** $\theta \neq$ *FAIL* **then return** $\theta \cup [\overline{y}/\overline{d}]$
15 **return** *FAIL*;

In lines 3 and 7, the algorithm needs to find tuples $\overline{d}$ such that respectively $I_\sigma \models \psi[\overline{y}/\overline{d}]$ and $I_\sigma \not\models \psi[\overline{y}/\overline{d}]$. If $\psi[\overline{y}]$ is an atom $P(\overline{y})$, this can be implemented by consulting the table $P^{I_\sigma}$. If $\psi$ is a kernel $\exists x\ \chi[x, \overline{y}]$, function `query` can be applied recursively to find the tuples. Indeed, any answer $(d', \overline{d})$ to $\chi[x, \overline{y}]$ provides a tuple $\overline{d}$ such that $I_\sigma \models \psi[\overline{y}/\overline{d}]$. Vice versa, $I_\sigma \not\models \psi[\overline{y}/\overline{d}]$ if $\chi[x, \overline{y}/\overline{d}]$ has no answer.

We illustrate the query algorithm on an example.

**Example 15.** Let $\varphi[x, y]$ be the BDD shown in figure 4, and let $\{a, b\}$ be the domain of $I_\sigma$, $P^{I_\sigma} = \{b\}$, $R^{I_\sigma} = \{\}$ and $Q^{I_\sigma} = \{(b, b)\}$. To find an answer for $\varphi[x, y]$, the query algorithm starts at the root $P(x)$. Since none of its children are equal to $\bot$, every domain constant is tried. Assume domain constant $a$ is tried first. Because $a \notin P^{I_\sigma}$, the algorithm continues with node $R(a) \twoheadrightarrow \top; \bot$. Because the "else" child of this node is $\bot$ and $a \notin R^{I_\sigma}$, the algorithm returns to the root and tries domain element $b$. Since $b \in P^{I_\sigma}$, it goes to node $Q(b, y) \multimap \top; \bot$. Since the "else" child of this node is $\bot$, the algorithm tries those substitutions $d$ for $y$ such that $(b, y/d) \in Q^{I_\sigma}$. Thus, $y$ is substituted by $b$. Finally, answer $[x/b, y/b]$ is returned.

#### 6.2.3 A Stop Criterion for the Refinement Algorithm

As shown in Section 4.4, the c-map refinement algorithm does not reach a fixpoint on certain inputs. Also, even in the case a fixpoint can be found, computing it may take a long time, and the bounds assigned by the fixpoint can be so complex that querying becomes very inefficient. Using such bounds may severely slow down grounding. This indicates the need for a good stop criterion.

**Simple Stop Criteria** A very simple stop criterion limits the number of one-step refinements that may be performed to a given maximum number $m$. This $m$ may depend on the theory $T$. For instance, $m$ can be set to $C \times$ (number of subformulas in $T$), where $C$ is some fixed constant.

A slightly less naive technique, which can be combined with the previous, limits the "complexity" of the bounds by putting a fixed upper bound $N$ on the number of nodes the BDD representation of a bound may have. If a one-step refinement would lead to a new bound with more nodes than $N$, this refinement is not performed. As this limits the number of applicable one-step refinements, the probability of reaching a fixpoint increases.

**Stop Criteria via Estimators** The experiments we present in Section 6.3 indicate that there exist appropriate values for $C$ and $N$ that produce positive results on most of the examples. Still, on some problems, grounding slows down severely, while the size of the produced grounding does not decrease. One of these problems is the following *clique problem* (entry 6 in Table 4).

**Example 16.** Recall that a clique is a maximally connected graph. Let

$$\sigma = \langle \{Edge/2\}, \emptyset \rangle,$$
$$\Sigma = \langle \sigma_P \cup \{Clique/1\}, \emptyset \rangle$$

and $T$ the theory

$$\forall x \forall y\ (Clique(x) \land Clique(y) \supset (x = y \lor Edge(x, y))).$$
$$\forall x\ ((\forall y\ (Clique(y) \land x \neq y \supset Edge(x, y))) \supset Clique(x)).$$

If $Edge^{I_\sigma}$ is symmetric, i.e., $I_\sigma$ represents an undirected graph, a model of $T$ expanding $I_\sigma$ is a clique in $I_\sigma$ that is not contained in a strictly larger clique in $I_\sigma$. Within a small number of iterations, the refinement algorithm finds for $Clique(x)$ the ct-bound $\forall x'\ x \neq x' \supset Edge(x, x')$. This formula expresses that $Clique(x)$ is certainly true in every solution if $x$ is directly connected to every other vertex in the input graph. Clearly, for most graphs, no vertex satisfies this condition. So, for most graphs, $\bot$ would be an equally precise ct-bound, but would allow much faster querying.

The situation is worse for the cf-bound for $Clique(x)$. Since for an undirected graph, every single vertex is a clique, and thus occurs in at least one of the solutions, the cf-bound is necessarily unsatisfiable with respect to $T$. Yet, our implementation of the refinement algorithm came up with $\exists x'\ (\neg Edge(x, x') \land x \neq x' \land (\forall x''\ (x' \neq x'' \supset Edge(x', x''))))$ as cf-bound. The query algorithm outlined above takes cubic time in the number of vertices to find out that no $x$ satisfies this formula.

To avoid the problems illustrated by the example above, one could estimate the reward of a bound versus the cost of evaluating it. Recall that more precise bounds yield smaller grounding sizes. Therefore, the reward of a bound $\psi$ is dictated by its precision. Given $I_\sigma$, it is possible to find a good estimate for the number of answers to $\psi$ in $I_\sigma$ (Demolombe, 1980), which is in turn a measure for the precision of $\psi$. For a fixed query algorithm, one can also estimate the cost $\text{cost}(\psi)$ of computing an answer in $I_\sigma$ to a query $\psi$. In the following, we assume that the reward of a bound is a positive real number, and its cost a strictly positive real number.

Given the reward and the cost of bounds, the complexity of a bound $\psi$ can be limited by restricting the ratio

$$r(\psi) := \frac{\text{cost}(\psi)}{\text{reward}(\psi) + 1}.$$

If a one-step refinement would replace a bound $\psi_1$ by $\psi_2$, but $r(\psi_1) < r(\psi_2)$, then this refinement is not performed. Clearly, for all bounds $\psi$ assigned by a c-map $\mathcal{C}$ computed according to this restriction, $r(\psi) \leq r(\bot)$ holds. Observe that to apply this restriction, an input structure $I_\sigma$ is needed. However, the obtained bounds are independent of $I_\sigma$.

It is beyond the scope of this paper to describe in detail estimators for the reward and cost of bounds. The fairly naive estimator used for the experiments in the next section assigns ratios of the order $\mathcal{O}(|D^{I_\sigma}|)$, respectively $\mathcal{O}(|D^{I_\sigma}|^3)$, to the ct-bound, respectively cf-bound, mentioned in Example 16. As such, if $|D^{I_\sigma}|$ is large enough, these bounds will be avoided.

#### 6.2.4 Implementation of the Refinement Algorithm

Our implementation of the refinement algorithm, including the heuristic for choosing refinement bounds (Section 4.4.1) and stop criterion, is presented by Algorithm 6. The algorithm maintains a queue $Q$ of one-step refinements that will be applied. Each of these is represented by a tuple $\langle r, \varphi \rangle$, where $r$ is the type of the refinement, e.g., axiom refinement, and $\varphi$ the formula on which $r$ will be applied.

**Algorithm 6**: Refinement Algorithm

```
1  Q := ∅; C := the trivial c-map for T;
2  for all sentences φ of T do  Q.push(⟨axiom, φ⟩);
3  for all subformulas φ of T over σ do
4      Q.push(⟨ct-input, φ⟩); Q.push(⟨cf-input, φ⟩);
5  while Q ≠ ∅ and the maximum number of refinements is not reached do
6      ⟨r, φ⟩ := Q.pop();
7      if r is a ct-refinement then
8          ψ := the r-refinement bound for φ with respect to C;
9          ψ := simplify(C^ct(φ) ∨ ψ);
10         if ψ ≠ C^ct(φ) and ψ is not too complex then
11             C^ct(φ) := ψ;
12             for all ⟨r, χ⟩ such that the r-refinement bound for χ contains C^ct(φ) do
13                 Q.push(⟨r, χ⟩);
14     else
15         ...  // Similar code for cf-refinements
16 return C;
```

As explained in Section 4.4.1, our implementation starts by scheduling all possible axiom- and input-refinements. If in a later stage a bound is changed (line 11), then all refinement bounds that contain this bound are scheduled to be applied (line 13). For example, assume that $T$ contains the formula $\varphi \wedge \psi$ and that the ct-bound of $\varphi$ is refined. Then bottom-up ct-refinement for $\varphi \wedge \psi$ is scheduled since the bottom-up ct-refinement bound for that formula is given by $\mathcal{C}^{\text{ct}}(\varphi) \wedge \mathcal{C}^{\text{ct}}(\psi)$, which contains $\mathcal{C}^{\text{ct}}(\varphi)$. For the same reason also top-down cf-refinement for $\psi$ is scheduled.

The algorithm applies all scheduled refinements, unless the maximum number of refinement steps is reached (line 5). The other part of the discussed stop criterion is applied in line 10. If the newly computed bound $\psi$ is too complex, i.e., its BDD representation contains too many nodes or the ratio $r(\psi)$ is above a certain threshold, $\psi$ is not used.

If BDDs are used to represent the bounds assigned by $\mathcal{C}$, line 8 can be implemented in linear time in the size of $\mathcal{C}$. If we use Goubault's simplification algorithm for BDDs for implementing line 9, the worst case complexity of this step is non-elementary in the size of $\mathcal{C}^{\mathrm{ct}}(\varphi) \vee \psi$ (Goubault, 1995). The estimators we used to implement line 10 take linear time in the size of $\psi$. It may seem that the complexity of the simplification method limits the practical applicability of Algorithm 6. However, since large BDDs usually do not pass the test in line 10, the simplification method is rarely applied on large BDDs. In the experiments of the next section, the running time of the refinement algorithm is negligible compared to the running time of the grounding algorithm.

### 6.3 Experiments

We implemented Algorithm 1 and Algorithm 6, using BDDs to represent bounds. The resulting grounder is called GidL. In this section, we present experiments, obtained with GidL, that show the impact of using bounds on grounding size and time.

As input for GidL, we used 37 benchmark problems, mainly taken from Asparagus.[6] The details about the experiments are available at `http://dtai.cs.kuleuven.be/krr/software.html`. We used four different versions of GidL:

**GidL$_{\mathrm{nb}}$:** Assigns $\langle \varphi, \neg\varphi \rangle$ as bound to every atomic subformula $\varphi$ over the input vocabulary, and $\langle \bot, \bot \rangle$ to every other subformula. As such, it creates the reduced grounding of the input theory.

**GidL$_{\mathrm{bu}}$:** Assigns $\langle \varphi, \neg\varphi \rangle$ as bound to every atomic subformula $\varphi$ over the input vocabulary and then applies bottom-up refinements to obtain a bottom-up c-map.

**GidL$_{\mathrm{mn}}$:** Limits the refinement algorithm to $4 \times$ (number of subformulas in $T$) one-step refinements and allows a maximum of 4 internal nodes in each BDD used to represent the bounds. According to previous experiments (Wittocx et al., 2008b), this is the best setting when limiting the number of nodes.

**GidL$_{\mathrm{r}}$:** Limits the refinement algorithm to $4 \times$ (number of subformulas in $T$) one-step refinements. It limits the complexity of the derived bounds by estimating the number of answers and the cost, as described in the previous section.

In Table 3, the influence of bounds on the grounding size is shown. The second and third column show the ratio of the grounding size obtained with GidL$_{\mathrm{mn}}$ and GidL$_{\mathrm{r}}$ compared to $\mathrm{Gr}_{\mathrm{red}}(T)$. For GidL$_{\mathrm{nb}}$ and GidL$_{\mathrm{bu}}$, this ratio is always equal to 1. When interpreting Table 3, it is important to note that small reductions in grounding size are not important. The reason being that all reductions that can be obtained by the refinement algorithm are also obtained by applying unit propagation on the grounding (see Section 7 for a discussion). Since there exist very efficient implementations of unit propagation, it is not beneficial to let the refinement algorithm find small reductions at a relatively high cost. We see that both GidL$_{\mathrm{mn}}$ and GidL$_{\mathrm{r}}$ reduce the grounding size with more than 50% in around 30% of the benchmarks. In 7, respectively 6, of the benchmarks there is a spectacular reduction of more than 95%.

More important than reductions in size are reductions in grounding time. Table 4 shows the running times of the different versions of GidL, and (between brackets) the ratio of the running time to the running time of GidL$_{\mathrm{nb}}$. The running time of the refinement algorithm is included (it never took more than 0.02 seconds). A time-out (###) of 600 seconds was used.

On many benchmarks, the reduction in grounding time with respect to GidL$_{\mathrm{nb}}$ is due to the reduction in grounding size. Yet there are also several benchmarks where time decreases a lot, while

6. `http://asp.haiti.cs.uni-potsdam.de/`

there is almost no reduction in size. This is mostly due to the creation of a bottom-up c-map, as can be seen from the running times of $\text{GidL}_{\text{bu}}$. Applying bottom-up refinements leads to the assignment of non-trivial bounds to non-atomic subformulas. This allows for earlier pruning by a top-down style grounder, and hence faster grounding.

From Table 4, we can see that $\text{GidL}_{\text{mn}}$ performs quite well. On half of the benchmarks, it is more than 44% faster than $\text{GidL}_{\text{nb}}$. It is also more than 20% faster than $\text{GidL}_{\text{bu}}$ on half of the benchmarks. There are some outliers however. On benchmarks 6 and 11, it is far slower than $\text{GidL}_{\text{bu}}$, while not producing a significantly smaller grounding. This indicates the use of a complex bound with relatively small reward. Compared to $\text{GidL}_{\text{mn}}$, $\text{GidL}_{\text{r}}$ is faster and more robust, indicating that using estimators for the reward and cost of bounds pays off in most cases. In only two of the benchmarks, our naive estimator makes a wrong guess. In benchmark 1, a bound with high cost and no reward is allowed, in benchmark 7, a bound with low cost and high reward is not allowed by $\text{GidL}_{\text{r}}$. It is part of future work to implement improved estimators.

We conclude from our experiments that grounding with bounds is applicable in practice. It often leads to smaller grounding sizes on standard benchmark problems, and if the bounds are carefully restricted, it yields a significant speed up. Since the time to compute bounds is small compared to the overall grounding time, computing them is essentially for free.

In general, a smaller grounding does not necessarily lead to faster propositional model generation. For example, grounding size (and time) increases when symmetry breaking formulas are added, but these formulas may drastically improve the overall solving time (Torlak & Jackson, 2007). Another example are clause-learning SAT solvers: the clauses learnt by these solvers are redundant, but may improve the solving time by orders of magnitude. The question arises whether our method of grounding with bounds may lead to slower overall model generation time compared to grounding without bounds. This is not the case. The experiments above show that in general, grounding with bounds is faster than grounding without bounds. Since grounding with bounds also produces smaller groundings, the subsequent initialization phase of the SAT solver is executed faster. If $T_1$ and $T_2$ are two groundings obtained by grounding the same input theory and structure with, respectively without bounds, it can be shown[7] that the typical simplification steps applied in this initialization phase transform $T_1$ and $T_2$ in exactly the same simplified theory $T_3$. Thus, after initialization, the SAT solver is applied on exactly the same theory, whether or not the grounder used bounds. It follows that in general, the overall model generation time does not increase when bounds are applied while grounding.

## 7. Related Work

In the previous sections we described a method to obtain fast and compact grounding. Several such methods have been described in the literature. Some of them are — like ours — preprocessing techniques that rewrite the input theory. Other techniques involve reasoning on the propositional level. In this section we provide an overview. We indicate which ones can be applied to improve GidL. We also give an overview of existing grounders.

### 7.1 Methods to Optimize Grounding

**Derivation of Bounds** To our knowledge, the methods proposed in the literature to derive bounds are less general than the one we presented in this paper. This is illustrated by Table 5, where we show for several grounders the impact of manually adding redundant information. For all the grounders in this table except GidL, manually adding redundancy may have a serious impact. For some grounders, the need to add redundancy can sometimes be avoided by writing the input theory in a specific format. For example, the grounder gringo (Gebser et al., 2007) uses a syntactic check to derive bounds: it derives that predicate `q` of the input vocabulary is a bound for predicate `p` if `p`

7. The exact formulation and the proof of the property are beyond the scope of this paper.

| Nr | Benchmark name | $\text{GIDL}_{mn}$ | $\text{GIDL}_{r}$ |
|---|---|---|---|
| 1 | 15puzzle | 1.00 | 1.00 |
| 2 | Battleship | 0.89 | 1.00 |
| 3 | Blocked N-queens | 0.02 | 0.02 |
| 4 | Blocksworld | 0.33 | 0.33 |
| 5 | Bounded spanningtree | 0.12 | 0.12 |
| 6 | Clique | 1.00 | 1.00 |
| 7 | Hierarchical clustering | 0.03 | 0.72 |
| 8 | Graph colouring | 1.00 | 1.00 |
| 9 | Debugging | 0.86 | 1.00 |
| 10 | Fastfood | 1.00 | 1.00 |
| 11 | FO-hamcircuit | 0.94 | 0.99 |
| 12 | Golomb ruler | 0.54 | 1.00 |
| 13 | Graph partitioning | 0.94 | 1.00 |
| 14 | Algebraic groups | 0.99 | 1.00 |
| 15 | Hamiltonian circuit | 0.01 | 0.01 |
| 16 | Tower of Hanoi | 1.00 | 1.00 |
| 17 | Knighttour | 0.00 | 0.00 |
| 18 | Labyrinth | 0.99 | 0.99 |
| 19 | Magic series | 1.00 | 1.00 |
| 20 | Maze generation | 0.90 | 0.90 |
| 21 | Mirror puzzle | 1.00 | 1.00 |
| 22 | Missionaries | 0.03 | 0.03 |
| 23 | N-queens | 1.00 | 1.00 |
| 24 | Pigeonhole | 1.00 | 1.00 |
| 25 | Disjunctive scheduling | 0.83 | 0.83 |
| 26 | Slitherlink | 0.04 | 0.04 |
| 27 | Social golfer | 1.00 | 1.00 |
| 28 | Sokoban | 0.59 | 0.59 |
| 29 | Solitaire | 1.00 | 0.73 |
| 30 | Spanningtree | 0.06 | 0.06 |
| 31 | Sudoku | 0.75 | 0.75 |
| 32 | Tarski | 1.00 | 1.00 |
| 33 | Toughnut | 0.00 | 0.00 |
| 34 | Train scheduling | 0.25 | 0.25 |
| 35 | Waterbucket | 0.36 | 0.36 |
| 36 | Weight bounded dominating set | 1.00 | 1.00 |
| 37 | Wire routing | 0.92 | 0.99 |
| | Average | 0.66 | 0.70 |
| | $\# < 1.00$ | 24 | 20 |
| | $\# < 0.50$ | 12 | 11 |
| | $\# < 0.05$ | 7 | 6 |

Table 3: Impact of bounds on grounding size

| Nr | GidL$_{nb}$ | GidL$_{bu}$ | | GidL$_{mn}$ | | GidL$_{r}$ | |
|---|---|---|---|---|---|---|---|
| 1 | 6.13 | 2.00 | (0.33) | 2.07 | (0.34) | 5.73 | (0.93) |
| 2 | 0.19 | 0.18 | (0.95) | 0.16 | (0.84) | 0.17 | (0.89) |
| 3 | 9.66 | 10.83 | (1.12) | 2.22 | (0.23) | 2.67 | (0.28) |
| 4 | 22.33 | 16.76 | (0.75) | 5.80 | (0.26) | 5.80 | (0.26) |
| 5 | 8.52 | 8.52 | (1.00) | 3.01 | (0.35) | 1.16 | (0.14) |
| 6 | 3.13 | 3.73 | (1.19) | 51.77 | (16.54) | 3.73 | (1.19) |
| 7 | 0.32 | 0.34 | (1.06) | 0.05 | (0.16) | 0.31 | (0.97) |
| 8 | 2.57 | 2.71 | (1.05) | 2.69 | (1.05) | 2.72 | (1.06) |
| 9 | 0.30 | 0.30 | (1.00) | 0.48 | (1.60) | 0.47 | (1.57) |
| 10 | ### | ### | (1.00) | 17.59 | (0.03) | 16.52 | (0.03) |
| 11 | ### | 5.87 | (0.01) | 37.86 | (0.06) | 6.06 | (0.01) |
| 12 | 14.05 | 3.54 | (0.25) | 4.13 | (0.29) | 3.40 | (0.24) |
| 13 | 0.03 | 0.04 | (1.33) | 0.03 | (1.00) | 0.02 | (0.67) |
| 14 | 9.68 | 9.58 | (0.99) | 11.20 | (1.16) | 9.60 | (0.99) |
| 15 | 70.75 | 71.50 | (1.01) | 2.56 | (0.04) | 1.81 | (0.03) |
| 16 | 2.32 | 1.83 | (0.79) | 1.96 | (0.84) | 1.83 | (0.79) |
| 17 | 12.22 | 10.35 | (0.85) | 0.06 | (0.00) | 0.10 | (0.01) |
| 18 | 8.80 | 8.83 | (1.00) | 8.83 | (1.00) | 8.73 | (0.99) |
| 19 | 1.83 | 1.76 | (0.96) | 1.79 | (0.98) | 1.81 | (0.99) |
| 20 | 2.77 | 2.80 | (1.01) | 0.51 | (0.18) | 0.17 | (0.06) |
| 21 | 0.12 | 0.11 | (0.92) | 0.12 | (1.00) | 0.10 | (0.83) |
| 22 | 17.4 | 18.08 | (1.04) | 2.29 | (0.13) | 2.68 | (0.15) |
| 23 | 4.62 | 4.60 | (1.00) | 4.62 | (1.00) | 4.64 | (1.00) |
| 24 | 4.92 | 5.01 | (1.02) | 4.90 | (1.00) | 4.90 | (1.00) |
| 25 | 151.15 | 151.66 | (1.00) | 172.50 | (1.14) | 171.54 | (1.13) |
| 26 | 0.25 | 0.13 | (0.52) | 0.02 | (0.08) | 0.02 | (0.08) |
| 27 | 5.47 | 5.47 | (1.00) | 5.37 | (0.98) | 5.41 | (0.99) |
| 28 | 2.78 | 2.66 | (0.96) | 1.57 | (0.56) | 1.54 | (0.55) |
| 29 | 0.43 | 0.43 | (1.00) | 0.46 | (1.07) | 0.49 | (1.14) |
| 30 | 6.86 | 6.79 | (0.99) | 0.59 | (0.09) | 0.57 | (0.08) |
| 31 | ### | 2.34 | (0.00) | 1.07 | (0.00) | 1.06 | (0.00) |
| 32 | 4.42 | 4.53 | (1.02) | 3.67 | (0.83) | 3.64 | (0.82) |
| 33 | 4.23 | 4.23 | (1.00) | 0.53 | (0.13) | 0.53 | (0.13) |
| 34 | 4.06 | 2.14 | (0.53) | 0.65 | (0.16) | 0.47 | (0.12) |
| 35 | 3.16 | 3.07 | (0.97) | 1.76 | (0.56) | 2.04 | (0.65) |
| 36 | 1.45 | 1.42 | (0.98) | 0.03 | (0.02) | 0.03 | (0.02) |
| 37 | 0.06 | 0.06 | (1.00) | 0.08 | (1.33) | 0.08 | (1.33) |
| Total | 2186.98 | 974.20 | (0.45) | 355.00 | (0.16) | 272.55 | (0.12) |
| Avg. gain | | | 12 % | | 0 % | | 40% |
| Median gain | | | 0 % | | 44 % | | 33% |

(For GidL$_{mn}$ and GidL$_{r}$, the time to compute the bounds is included.)

Table 4: Impact of bounds on grounding time

| | constr | redun | defin |
|---|---|---|---|
| GRINGO | 76.33 | 1.59 | 0.60 |
| DLV | 339.37 | 4.23 | 2.81 |
| LPARSE | 63.25 | 0.78 | 63.58 |
| PSGRND | 44.79 | 0.72 | n/a |
| GIDL | 0.26 | 0.42 | n/a |

Table 5: Grounding times (in seconds) for the Hamiltonian circuit problem with an input graph of 200 nodes and 1800 edges. Encoding *constr* uses a constraint to state that each edge in the cycle should be an edge of the graph. Encoding *redun* adds redundancy to include this bound in all rules and constraints. Encoding *defin* contains no redundancy, but limits the possible edges in the cycle to the edges in the graph while defining the search space for the cycle.

is defined by a choice rule of the form, e.g., `{p(X)} :- q(X)`. However, if this rule is replaced by `{p(X)} :- dom(X)`, where `dom` denotes the domain, and the constraint `:- p(X),not q(X),dom(X)` is added, `q` is still a bound for `p`, but this is not detected by GRINGO, as can be seen in Table 5.

The grounder of the DLV system (Perri et al., 2007) may derive bounds by reasoning on the propositional level. As we explain below, the order in which rules and constraints are grounded is of crucial importance for such a method to pay off. Since DLV grounds rules before constraints, using a constraint to state that `q` is a bound for `p` does not improve grounding time.

**Propagation on the Propositional Level** One of the techniques to produce smaller groundings consists of applying a constraint propagation method on the ground theory $T_{\mathrm{g}}$ and replacing by $\top$, respectively $\bot$, every ground literal that is derived to be true, respectively false. The resulting theory is then simplified. This technique is applied by the grounder PSGRND (East et al., 2006), which uses *unit propagation* (Davis & Putnam, 1960) and complete *one-atom lookahead* (Li & Anbulagan, 1997) as propagation methods. The latter is performed once the grounding is finished, the former is triggered each time a unit clause is added to the grounding. If an inconsistency is detected by unit propagation, the grounding process is terminated immediately. Observe that this technique yields small groundings but does not improve grounding speed, except for the (rare) case where the propagation method detects an inconsistency during grounding. Indeed, it does not avoid computing all ground instances of the formulas in the input theory.

If a propositional constraint propagation method is applied while the grounding is being constructed, the derived information could be used to refine bounds. For instance, if unit-propagation derives that the domain atom $P(d_1, \ldots, d_n)$ is true, then $x_1 = d_1 \wedge \ldots \wedge x_n = d_n$ is a ct-bound for $P(x_1, \ldots, x_n)$. These bounds could be used to speed up the construction of the rest of the grounding. For this method to be effective, however, some careful fine-tuning of the order in which sentences are grounded is required. It may even be necessary to alternatingly compute partial groundings of different sentences. To the best of our knowledge, this process has not been worked out or implemented with unit-propagation or one-atom lookahead as underlying propagation method. On the other hand, most ASP grounders apply it for the following limited propagation method: if all rules defining a predicate $P$ are grounded, it is concluded that a domain atom $P(\overline{d})$ is certainly true if it occurs in a ground rule of the form $P(\overline{d}) \leftarrow \top$, and certainly false if it does not occur in the head of any ground rule. In this case, a good grounding order can be derived from the dependency graph of the input theory (e.g., Cadoli & Schaerf, 2005; Perri et al., 2007). In GIDL, this strategy is implemented for grounding definitions.

**Sharing** A second technique is called *sharing* and consists of detecting subformulas in the ground theory $T_{\mathrm{g}}$ that occur more than once. If such a subformula $\varphi$ is detected, all its occurrences in $T_{\mathrm{g}}$ are replaced by a new atom $P$, and the sentence $P \equiv \varphi$ is added. If $\varphi$ is a large formula and occurs often in $T_g$, this may result in a significant grounding size reduction. Also, sharing improves the propagation in SAT solvers.

Shlyakhter, Sridharan, Seater, and Jackson (2003) present an algorithm to detect identical subformulas on the first-order level, Torlak and Jackson (2007) for the propositional level. In GIDL, we implemented a simple sharing technique using dynamic programming. We adapted function `groundConj` so that instead of returning a conjunction $\bigwedge C$, it creates a new atom $P$, adds the sentence $P \equiv \bigwedge C$ to the grounding, and returns $P$. If `groundConj` is applied multiple times on the same input $\varphi$, the same predicate $P$ is returned each time, but $P \equiv \bigwedge C$ is added only once. Function `groundDisj` is adapted in a similar fashion.

**Clause splitting** Clause splitting is a well-known rewriting technique applied in MACE style model generation (McCune, 2003). It consists of splitting a first-order clause

$$\forall x \forall y \forall \overline{z}\ (\varphi_1[x, \overline{z}_1] \vee \varphi_2[y, \overline{z}_2]) \tag{20}$$

where $x \notin \overline{z}_2$, $y \notin \overline{z}_1$ and $\overline{z} = \overline{z}_1 \cup \overline{z}_2$ into two new clauses

$$\forall x \forall \overline{z}_1\ (\varphi_1[x, \overline{z}_1] \vee S(\overline{z}_1 \cap \overline{z}_2)) \tag{21}$$

$$\forall y \forall \overline{z}_2\ (\neg S(\overline{z}_1 \cap \overline{z}_2) \vee \varphi_2[y, \overline{z}_2]). \tag{22}$$

Here, $S$ is a new predicate symbol. The full grounding of (20) is of the size $\mathcal{O}(|D|^3)$, while the full grounding of (21) and (22) has only size $\mathcal{O}(|D|^2)$.

If sharing is implemented by adapting the functions `groundConj` and `groundDisj` as explained above, the effect of clause splitting can be obtained by moving quantifiers according to the equivalences (4), (5), (8) and (9) of Section 2.2. For instance, we can apply equivalences (4) and (8) to replace (20) by $\forall x \forall \overline{z}\ (\varphi_1 \vee (\forall y\ \varphi_2))$. Grounding the latter while applying sharing has the same effect as clause splitting. Similarly, the grounding size of $\exists x \exists y \exists \overline{z}\ (\varphi_1[x, \overline{z}_1] \wedge \varphi_2[y, \overline{z}_2])$ can be reduced by replacing this formula by $\exists x \exists \overline{z}\ (\varphi_1 \wedge (\exists y\ \varphi_2))$.

The simple heuristic to guide clause splitting described by Claessen and Sörensson (2003) can directly be applied to choose which quantifiers to move inside. We conclude that clause splitting could easily be incorporated in GIDL.

**Database Techniques** Several techniques for optimizing querying in databases can be used to optimize grounding. Examples are join-ordering strategies, backjumping and indexing techniques.

One of the most basic techniques to improve grounding speed consists of reordering (long) conjunctions or disjunctions of literals to speed up grounding. Which order is best depends on the grounding algorithm. Different strategies are described by, e.g, Leone, Perri, and Scarcello (2001), Syrjänen (1998, 2009) and in the database literature (Garcia-Molina, Ullman, & Widom, 2000). There is no problem implementing a similar technique in GIDL. Also, reordering the nodes in the BDD representation of the bounds could optimize querying. It is part of future work to investigate such reordering strategies for BDDs.

One of the important methods in the DLV grounder is the use of a backjumping technique (Perri et al., 2007) to efficiently find all instances of a conjunction $\varphi_1 \wedge \ldots \wedge \varphi_n$ that are possibly true, given (an overestimation of) the possibly true instances of each of the conjuncts $\varphi_i$. In GIDL, this backjumping technique is applied to implement line 12 of function `groundDisj`. Indeed, if $\varphi$ is the formula $\varphi_1 \wedge \ldots \wedge \varphi_n$, then line 12 amounts to finding all possible instances of $\varphi$, while the cf-bounds for $\varphi_1, \ldots, \varphi_n$ provide an overestimation of the possibly true instances of these conjuncts. Similarly, the backjumping technique is applied to improve line 12 of `groundConj`, where all possibly false instances of a disjunction are calculated.

Catalano, Leone, and Perri (2008) present an adaptation of indexing strategies for grounding.

**Partition-Based Reasoning** Ramachandran and Amir (2005) describe a sophisticated grounding technique that can reduce the grounding size of FO theories, depending on the availability of some graphical structure in these theories. This technique is not directly applicable in our case, since it produces groundings that are not necessarily $I_\sigma$-equivalent to the input theory. The only guarantee is that the ground theory is satisfiable iff the input problem is satisfiable.

### 7.2 Grounders

A non-native approach to ground an MX(FO(ID)) problem consists of first translating it to an equivalent normal logic program under the well-founded semantics. This translation is described by Mariën et al. (2004). Next, a (slightly adapted) grounder for ASP is used to ground the logic program. This is the approach taken by MXIDL (Mariën, Wittocx, & Denecker, 2006).

The first native grounding algorithm for MX(FO) and MX(FO(ID)) was described by Patterson, Liu, Ternovska, and Gupta (2007). It is based on relational algebra and takes a "bottom-up approach" (see Section 3.2.1). To construct a grounding of a sentence $\varphi$, it first creates all possible groundings of the atomic subformulas. Then it combines these groundings using relational algebra operations, working its way up the syntax tree. Finally, a grounding for $\varphi$ is obtained. Mitchell et al. (2006) report on an implementation, called MXG, of the algorithm.

KODKOD (Torlak & Jackson, 2007) is an MX grounder for a syntactic variant of FO. Like MXG, it works in a bottom-up way. It represents intermediate groundings by (sparse) matrices. One of the features of KODKOD is that it allows a user to give part of a solution to an MX problem as a three-valued structure. Specifically, the user can force that some atoms $P(\overline{d})$, where $P$ is an expansion predicate, are certainly true (or certainly false). KODKOD then takes advantage of this information to produce smaller groundings. GIDL also allows for a three-valued structure as input. When applying the refinement algorithm, the set of tuples $\overline{d}$ for which the user indicates that $P$ should be true is then used as initial ct-bound for $P$ instead of $\bot$. Similarly for the cf-bound. This leads to more efficient and compact groundings.

MACE (McCune, 2003) and PARADOX (Claessen & Sörensson, 2003) are finite model generators for FO. They work by choosing a domain and grounding the input theory to SAT. If the resulting grounding is unsatisfiable, the domain size is increased and the process is repeated. The grounding algorithm in MACE and PARADOX basically constructs the full grounding and simplifies it afterwards. Small groundings are obtained by first rewriting the input theory using, e.g., clause splitting. Also methods that build the grounding *incrementally* are applied in these systems to avoid recomputing every grounding from scratch.

East et al. (2006) developed the grounder PSGRND for $MX(PS^{pb})$. $PS^{pb}$ is a fragment of FO(ID), extended with pseudo-boolean constraints. As explained above, PSGRND performs reasoning on the ground theory to reduce memory usage and grounding size. The experiments performed by East et al. (2006) show that carefully designed data structures are of key importance to build an efficient grounder.

ASP grounders take as input a normal logic program and transform it into an equivalent ground normal logic program. As such, these grounders do not deal with (deeply) nested formulas. Currently, there are three ASP grounders: LPARSE (Syrjänen, 2000; Syrjänen, 2009), GRINGO (Gebser et al., 2007) and the grounding component of DLV (Perri et al., 2007). All of them use techniques from database theory to perform grounding efficiently.

Finally, we mention the grounder SPEC2SAT (Cadoli & Schaerf, 2005). Its input theories are in the NP-SPEC language, a language with Datalog-like syntax and semantics based on model minimality. The grounding algorithm implemented in SPEC2SAT is basically a simplified version of the grounding algorithm of DLV.

It would be interesting to compare the efficiency of the above mentioned grounders experimentally. However, it is currently not possible to conduct such an experiment in a scientifically fair way. There are several reasons for this. First, all grounders have a different input language, making it impossible to run them on the same input. Also, there are several output languages for grounders. A richer output language leads to more compact and fast grounding. For instance, for some prob-lems, lparse's output size is necessarily cubic in the input domain size, while GidL's output format allows for quadratic size. Thirdly, even if the input and output languages of all grounders were the same, an expert could easily manipulate experiments by carefully choosing his modelling style. For example, if he does not manually add bounds to the input theories, GidL has an advantage. If bodies of rules are not ordered, dlv is more likely to produce good results. Etc. Finally, because of the large amount of data processed by grounders, carefully designed data structures and an optimized implementation of the core grounding algorithm is very important to achieve fast grounding (East et al., 2006). However, several of the above mentioned grounders are not yet optimized in that sense. As such, it is difficult to derive conclusions about grounding *algorithms* by experimentally comparing the efficiency of current *implementations* of these algorithms.

## 8. Conclusions

We presented a method to compute for a given theory, upper and lower bounds for all subformulas of that theory. We showed how these bounds can be used for efficiently creating small groundings in the context of Model Expansion for FO and FO(ID). Our method frees a user from manually discovering bounds and adding them to a theory.

We presented a top-down style grounding algorithm that incorporates bounds. We discussed implementation issues and showed by experiments that our method works in practice: on many benchmark problems, it leads to significant reductions in grounding size and time.

Future work includes the extension of our algorithm to compute bounds for richer logics, such as, e.g., extensions of FO with aggregates and arithmetic. On the implementation side, we plan to use more sophisticated estimators to evaluate whether a computed bound is beneficial for grounding.

## Acknowledgments

Research supported by Research Foundation-Flanders (FWO-Vlaanderen) and by GOA 2003/08 "Inductive Knowledge Bases". Johan Wittocx is research assistant of the Research Foundation-Flanders (FWO-Vlaanderen).